\documentclass[aps,prd,showpacs,onecolumn,twoside]{revtex4-2}
\usepackage{amssymb}
\usepackage{mathrsfs}
\usepackage{amsfonts}
\usepackage{epstopdf}
\usepackage{bm,color,bbm}
\usepackage{amsmath}
\usepackage{graphicx}
\usepackage{natbib}
\usepackage[hidelinks]{hyperref}
\usepackage{caption}
\usepackage{subcaption}

\usepackage{fancyhdr}
\providecommand{\U}[1]{\protect\rule{.1in}{.1in}}

\begin{document}
\title{\textbf{Geometric Aspects of Entanglement Generating Hamiltonian Evolutions}}
\author{\textbf{Carlo Cafaro}$^{1}$ and \textbf{James Schneeloch}$^{2}$}
\affiliation{$^{1}$University at Albany-SUNY, Albany, NY 12222, USA}
\affiliation{$^{2}$Air Force Research Laboratory, Information Directorate, Rome, New York,
13441, USA}

\begin{abstract}
We examine the pertinent geometric characteristics of entanglement that arise
from stationary Hamiltonian evolutions transitioning from separable to
maximally entangled two-qubit quantum states. From a geometric perspective,
each evolution is characterized by means of geodesic efficiency, speed
efficiency, and curvature coefficient. Conversely, from the standpoint of
entanglement, these evolutions are quantified using various metrics, such as
concurrence, entanglement power, and entangling capability.

Overall, our findings indicate that time-optimal evolution trajectories are
marked by high geodesic efficiency, with no energy resource wastage, no
curvature (i.e., zero bending), and an average path entanglement that is less
than that observed in time-suboptimal evolutions. Additionally, when analyzing
separable-to-maximally entangled evolutions between nonorthogonal states,
time-optimal evolutions demonstrate a greater short-time degree of nonlocality
compared to time-suboptimal evolutions between the same initial and final
states. Interestingly, the reverse is generally true for
separable-to-maximally entangled evolutions involving orthogonal states. Our
investigation suggests that this phenomenon arises because suboptimal
trajectories between orthogonal states are characterized by longer path
lengths with smaller curvature, which are traversed with a higher energy
resource wastage compared to suboptimal trajectories between nonorthogonal
states. Consequently, a higher initial degree of nonlocality in the unitary
time propagators appears to be essential for achieving the maximally entangled
state from a separable state. Furthermore, when assessing optimal and
suboptimal evolutions from initial to final states while maintaining a
constant degree of entanglement change, irrespective of whether they are
orthogonal, the average speed of path entanglement is inversely related to the
duration of travel and, additionally, directly related to the variation in
concurrence. In summary, dynamic paths that exhibit greater (instantaneous)
evolution speeds are also defined by higher (average) path entanglement speeds.

\end{abstract}

\pacs{Quantum Computation (03.67.Lx), Quantum Information (03.67.Ac), Quantum
Mechanics (03.65.-w).}
\maketitle
\thispagestyle{fancy}

\section{Introduction}

The relationship between nonlocality and entanglement in quantum theory is
nontrivial \cite{peres91,bennett99,halder19,banik20}. For instance, entangled
states exhibit nonlocality as they violate Bell-type inequalities (known as
Bell nonlocality \cite{bell64,clauser69,aspect82,cafaroijtp}). Nevertheless,
nonlocal characteristics are not exclusive to entangled systems. Specifically,
one might anticipate that if the states of a quantum system were confined to a
set of orthogonal product states, the system would function entirely in a
classical manner and would not display any nonlocality. Moreover, it should be
feasible to ascertain the state of the system using only local measurements.
Surprisingly, this is not the case. In Ref. \cite{bennett99}, Bennett and his
colleagues devised sets of product states that cannot be precisely
distinguished through local operations and classical communication (LOCC),
even though their mutual orthogonality guarantees perfect global
discrimination. The observation that nonlocal behavior in quantum theory does
not always require entanglement also arises when examining quantum operations.
Notably, while the SWAP operation transforms product states into product
states, its nonlocal content is maximal \cite{collins01}. When concentrating
on quantum evolutions defined by nonlocal Hamiltonians \cite{dur01}, it
becomes particularly important to comprehend how the level of nonlocality of
the operation influences its capacity for entanglement (i.e., its potential to
transition states with a lower degree of entanglement to states with a higher
degree of entanglement).

\medskip

The significance of entanglement in speeding up the dynamical evolutions of
composite quantum systems in pure states was initially documented in Refs.
\cite{maccone03,maccone03b,maccone04}. Regardless of whether the evolution
occurs between orthogonal \cite{maccone03} or nonorthogonal \cite{maccone03b}
configurations, it has been determined that in the absence of interactions
among subsystems, and when the initial state of the composite system is
separable, the quantum speed limit can only be attained in the asymmetric
scenario where only one of the subsystems undergoes evolution over time and
possesses all the energy resources of the system. Conversely, the existence of
entanglement in the initial state facilitates a dynamical speedup even when
the energy resources are uniformly distributed across the subsystems. In
particular, Hamiltonians that generate entanglement can speed up the dynamics,
even when beginning from separable configurations \cite{maccone03}. The
phenomenon of entanglement accelerating quantum dynamics was further
investigated in Refs.
\cite{batle05,batle06,borras06,curilef06,zander07,curilef08}. In particular,
this phenomenon was analyzed for two-qubit systems and systems of two
identical bosons in Ref. \cite{batle05}. Conversely, Ref. \cite{curilef06}
examined the relationship between entanglement and the speed of quantum
evolution for two quantum particles in a one-dimensional double well. Although
many of these findings are only partially conclusive \cite{chau10}, as they
stem from numerical methods and analytic calculations for specific scenarios,
the complexity increases for mixed states. For instance, in the case of pure
states, a higher degree of entanglement in the initial state correlates with a
quicker evolution to an orthogonal state. Nevertheless, in the case of mixed
states, there is no conclusive proof that the rate of evolution speeds up
solely due to increased entanglement. In fact, the speed may increase with
rising entanglement, but this is also observed in other circumstances
\cite{reznik08}. In Ref. \cite{frowis12}, it was demonstrated that when
examining the unitary evolution defined by stationary Hamiltonians of
generally mixed states, not all forms of entanglement are beneficial for
accelerating time evolution in comparison to non-entangled states.
Specifically, it was noted that a significant (time-independent) quantum
Fisher information is required to speed up the time evolution relative to
non-entangled states. In Ref. \cite{rudniki21}, the geometric measure of
entanglement was associated with the minimum time required to render a pure
state separable through a unitary evolution defined by a time-independent
nonlocal Hamiltonian. In Ref. \cite{pati22}, the notion of entanglement
capacity was established for any bipartite pure state as the variance of the
modular Hamiltonian within the reduced state of any subsystem of the composite
quantum system. Specifically, it was demonstrated that the quantum speed limit
for generating entanglement is influenced not only by the fluctuations in the
nonlocal Hamiltonian but also inversely related to the time average of the
square root of the entanglement capacity. Lastly, we recommend Refs.
\cite{pati23,pandey24} for recent compelling studies regarding the speed limit
of entanglement dynamics, which is considered the maximum rate at which
entanglement can be created or diminished in bipartite quantum systems.

\medskip

Intriguing studies focusing on linking the geometry of quantum state spaces to
entanglement measures defined by negativity, can be found in Refs.
\cite{deb16,deb19}. Specifically, when an arbitrary qubit is examined and
entanglement is established between the qubit and an ancilla qubit through the
application of interaction, the negativity of entanglement is, up to a
constant factor, equivalent to the square root of a Riemannian metric. For
instance, in the specific case of Ref. \cite{deb16}, the metric in question is
the Wigner-Yanase skew-information \cite{luo03}. In Ref. \cite{kuzmak19}, the
geometric characteristics of two-qubit quantum state spaces utilizing the
Fubini-Study metric were examined. More precisely, these state spaces were
produced through the application of unitary time propagators that correspond
to time-independent parametric Hamiltonians acting on specified initial
states. Particular attention was given to the impact on the scalar curvature
of the state spaces, along with the level of entanglement (measured through
the concurrence) of the states. In Ref. \cite{saleem25}, the concept of
\textquotedblleft curvature of entanglement\textquotedblright\ was
characterized as the second derivative of the concurrence with respect to the
coupling parameter. This parameter was presumed to be independent of both
space and time and, furthermore, delineated the interaction between two
qubits. The primary finding of Ref. \cite{saleem25} was a fascinating
connection between the instantaneous quantum Fisher information and the
curvature of entanglement of a quantum probe characterized by the interaction
of two qubits through a Jaynes-Cummings Hamiltonian, while also being
influenced by Markovian noise.

\medskip

Motivated by the complex relationship between nonlocality and entanglement,
driven by the absence of a comprehensive geometric description of entanglement
in quantum evolutions, and spurred by the increasing interest in linking the
geometry of quantum state spaces to measures of entanglement, this paper seeks
to investigate the relevant geometric properties of entanglement that emerge
from stationary Hamiltonian evolutions transitioning from separable to
maximally entangled two-qubit quantum states through the lens of geodesic
efficiency \cite{anandan90,carlini06,carlini07,carlocqg23,rossetti24}, speed
efficiency \cite{uzdin12,rossetti25}, and the curvature coefficient of quantum
evolution \cite{alsing24A,alsing24B,cafaropra25}. A partial list of questions
being addressed includes:

\begin{enumerate}
\item[{[i]}] Is it possible to create a parametric family of both optimal and
suboptimal stationary Hamiltonians that connect any separable quantum state to
a maximally entangled quantum state, given that the evolution is confined to
the two-dimensional subspace defined by the source and target states?

\item[{[ii]}] If suboptimal evolutions with constant Hamiltonians between
orthogonal states are not permitted in two dimensions, is it possible to
create ad hoc examples of such evolutions in higher-dimensional subspaces of
the complete Hilbert space for two-qubit quantum states?

\item[{[iii]}] In the context of evolutions between identical pairs of
separable and maximally entangled quantum states, are time optimal evolutions
typically defined by unitary time propagators that exhibit a level of
nonlocality exceeding that of their suboptimal counterparts?

\item[{[iv]}] Do dynamical trajectories that exhibit a higher level of
geodesic and speed efficiency also display a greater degree of average path entanglement?

\item[{[v]}] Do transitions between orthogonal states necessitate greater
levels of entanglement in comparison to transitions between nonorthogonal
states under the same energy constraints?

\item[{[vi]}] What are the primary distinctions between suboptimal evolutions
in lower-dimensional and higher-dimensional subspaces regarding curvature,
energy dissipation, and entanglement capability?
\end{enumerate}

\medskip

The significance of this analysis is twofold concerning quantum control
strategies. Firstly, it broadens the geometric characterization of quantum
evolutions by incorporating concepts of efficiency and curvature beyond the
dynamics of single qubits, thereby allowing for the examination of the
influence of nonlocality and entangling capability on these metrics. Secondly,
it provides a framework for comprehending the entanglement properties of both
quantum states and unitary time propagators, focusing on their nonlocal
characteristics and entangling capabilities, by exploring their
interrelationship in connection to their impacts on time-optimality, energy
consumption, and curvature of quantum evolutions.

\medskip

The remainder of this paper is structured as follows. In Section II, we
present the fundamental components utilized to characterize the entanglement
of a two-qubit quantum state, the nonlocal nature of the unitary time
propagator, and its overall ability to generate entanglements from separable
states. Specifically, we examine the concepts of concurrence \cite{wootters98}%
, geometric measure of entanglement \cite{shimony95}, entanglement production
\cite{yukalov15}, and entangling power \cite{zanardi00}. In Section III, we
introduce the tools employed to describe certain geometric aspects of quantum
evolutions, including geodesic efficiency \cite{anandan90}, speed efficiency
\cite{uzdin12}, and curvature coefficient \cite{alsing24A,alsing24B}. In
Section IV, we present the concepts of time optimal and time suboptimal
Hamiltonian evolutions between any pair of quantum states that define any
finite-dimensional quantum systems. Notably, we emphasize that while the time
optimal scheme is applicable to both orthogonal and nonorthogonal pairs of
initial and final states, the suboptimal scheme is limited to pairs of
nonorthogonal states. In Section V, concentrating on two-qubit stationary
evolutions between separable and maximally entangled states, we investigate
four types of Hamiltonian evolutions using both geometric and entanglement
quantifiers. The four scenarios we examine are: i) Time optimal evolutions
between nonorthogonal quantum states; ii) Time suboptimal quantum evolutions
between nonorthogonal states; iii) Time optimal evolutions between orthogonal
quantum states; iv) Time suboptimal quantum evolutions between orthogonal
states. In Section VI, we focus on links among time optimality, nonlocal
character of unitary time propagators, and entanglement capabilities of time
optimal evolutions with a different degree of energy wastefulness. Finally,
our summary of findings along with our concluding remarks are presented in
Section VII.

\section{Entanglement measures}

In this section, we detail the essential elements that we utilize to
characterize the entanglement of a two-qubit quantum state, the nonlocal
attributes of the unitary time propagator, and its overall capacity to produce
entanglements from separable states. In particular, we explore the notions of
concurrence, geometric measure of entanglement, entanglement production, and
entangling power.

\subsection{Concurrence}

To measure the extent of entanglement in a pure two-qubit quantum state
$\left\vert \psi\right\rangle $, we employ the concept of concurrence
\cite{wootters97,wootters98,wootters01}. This measure of entanglement is
defined as,%
\begin{equation}
\mathrm{C}\left[  \left\vert \psi\right\rangle \right]  \overset{\text{def}%
}{=}\left\langle \psi\left\vert \sigma_{y}\otimes\sigma_{y}\right\vert
\psi^{\ast}\right\rangle \text{.} \label{conc1}%
\end{equation}
It is important to note that $\left\vert \psi^{\ast}\right\rangle $ in Eq.
(\ref{conc1}) represents the complex conjugate of $\left\vert \psi
\right\rangle $ within the computational basis, while $\sigma_{y}$ refers to
the standard Pauli operator. It is noteworthy that if the normalized two-qubit
state $\left\vert \psi\right\rangle $ is expressed in terms of the
computational basis $\left\{  \left\vert 00\right\rangle \text{, }\left\vert
01\right\rangle \text{, }\left\vert 10\right\rangle \text{, }\left\vert
11\right\rangle \right\}  $ as $\left\vert \psi\right\rangle =\alpha\left\vert
00\right\rangle +\beta\left\vert 01\right\rangle +\gamma\left\vert
10\right\rangle +\delta\left\vert 11\right\rangle $, the concurrence
$\mathrm{C}\left[  \left\vert \psi\right\rangle \right]  $ in Eq.
(\ref{conc1}) simplifies to $\mathrm{C}\left[  \left\vert \psi\right\rangle
\right]  =2\left\vert \alpha\delta-\beta\gamma\right\vert $. Alternatively, if
the state $\left\vert \psi\right\rangle $ is reformulated in terms of the
magic basis defined as \cite{collins11}%
\begin{equation}
\left\{  \left\vert \Phi_{1}\right\rangle \overset{\text{def}}{=}\left\vert
\Phi^{+}\right\rangle \text{, }\left\vert \Phi_{2}\right\rangle \overset
{\text{def}}{=}-i\left\vert \Phi^{-}\right\rangle \text{, }\left\vert \Phi
_{3}\right\rangle \overset{\text{def}}{=}\left\vert \Psi^{-}\right\rangle
\text{, }\left\vert \Phi_{4}\right\rangle \overset{\text{def}}{=}-i\left\vert
\Psi^{+}\right\rangle \right\}  \text{,}%
\end{equation}
where $\left\vert \Phi^{\pm}\right\rangle \overset{\text{def}}{=}(1/\sqrt
{2})\left(  \left\vert 00\right\rangle \pm\left\vert 11\right\rangle \right)
$ and $\left\vert \Psi^{\pm}\right\rangle \overset{\text{def}}{=}(1/\sqrt
{2})\left(  \left\vert 01\right\rangle \pm\left\vert 10\right\rangle \right)
$ represent the four maximally entangled Bell states, expressed as $\left\vert
\psi\right\rangle =\mu_{1}\left\vert \Phi_{1}\right\rangle +\mu_{2}\left\vert
\Phi_{2}\right\rangle +\mu_{3}\left\vert \Phi_{3}\right\rangle +\mu
_{4}\left\vert \Phi_{4}\right\rangle $, the concurrence $\mathrm{C}\left[
\left\vert \psi\right\rangle \right]  $ as defined in Eq. (\ref{conc1})
simplifies to $\mathrm{C}\left[  \left\vert \psi\right\rangle \right]
=\left\vert \mu_{1}^{2}+\mu_{2}^{2}+\mu_{3}^{2}+\mu_{4}^{2}\right\vert $.
Furthermore, let us assume that the Schmidt decomposition of $\left\vert
\psi\right\rangle $ is provided by
\begin{equation}
\left\vert \psi\right\rangle =\sum_{k=1}^{2}\sqrt{\lambda_{k}}\left\vert
v_{k}^{\left(  A\right)  }\right\rangle \otimes\left\vert v_{k}^{\left(
B\right)  }\right\rangle \text{,} \label{se}%
\end{equation}
with $\left\{  \sqrt{\lambda_{k}}\right\}  _{k=1,2}$ representing the Schmidt
coefficients, where $\lambda_{1}+\lambda_{2}=1$ and $\lambda_{k}\in%
\mathbb{R}
_{+}$ for any $k$. Consequently, for two qubits, the concurrence
$\mathrm{C}\left[  \left\vert \psi\right\rangle \right]  $ in Eq.
(\ref{conc1}) simplifies to $\mathrm{C}\left[  \left\vert \psi\right\rangle
\right]  =2\Lambda_{\max}\sqrt{1-\Lambda_{\max}^{2}}$, where $\Lambda_{\max}$
denotes the maximum Schmidt coefficient. Notably, $\left\{  \lambda
_{k}\right\}  _{k=1,2}$ are the eigenvalues of the reduced density operators
$\rho_{A}=\mathrm{Tr}_{B}\left(  \left\vert \psi\right\rangle \left\langle
\psi\right\vert \right)  $ and $\rho_{B}=\mathrm{Tr}_{A}\left(  \left\vert
\psi\right\rangle \left\langle \psi\right\vert \right)  $. Lastly, it is
important to mention that the number of terms in the expansion in Eq.
(\ref{se}) is referred to as the Schmidt number. For a separable state, the
Schmidt number is consistently $1$. In contrast, an entangled state has a
Schmidt number greater than $1$.

In the next subsection, we introduce an alternative measure of entanglement of
pure states.

\subsection{Geometric measure of entanglement}

The geometric measure of entanglement for pure states, as introduced by
Shimony in Ref. \cite{shimony95}, employs fundamental concepts of Hilbert
space geometry to assess the extent of entanglement in pure states. It is
characterized as the minimal squared distance between an entangled state
$\left\vert \psi\right\rangle $ and the collection of separable states
$\left\vert \phi\right\rangle $,%
\begin{equation}
\mathrm{E}_{\mathrm{geo}}\left[  \left\vert \psi\right\rangle \right]
\overset{\text{def}}{=}\underset{\left\vert \phi\right\rangle }{\min
}\left\Vert \left\vert \psi\right\rangle -\left\vert \phi\right\rangle
\right\Vert ^{2}\text{.} \label{geom}%
\end{equation}
With respect to the squared Fubini-Study distance \cite{caves94},
$\mathrm{E}_{\mathrm{geo}}\left[  \left\vert \psi\right\rangle \right]  $ in
Eq. (\ref{geom}) can be reformulated as%
\begin{equation}
\mathrm{E}_{\mathrm{geo}}\left[  \left\vert \psi\right\rangle \right]
=\underset{\left\vert \phi\right\rangle }{\min}\left[  1-\left\vert
\left\langle \psi\left\vert \phi\right.  \right\rangle \right\vert
^{2}\right]  =1-\underset{\left\vert \phi\right\rangle }{\max}\left[
\left\vert \left\langle \psi\left\vert \phi\right.  \right\rangle \right\vert
^{2}\right]  \text{.} \label{geom2}%
\end{equation}
Although its definition is straightforward, calculating $\mathrm{E}%
_{\mathrm{geo}}\left[  \left\vert \psi\right\rangle \right]  $ in
Eq.(\ref{geom2}) necessitates a complex minimization process across all
separable states. Notably, when we define the entanglement eigenvalue
$\Lambda_{\max}$ as $\Lambda_{\max}\overset{\text{def}}{=}\underset{\left\vert
\phi\right\rangle }{\max}\left[  \left\vert \left\langle \psi\left\vert
\phi\right.  \right\rangle \right\vert \right]  $, it is revealed that for
bipartite systems, $\Lambda_{\max}$ corresponds to the largest Schmidt
coefficient found in the Schmidt decomposition of $\left\vert \psi
\right\rangle $ \cite{gold03}. Consequently, in this scenario, $\mathrm{E}%
_{\mathrm{geo}}\left[  \left\vert \psi\right\rangle \right]  $ simplifies to
$\mathrm{E}_{\mathrm{geo}}\left[  \left\vert \psi\right\rangle \right]
=1-\Lambda_{\max}^{2}$. Lastly, considering that the concurrence
\textrm{C}$\left[  \left\vert \psi\right\rangle \right]  $ of $\left\vert
\psi\right\rangle $ is represented as $2\Lambda_{\max}\sqrt{1-\Lambda_{\max
}^{2}}$, the connection between $\mathrm{E}_{\mathrm{geo}}\left[  \left\vert
\psi\right\rangle \right]  $ and \textrm{C}$\left[  \left\vert \psi
\right\rangle \right]  $ is expressed as $\mathrm{E}_{\mathrm{geo}}\left[
\left\vert \psi\right\rangle \right]  =(1/2)\left[  1-\sqrt{1-\mathrm{C}%
^{2}\left[  \left\vert \psi\right\rangle \right]  }\right]  $.

Considering the evident connection between concurrence and the geometric
measure of entanglement, we will utilize concurrence to quantify the degree of
entanglement in a pure state in the subsequent sections. Conversely, in the
following subsection, we will introduce an entanglement measure that is
appropriate for assessing the propensity of an operator to produce entangled
states from separable states.

\subsection{Entanglement production}

In the following subsection, we will provide a brief introduction to the
concept of entanglement production, as originally presented by Yukalov and
collaborators in Refs. \cite{yukalov03,yukalov03b,yukalov15,yukalov17}.
Specifically, our focus here will be on Ref. \cite{yukalov15}. We will examine
a bipartite quantum system defined by the Hilbert space $\mathcal{H}%
\overset{\text{def}}{=}\mathcal{H}_{A}\otimes\mathcal{H}_{B}$, where
$\mathcal{H}_{A}\overset{\text{def}}{=}\mathrm{Span}\left\{  \left\vert
n\right\rangle \right\}  _{n=1\text{,..., }n_{A}}$, $\mathcal{H}_{B}%
\overset{\text{def}}{=}\mathrm{Span}\left\{  \left\vert \alpha\right\rangle
\right\}  _{\alpha=1\text{,..., }n_{B}}$, and $\mathcal{H}\overset{\text{def}%
}{=}\mathrm{Span}\left\{  \left\vert n\right\rangle \otimes\left\vert
\alpha\right\rangle \right\}  _{n=1\text{,..., }n_{A}}^{\alpha=1\text{,...,
}n_{B}}$. The dimensions are given by $\dim\mathcal{H}_{A}\overset{\text{def}%
}{=}n_{A}$, $\dim\mathcal{H}_{B}\overset{\text{def}}{=}n_{B}$, and
$\dim\mathcal{H}\overset{\text{def}}{=}n_{A}n_{B}$. We will consider a unitary
evolution operator represented by $U\left(  t\right)  =e^{-\frac{i}{\hslash
}\mathrm{H}t}$ that operates on $\mathcal{H}\overset{\text{def}}{=}%
\mathcal{H}_{A}\otimes\mathcal{H}_{B}$, where \textrm{H} denotes a
time-independent Hamiltonian for the composite quantum system. Consequently,
the entanglement production of $U(t)$ is defined as%
\begin{equation}
\varepsilon_{\mathrm{EP}}^{\mathrm{Yukalov}}\left[  U(t)\right]
\overset{\text{def}}{=}\log\left[  \frac{\left\Vert U(t)\right\Vert
}{\left\Vert U^{\otimes}(t)\right\Vert }\right]  \text{,} \label{russia}%
\end{equation}
where $\left\Vert U(t)\right\Vert $ is the Hilbert-Schmidt norm of $U(t)$
given by $\left\Vert U(t)\right\Vert \overset{\text{def}}{=}\sqrt
{\mathrm{Tr}_{\mathcal{H}}\left[  U^{\dagger}(t)U(t)\right]  }$. To ensure
thoroughness, we highlight that\textbf{ }$\left\Vert U(t)\right\Vert $ is
equal to one for unitary operators $U(t)$. Nevertheless, the definition of
$\varepsilon_{\mathrm{EP}}^{\mathrm{Yukalov}}\left[  U(t)\right]  $ is
applicable to broader scenarios where unitarity is not a strict requirement.
The so-called nonentangling operator $U^{\otimes}(t)$ in Eq. (\ref{russia}) is
defined as%
\begin{equation}
U^{\otimes}(t)\overset{\text{def}}{=}\frac{U_{A}(t)\otimes U_{B}%
(t)}{\mathrm{Tr}_{\mathcal{H}}\left[  U(t)\right]  }=\frac{\mathrm{Tr}%
_{\mathcal{H}_{B}}\left[  U(t)\right]  \otimes\mathrm{Tr}_{\mathcal{H}_{A}%
}\left[  U(t)\right]  }{\mathrm{Tr}_{\mathcal{H}}\left[  U(t)\right]
}\text{.} \label{russia1}%
\end{equation}
The operators $U_{A}(t)$ and $U_{B}(t)$ in\ Eq. (\ref{russia1}) are the
partial evolution operators used to construct the nonentangling operator
$U^{\otimes}(t)$. Note that the logarithm in Eq. (\ref{russia}) can be taken
with respect to any convenient base.

More broadly, the concept underlying the definition of this measure in Eq.
(\ref{russia}) is to evaluate the action of a specific operator $A$ in
comparison to that of its non-entangling equivalent $A^{\otimes}$. A
non-entangling operator applied to a subset $\mathcal{D}$ of the complete
Hilbert space $\mathcal{H}$ is defined as an operator that transforms any
separable state within $\mathcal{D}$ into a separable state also within
$\mathcal{D\subseteq H}$. Conversely, an entangling operator on $\mathcal{D}$
is characterized by the existence of at least one element in $\mathcal{D}$
that is transformed into an entangled state in $\mathcal{H}\backslash
\mathcal{D}$. A universal entangling operator on $\mathcal{D}$ is defined as
one that causes any element of $\mathcal{D}$ to become entangled when acted
upon by the universal entangling operator. It is evident that the SWAP
operator and tensor products of local operators serve as examples of operators
that maintain separability. The idea of establishing this measure for
entanglement production in Eq. (\ref{russia}) stems from the notion of
operator order indices within the realm of statistical mechanics
\cite{coleman00,yukalov02}. These indices are articulated through the norms of
operators. Furthermore, given that the order structure of composite quantum
systems displaying interparticle correlations aligns well with entangled
systems, entanglement production was initially introduced by Yukalov in Ref.
\cite{yukalov03}. Considering the difference between the \textquotedblleft
static\textquotedblright\ concepts of state entanglement and the nonlocal
characteristics of an operator, Yukalov's concept of entanglement production
is a \textquotedblleft dynamic\textquotedblright\ idea that illustrates how an
operator can generate entangled states from separable ones. Its operational
significance demonstrates the capacity of an operator to create entangled
state vectors within the Hilbert space on which it operates. This concept is
not restricted to unitary operators, bipartite systems, and pure states;
rather, it also encompasses nonunitary operators, multipartite systems, and
mixed states. The measure of entanglement production is zero for non
entangling operators. Furthermore, it is continuous in terms of norm
convergence, additive, invariant under local unitary transformations, and
semipositive. In particular, the characteristics of the Hilbert-Schmidt norm
play an important role in demonstrating that the entanglement production meets
all the fundamental criteria of a well-defined entanglement measure. For
completeness, we point out here that the Hilbert-Schmidt norm of operators is
extensively utilized in the literature for characterizing entanglement
\cite{witte99,ovrum06,yuka20,pandya20,siewert22}. For example, in Ref.
\cite{witte99}, the Hilbert-Schmidt entanglement of a state is defined as the
minimum distance, as determined by the Hilbert-Schmidt norm, between the
density matrix representing the state and the collection of all disentangled
states. Due to the practical challenges associated with calculating the values
of most entanglement measures, Ref. \cite{pandya20} employed the
Hilbert-Schmidt distance to develop an entanglement witness aimed at simply
certifying the presence of entanglement in the state, while disregarding its
specific entanglement content. For further details on entanglement production,
we suggest Refs. \cite{yukalov03,yukalov03b,yukalov15,yukalov17}.

In the next subsection, we introduce an alternative to Yukalov's entanglement production.

\subsection{Entangling power}

\begin{table}[t]
\centering
\begin{tabular}
[c]{c|c|c|c}\hline\hline
\textbf{Operator}, $U$ & \textbf{Weyl chamber location}, $\left(  c_{1}\text{,
}c_{2}\text{, }c_{3}\right)  $ & \textbf{Entangling power}, $\varepsilon
_{\mathrm{EP}}^{\mathrm{Zanardi}}\left(  U\right)  $ & \textbf{Schmidt
number}, \textrm{Sch}$\left(  U\right)  $\\\hline
\textrm{Identity} & $\left(  0\text{, }0\text{, }0\right)  $ & $0$ &
$1$\\\hline
\textrm{CNOT} & $\left(  \pi/4\text{, }0\text{, }0\right)  $ & $2/9$ &
$2$\\\hline
\textrm{SWAP} & $\left(  \pi/4\text{, }\pi/4\text{, }\pi/4\right)  $ & $0$ &
$4$\\\hline
$\sqrt{\mathrm{SWAP}}$ & $\left(  \pi/8\text{, }\pi/8\text{, }\pi/8\right)  $
& $1/6$ & $2$\\\hline
\textrm{DCNOT} & $\left(  \pi/4\text{, }\pi/4\text{, }0\right)  $ & $2/9$ &
$4$\\\hline
\end{tabular}
\caption{Tabular summary illustrating the the Weyl chamber position, the
entangling capability, and the Schmidt number associated with standard
two-qubit quantum gates.}%
\end{table}

Zanardi's entangling power defines the entangling ability of unitary
operators. It is characterized as the average entanglement that the unitary
operator can generate when applied to all separable states, which are
distributed according to a specific probability density over the manifold of
product states.

Consider a bipartite quantum system characterized by the state space
$\mathcal{H}=\mathcal{H}_{A}\otimes\mathcal{H}_{B}$, where $\dim_{%
\mathbb{C}
}\mathcal{H}_{A}=d_{A}$ and $\dim_{%
\mathbb{C}
}\mathcal{H}_{B}=d_{B}$. Note that the subscript \textquotedblleft$%
\mathbb{C}
$\textquotedblright\ denotes the field of complex numbers. Let $U$ represent a
unitary operator that acts on $\mathcal{H}$, and let $\mathcal{E}$ denote an
entanglement measure defined over $\mathcal{H}$. Consequently, in relation to
$\mathcal{E}$, the entangling power of $U$ is defined as%
\begin{equation}
\varepsilon_{\mathrm{EP}}^{\mathrm{Zanardi}}\left(  U\right)  \overset
{\text{def}}{=}\overline{\mathcal{E}\left[  U\left(  \left\vert \psi
_{A}\right\rangle \otimes\right)  \left\vert \psi_{B}\right\rangle \right]
}\text{,} \label{epower}%
\end{equation}
where the bar in Eq. (\ref{epower}) signifies the average across all product
states $\left\vert \psi_{A}\right\rangle \otimes\left\vert \psi_{B}%
\right\rangle $ that are distributed according to a certain probability
density $p=p\left(  \left\vert \psi_{A}\right\rangle \text{, }\left\vert
\psi_{B}\right\rangle \right)  $ over the manifold of product states
\cite{zanardi00,zanardi01,wang02}. The selected entanglement measure for
$\left\vert \psi\right\rangle \in\mathcal{H}$ is the linear entropy
$\mathcal{E}\left[  \left\vert \psi\right\rangle \right]  $ defined as
$\mathcal{E}\left[  \left\vert \psi\right\rangle \right]  \overset{\text{def}%
}{=}1-\mathrm{Tr}\left[  \rho_{A}^{2}\right]  =1-\mathrm{Tr}\left[  \rho
_{B}^{2}\right]  $, where $\rho_{A}\overset{\text{def}}{=}\mathrm{Tr}%
_{B}\left[  \left\vert \psi\right\rangle \left\langle \psi\right\vert \right]
$ and $\rho_{B}\overset{\text{def}}{=}\mathrm{Tr}_{A}\left[  \left\vert
\psi\right\rangle \left\langle \psi\right\vert \right]  $. It is important to
note that $\mathcal{E}\left[  \left\vert \psi\right\rangle \right]  $
quantifies the lack of purity or, alternatively, the degree of mixedness of
the reduced density matrix $\rho_{A}$, with $\mathrm{Tr}\left[  \rho_{A}%
^{2}\right]  $ being the purity of\textbf{ }$\rho_{A}$.

For systems involving two qubits, any unitary operator $U$ belonging to
$\mathrm{SU}\left(  4\right)  $ can be expressed in a canonical form as
\cite{reza04,bala10}%
\begin{equation}
U=\left(  A_{1}\otimes B_{1}\right)  e^{-i\left(  c_{1}\sigma_{1}\otimes
\sigma_{1}+c_{2}\sigma_{2}\otimes\sigma_{2}+c_{3}\sigma_{3}\otimes\sigma
_{3}\right)  }\left(  A_{2}\otimes B_{2}\right)  \text{,} \label{Q1}%
\end{equation}
where $\mathbf{\boldsymbol{\sigma}}\overset{\text{def}}{=}\left(  \sigma
_{1}\text{, }\sigma_{2}\text{, }\sigma_{3}\right)  $ is the vector of Pauli
spin matrices, and $A_{i}$, $B_{i}\in\mathrm{SU}\left(  2\right)  $ are
single-qubit unitary operators. Interestingly, two operators $U_{1}$ and
$U_{2}$ $\in\mathrm{SU}\left(  4\right)  $ are locally equivalent if there
exist some single-qubit unitary operators $\tilde{A}_{1}$, $\tilde{B}_{1}$,
$\tilde{A}_{2}$, $\tilde{B}_{2}$ such that%
\begin{equation}
U_{2}=\left(  \tilde{A}_{1}\otimes\tilde{B}_{1}\right)  U_{1}\left(  \tilde
{A}_{2}\otimes\tilde{B}_{2}\right)  \text{.} \label{Q2}%
\end{equation}
From Eqs. (\ref{Q1}) and (\ref{Q2}), it can be inferred that any operator
within $\mathrm{SU}\left(  4\right)  $ is locally equivalent to $e^{-i\left(
c_{1}\sigma_{1}\otimes\sigma_{1}+c_{2}\sigma_{2}\otimes\sigma_{2}+c_{3}%
\sigma_{3}\otimes\sigma_{3}\right)  }$ for an appropriate selection of the
geometrical point that defines the two-qubit gate in the Weyl chamber, where
$c_{1}\geq c_{2}\geq c_{3}\geq0$. In the standard computational basis, the
matrix representation of $e^{-i\left(  c_{1}\sigma_{1}\otimes\sigma_{1}%
+c_{2}\sigma_{2}\otimes\sigma_{2}+c_{3}\sigma_{3}\otimes\sigma_{3}\right)  }$
is expressed as%
\begin{equation}
U=\left(
\begin{array}
[c]{cccc}%
e^{-ic_{3}}c_{-} & 0 & 0 & -ie^{-ic_{3}}s_{-}\\
0 & e^{ic_{3}}c_{+} & -ie^{ic_{3}}s_{-} & 0\\
0 & -ie^{ic_{3}}s_{-} & e^{ic_{3}}c_{+} & 0\\
-ie^{-ic_{3}}s_{-} & 0 & 0 & e^{-ic_{3}}c_{-}%
\end{array}
\right)  \text{,} \label{Q3}%
\end{equation}
where $c_{\pm}\overset{\text{def}}{=}\cos\left(  c_{1}\pm c_{2}\right)  $ and
$s_{\pm}\overset{\text{def}}{=}\sin\left(  c_{1}\pm c_{2}\right)  $. It has
been determined that the entangling power $\varepsilon\left(  U\right)  $ as
stated in Eq. (\ref{epower}) for the unitary $U$ presented in Eq. (\ref{Q3})
is equivalent to \cite{reza04,bala10},%
\begin{equation}
\varepsilon_{\mathrm{EP}}^{\mathrm{Zanardi}}\left(  U\right)  =\frac{1}%
{18}\left\{  3-\left[  \cos\left(  4c_{1}\right)  \cos(4c_{2})+\cos\left(
4c_{2}\right)  \cos(4c_{3})+\cos\left(  4c_{3}\right)  \cos(4c_{1})\right]
\right\}  \text{,} \label{cami}%
\end{equation}
where it can be noted that $\varepsilon\left(  U\right)  $ is determined
solely by the non-local component of $U$, which is defined through the vector
$\mathbf{c}\overset{\text{def}}{\mathbf{=}}\left(  c_{1}\text{, }c_{2}\text{,
}c_{3}\right)  $. In Table I, we present a tabular summary that illustrates
the Weyl chamber position, the entangling capability, and the Schmidt number
related to standard two-qubit quantum gates. For completeness, we remark that
the Schmidt number of a quantum gate is the number of nonzero Schmidt
coefficients in its operator-Schmidt decomposition \cite{mike98}. In Table II,
on the other hand, we provide a schematic overview of the entangling
capabilities, nonlocal properties, and Schmidt number associated with the
\textrm{SWAP}, \textrm{DCNOT}, and \textrm{CNOT} gates \cite{stefano}. It is
crucial to emphasize that the \textrm{SWAP} gate shows no entangling power,
despite exhibiting a maximal nonlocal character. Moreover, both the
\textrm{DCNOT} and \textrm{CNOT} gates display the same entangling power, even
though they possess different levels of nonlocal character.

\medskip

In this paper, we focus our research on two-qubit systems and stationary
Hamiltonian evolutions. Nevertheless, we emphasize that both Yukalov's
entanglement production and Zanardi's entangling power can also be formally
extended to multi-qubit quantum systems, where the dynamics are dictated by
time-dependent Hamiltonians. In particular, especially when non-commuting
Hamiltonians at different times are present, one anticipates a more complex
entanglement dynamics, where analytical results are less common and numerical
calculations are typically required. Although both measures are applicable to
both multiqubit and qudit systems, Zanardi's approach, in contrast to that of
Yukalov, necessitates the selection of a bipartition of the composite system.
Furthermore, while Zanardi's entangling power relies on the average
entanglement across pure product states, Yukalov's method of entanglement
production is founded on operator norms and the factorization of general
operators, rather than solely unitaries. Additionally, unlike Yukalov's
measure, Zanardi's measure is not applicable to systems that are in mixed
quantum states. Most relevant to our analysis are the distinct features of the
unitary time propagators highlighted by the two measures. While Zanardi's
entangling power reflects the propagator's ability to generate entanglement,
Yukalov's entanglement production instead characterizes its inherent nonlocal
structure. The distinct characteristics of these two measures can be partially
understood by examining their effects on basic unitary gates such as the
\textrm{SWAP} and the $\sqrt{\mathrm{SWAP}}$ gates. Regarding $\varepsilon
_{\mathrm{EP}}^{\mathrm{Zanardi}}$, we find that $\varepsilon_{\mathrm{EP}%
}^{\mathrm{Zanardi}}\left[  \mathrm{SWAP}\right]  =0\leq1/6=\varepsilon
_{\mathrm{EP}}^{\mathrm{Zanardi}}\left[  \sqrt{\mathrm{SWAP}}\right]  $. This
aligns with the observation that the \textrm{SWAP} gate, in contrast to the
$\sqrt{\mathrm{SWAP}}$ gate, lacks any entangling capability. In terms of
$\varepsilon_{\mathrm{EP}}^{\mathrm{Yukalov}}$, we observe that $\varepsilon
_{\mathrm{EP}}^{\mathrm{Yukalov}}\left[  \mathrm{SWAP}\right]  =\log
(2)\geq\log(\sqrt{8/5})=\varepsilon_{\mathrm{EP}}^{\mathrm{Yukalov}}\left[
\sqrt{\mathrm{SWAP}}\right]  $. This is also consistent with the fact that the
$\mathrm{SWAP}$ gate, which has a Schmidt number of $4$, possesses an inherent
nonlocal structure that is more robust than that of the $\sqrt{\mathrm{SWAP}}$
gate, which has a Schmidt number of\textbf{ }$2$.

\medskip

After discussing both static and dynamic entanglement quantifiers, we will now
present the measures we employ to describe quantum evolutions from a geometric
perspective.\begin{table}[t]
\centering
\begin{tabular}
[c]{c|c|c|c}\hline\hline
\textbf{Quantum} \textbf{Gate, }$U$ & \textbf{Entangling power},
$\varepsilon_{\mathrm{EP}}^{\mathrm{Zanardi}}\left(  U\right)  $ &
\textbf{Nonlocal character} & \textbf{Schmidt number}, \textrm{Sch}$\left(
U\right)  $\\\hline
\textrm{SWAP} & No, $0$ & Yes, maximal & $4$\\\hline
\textrm{DCNOT} & Yes, $2/9$ & Yes, maximal & $4$\\\hline
\textrm{CNOT} & Yes, $2/9$ & Yes, non-maximal & $2$\\\hline
\end{tabular}
\caption{A schematic overview of the entangling capabilities, nonlocal
properties, and Schmidt number associated with the SWAP, DCNOT, and CNOT gates
is presented. It is important to highlight that the SWAP gate demonstrates no
entangling power, even though it displays a maximal nonlocal character.
Furthermore, both the DCNOT and CNOT gates exhibit identical entangling power,
despite having distinct levels of nonlocal character.}%
\end{table}

\section{Geometric aspects of quantum evolutions}

In this section, we present the tools utilized to articulate certain geometric
features of quantum evolutions, which encompass geodesic efficiency, speed
efficiency, and curvature coefficient.

\subsection{Efficiency}

\subsubsection{Geodesic efficiency}

We start with the concept of geodesic efficiency. Examine the progression of a
state vector $\left\vert \psi\left(  t\right)  \right\rangle $ as articulated
by the time-dependent Schr\"{o}dinger equation, $i\hslash\partial
_{t}\left\vert \psi\left(  t\right)  \right\rangle =\mathrm{H}\left(
t\right)  \left\vert \psi\left(  t\right)  \right\rangle $, over the interval
$t_{A}\leq t\leq t_{B}$. As a result, the geodesic efficiency $\eta
_{\mathrm{GE}}$ for this quantum evolution is a scalar value that remains
invariant over time (global) and is characterized within the limits of
$0\leq\eta_{\mathrm{GE}}\leq1$. This is defined as \cite{anandan90,cafaro20}%
\begin{equation}
\eta_{\mathrm{GE}}\overset{\text{def}}{=}\frac{s_{0}}{s}=\frac{2\arccos\left[
\left\vert \left\langle A|B\right\rangle \right\vert \right]  }{2\int_{t_{A}%
}^{t_{B}}\frac{\Delta E\left(  t\right)  }{\hslash}dt}\text{.}
\label{efficiency11}%
\end{equation}
The quantity $s_{0}$ denotes the distance along the shortest geodesic path
that links the initial state $\left\vert A\right\rangle \overset{\text{def}%
}{=}$ $\left\vert \psi\left(  t_{A}\right)  \right\rangle $ to the final state
$\left\vert B\right\rangle \overset{\text{def}}{=}\left\vert \psi\left(
t_{B}\right)  \right\rangle $ within the complex projective Hilbert space.
Furthermore, the quantity $s$ in Eq. (\ref{efficiency11}) represents the
distance along the dynamical trajectory $\gamma\left(  t\right)
:t\mapsto\left\vert \psi\left(  t\right)  \right\rangle $ that corresponds to
the evolution of the state vector $\left\vert \psi\left(  t\right)
\right\rangle $ for $t_{A}\leq t\leq t_{B}$. It is evident that a geodesic
quantum evolution characterized by $\gamma\left(  t\right)  =\gamma
_{\mathrm{geo}}\left(  t\right)  $ is defined by the equation $\eta
_{\mathrm{GE}}^{(\gamma_{\mathrm{geo})}}=1$. The term $\Delta E\left(
t\right)  \overset{\text{def}}{=}\left[  \left\langle \psi|\mathrm{H}%
^{2}\left(  t\right)  |\psi\right\rangle -\left\langle \psi|\mathrm{H}\left(
t\right)  |\psi\right\rangle ^{2}\right]  ^{1/2}$ indicates the energy
uncertainty of the system, expressed as the square root of the dispersion of
$\mathrm{H}\left(  t\right)  $. To ensure thoroughness, we stress that the
time-optimal trajectories defined by the condition $\eta_{\mathrm{GE}}=1$ are
associated with quantum evolutions where the Mandelstam-Tamm bound is
saturated \cite{MT}. For example, concentrating on quantum evolutions
characterized by stationary Hamiltonians, where the initial and final states
$\left\vert A\right\rangle $ and $\left\vert B\right\rangle $ are orthogonal,
we find that Eq. (\ref{efficiency11}) indicates that $\eta_{\mathrm{GE}}\leq1$
if and only if $\Delta t\geq\left(  \pi\hslash\right)  /2\Delta E$. This
latter inequality precisely represents the Mandelstam-Tamm bound, with\textbf{
}$\Delta t\overset{\text{def}}{=}t_{B}-t_{A}$ and $\Delta E$ denoting the
(constant) energy uncertainty of the system. Importantly, Anandan and Aharonov
established that the infinitesimal distance $ds\overset{\text{def}}{=}2\left[
\Delta E\left(  t\right)  /\hslash\right]  dt$ is connected to the
Fubini-Study infinitesimal distance $ds_{\text{\textrm{FS}}}$ through the
relationship \cite{anandan90},%
\begin{equation}
ds_{\text{\textrm{FS}}}^{2}\left(  \left\vert \psi\left(  t\right)
\right\rangle \text{, }\left\vert \psi\left(  t+dt\right)  \right\rangle
\right)  \overset{\text{def}}{=}4\left[  1-\left\vert \left\langle \psi\left(
t\right)  |\psi\left(  t+dt\right)  \right\rangle \right\vert ^{2}\right]
=4\frac{\Delta E^{2}\left(  t\right)  }{\hslash^{2}}dt^{2}+\mathcal{O}\left(
dt^{3}\right)  \text{,} \label{relation11}%
\end{equation}
with $\mathcal{O}\left(  dt^{3}\right)  $ being an infinitesimal term of an
order that equals or is greater than $dt^{3}$. From the connection between
$ds_{\mathrm{FS}}$ and $ds$, it can be inferred that $s$ is directly
proportional to the time integral of $\Delta E$. Additionally, $s$ denotes the
distance computed using the Fubini-Study metric during the evolution of the
quantum system in ray space (i.e., the projective Hilbert space
$\mathcal{P(H)}$ of a complex Hilbert space $\mathcal{H}$ defined by the set
of equivalence classes of non-zero state vectors in $\mathcal{H}$
\cite{anandan90}). It is crucial to highlight that when the actual dynamical
trajectory aligns with the shortest geodesic path linking $\left\vert
A\right\rangle $ and $\left\vert B\right\rangle $, $s$ is equivalent to
$s_{0}$, and the geodesic efficiency $\eta_{\mathrm{GE}}$ in Eq.
(\ref{efficiency11}) is equal to one. Clearly, $\pi$ (i.e., $2\arccos
(0)$\textbf{)} signifies the minimal distance separating two orthogonal pure
states in ray space.

\subsubsection{Speed efficiency}

We proceed here with the concept of speed efficiency. We start by recalling
that appropriate families of nonstationary Hamiltonians capable of generating
predetermined dynamical trajectories with minimal energy resource expenditure
were first introduced in Ref. \cite{uzdin12}. While these trajectories are
energy-efficient, they do not usually correspond to geodesic paths of the
shortest length. The criterion for minimal energy expenditure is met when no
energy is wasted on segments of the Hamiltonian $\mathrm{H}=\mathrm{H}\left(
t\right)  $ that do not effectively direct the system. In other words, all
available energy, as represented by the spectral norm of the Hamiltonian
$\left\Vert \mathrm{H}\right\Vert _{\mathrm{SP}}$, is transformed into the
system's evolution speed $v_{\mathrm{H}}(t)$ $\overset{\text{def}}%
{=}(2/\hslash)\Delta E\left(  t\right)  $, where $\Delta E\left(  t\right)  $
signifies the energy uncertainty.

More specifically, Uzdin's speed efficiency $\eta_{\mathrm{SE}}$ in Ref.
\cite{uzdin12}\textbf{ }denotes a time-dependent (local) scalar quantity that
fulfills the condition $0\leq\eta_{\mathrm{SE}}\leq1$. It is identified
as\textbf{ }%
\begin{equation}
\eta_{\mathrm{SE}}\left(  t\right)  \overset{\text{def}}{=}\frac
{\Delta\mathrm{H}_{\rho}}{\left\Vert \mathrm{H}\right\Vert _{\mathrm{SP}}%
}=\frac{\sqrt{\mathrm{tr}\left(  \rho\mathrm{H}^{2}\right)  -\left[
\mathrm{tr}\left(  \rho\mathrm{H}\right)  \right]  ^{2}}}{\max\left[
\sqrt{\mathrm{eig}\left(  \mathrm{H}^{\dagger}\mathrm{H}\right)  }\right]
}\text{.} \label{se111}%
\end{equation}
While $\Delta\mathrm{H}_{\rho}=\Delta E\left(  t\right)  $ and $\rho
=\rho\left(  t\right)  $ denotes the density operator that defines the quantum
system at time $t$, the quantity $\left\Vert \mathrm{H}\right\Vert
_{\mathrm{SP}}$ present in the denominator of Eq. (\ref{se111}) is given by
$\left\Vert \mathrm{H}\right\Vert _{\mathrm{SP}}\overset{\text{def}}{=}%
\max\left[  \sqrt{\mathrm{eig}\left(  \mathrm{H}^{\dagger}\mathrm{H}\right)
}\right]  $. Note that if $\mathrm{H}$ is Hermitian, $\left\Vert
\mathrm{H}\right\Vert _{\mathrm{SP}}$ is simply the magnitude of the
maximum-magnitude eigenvalue of $\mathrm{H}$. This quantity is referred to as
the spectral norm $\left\Vert \mathrm{H}\right\Vert _{\mathrm{SP}}$ of the
Hamiltonian operator \textrm{H}, which acts as a measure of the magnitude of
bounded linear operators. It is computed as the square root of the maximum
eigenvalue of the operator $\mathrm{H}^{\dagger}\mathrm{H}$, with
$\mathrm{H}^{\dagger}$ representing the Hermitian conjugate of $\mathrm{H}$.

After implementing methods to measure the length and energy dissipation of
dynamical trajectories, we will detail in the following subsection how these
trajectories \textquotedblleft bend\textquotedblright\ during quantum evolutions.

\subsection{Curvature}

In the broadest context, we examine a nonstationary Hamiltonian evolution as
described by Schr\"{o}dinger's equation $i\hslash\partial_{t}\left\vert
\psi\left(  t\right)  \right\rangle =\mathrm{H}\left(  t\right)  \left\vert
\psi\left(  t\right)  \right\rangle $, where $\left\vert \psi\left(  t\right)
\right\rangle $ represents an element within an arbitrary $N$-dimensional
complex Hilbert space $\mathcal{H}_{N}$. Typically, the normalized state
vector $\left\vert \psi\left(  t\right)  \right\rangle $ adheres to the
condition $\left\langle \psi\left(  t\right)  \left\vert \dot{\psi}\left(
t\right)  \right.  \right\rangle =(-i/\hslash)\left\langle \psi\left(
t\right)  \left\vert \mathrm{H}\left(  t\right)  \right\vert \psi\left(
t\right)  \right\rangle \neq0$. For the state $\left\vert \psi\left(
t\right)  \right\rangle $, we introduce the parallel transported unit state
vector $\left\vert \Psi\left(  t\right)  \right\rangle \overset{\text{def}}%
{=}e^{i\beta\left(  t\right)  }\left\vert \psi\left(  t\right)  \right\rangle
$, with the phase $\beta\left(  t\right)  $ defined such that $\left\langle
\Psi\left(  t\right)  \left\vert \dot{\Psi}\left(  t\right)  \right.
\right\rangle =0$. It is important to note that $i\hslash\left\vert \dot{\Psi
}\left(  t\right)  \right\rangle =\left[  \mathrm{H}\left(  t\right)
-\hslash\dot{\beta}\left(  t\right)  \right]  \left\vert \Psi\left(  t\right)
\right\rangle $. As a result, the condition $\left\langle \Psi\left(
t\right)  \left\vert \dot{\Psi}\left(  t\right)  \right.  \right\rangle =0$ is
equivalent to establishing $\beta\left(  t\right)  $ as $\beta\left(
t\right)  \overset{\text{def}}{=}(1/\hslash)\int_{0}^{t}\left\langle
\psi\left(  t^{\prime}\right)  \left\vert \mathrm{H}\left(  t^{\prime}\right)
\right\vert \psi\left(  t^{\prime}\right)  \right\rangle dt^{\prime}$.
Therefore, $\left\vert \Psi\left(  t\right)  \right\rangle $ can be simplified
to%
\begin{equation}
\left\vert \Psi\left(  t\right)  \right\rangle =e^{(i/\hslash)\int_{0}%
^{t}\left\langle \psi\left(  t^{\prime}\right)  \left\vert \mathrm{H}\left(
t^{\prime}\right)  \right\vert \psi\left(  t^{\prime}\right)  \right\rangle
dt^{\prime}}\left\vert \psi\left(  t\right)  \right\rangle \text{,}
\label{reasonaaa}%
\end{equation}
and satisfies the evolution equation $i\hslash\left\vert \dot{\Psi}\left(
t\right)  \right\rangle =\Delta\mathrm{H}\left(  t\right)  \left\vert
\Psi\left(  t\right)  \right\rangle $ where $\Delta\mathrm{H}\left(  t\right)
\overset{\text{def}}{=}\mathrm{H}\left(  t\right)  -\left\langle
\mathrm{H}\left(  t\right)  \right\rangle $. It is important to note that the
speed $v(t)$ of quantum evolution changes when the Hamiltonian is
time-dependent. Specifically, $v(t)$ is defined such that $v^{2}\left(
t\right)  =\left\langle \dot{\Psi}\left(  t\right)  \left\vert \dot{\Psi
}\left(  t\right)  \right.  \right\rangle =\left\langle \left(  \Delta
\mathrm{H}\left(  t\right)  \right)  ^{2}\right\rangle /\hslash^{2}$. For ease
of reference, we define the arc length $s=s\left(  t\right)  $ in relation to
$v(t)$ as $s\left(  t\right)  \overset{\text{def}}{=}\int_{0}^{t}v(t^{\prime
})dt^{\prime}$, with $ds=v(t)dt$ (which consequently indicates that
$\partial_{t}=v(t)\partial_{s}$), where $\partial_{t}\overset{\text{def}}%
{=}\partial/\partial t$ and $\partial_{s}\overset{\text{def}}{=}%
\partial/\partial s$. To clarify, we emphasize that the variable $s$ in Eq.
(\ref{efficiency11}) represents the distance along the dynamical trajectory of
the quantum system within projective Hilbert space. In contrast, the quantity
$s\left(  t\right)  $ utilized in the formulation of the curvature coefficient
for quantum evolution is a time-dependent arc length function, which is
employed for the temporal parametrization of pertinent vectors that produce
the curvature. In summary, by introducing the dimensionless operator%
\begin{equation}
\Delta h\left(  t\right)  \overset{\text{def}}{=}\frac{\Delta\mathrm{H}\left(
t\right)  }{\hslash v(t)}=\frac{\Delta\mathrm{H}\left(  t\right)  }%
{\sqrt{\left\langle \left(  \Delta\mathrm{H}\left(  t\right)  \right)
^{2}\right\rangle }}\text{,}%
\end{equation}
the normalized tangent vector $\left\vert T\left(  s\right)  \right\rangle
\overset{\text{def}}{=}\partial_{s}\left\vert \Psi\left(  s\right)
\right\rangle =\left\vert \Psi^{\prime}\left(  s\right)  \right\rangle $
becomes $\left\vert T\left(  s\right)  \right\rangle =-i\Delta h\left(
s\right)  \left\vert \Psi\left(  s\right)  \right\rangle $. It is crucial to
acknowledge that $\left\langle T\left(  s\right)  \left\vert T\left(
s\right)  \right.  \right\rangle =1$ by construction. Additionally, we have
$\partial_{s}\left\langle \Delta h(s)\right\rangle =\left\langle \Delta
h^{\prime}(s)\right\rangle $. From the tangent vector $\left\vert T\left(
s\right)  \right\rangle =-i\Delta h\left(  s\right)  \left\vert \Psi\left(
s\right)  \right\rangle $, we can derive $\left\vert T^{\prime}\left(
s\right)  \right\rangle \overset{\text{def}}{=}\partial_{s}\left\vert T\left(
s\right)  \right\rangle $. Through algebraic manipulation, we find that
$\left\vert T^{\prime}\left(  s\right)  \right\rangle =-i\Delta h(s)\left\vert
\Psi^{\prime}\left(  s\right)  \right\rangle -i\Delta h^{\prime}(s)\left\vert
\Psi\left(  s\right)  \right\rangle $, where $\left\langle T^{\prime}\left(
s\right)  \left\vert T^{\prime}\left(  s\right)  \right.  \right\rangle
=\left\langle \left(  \Delta h^{\prime}(s)\right)  ^{2}\right\rangle
+\left\langle \left(  \Delta h(s)\right)  ^{4}\right\rangle
-2i\operatorname{Re}\left[  \left\langle \Delta h^{\prime}(s)\left(  \Delta
h(s)\right)  ^{2}\right\rangle \right]  \neq1$ in most instances. We are now
prepared to introduce the curvature coefficient for quantum evolutions
generated by nonstationary Hamiltonians.

Following the introduction of the vectors $\left\vert \Psi\left(  s\right)
\right\rangle $, $\left\vert T\left(  s\right)  \right\rangle $, and
$\left\vert T^{\prime}\left(  s\right)  \right\rangle $, we are now able to
define the curvature coefficient as initially outlined in Refs.
\cite{alsing24A,alsing24B,alsing24C}. This coefficient is represented as
$\kappa_{\mathrm{AC}}^{2}\left(  s\right)  \overset{\text{def}}{=}\left\langle
\tilde{N}_{\ast}\left(  s\right)  \left\vert \tilde{N}_{\ast}\left(  s\right)
\right.  \right\rangle $. It is essential to recognize that $\left\vert
\tilde{N}_{\ast}\left(  s\right)  \right\rangle \overset{\text{def}}%
{=}\mathrm{P}^{\left(  \Psi\right)  }\left\vert T^{\prime}\left(  s\right)
\right\rangle $, where the projection operator $\mathrm{P}^{\left(
\Psi\right)  }$ onto states orthogonal to $\left\vert \Psi\left(  s\right)
\right\rangle $ is given by $\mathrm{P}^{\left(  \Psi\right)  }\overset
{\text{def}}{=}\mathrm{I}-\left\vert \Psi\left(  s\right)  \right\rangle
\left\langle \Psi\left(  s\right)  \right\vert $. In this context,
\textquotedblleft$\mathrm{I}$\textquotedblright denotes the identity operator
in $\mathcal{H}_{N}$. The subscript \textquotedblleft\textrm{AC}%
\textquotedblright\ signifies Alsing and Cafaro. It is significant to mention
that the curvature coefficient $\kappa_{\mathrm{AC}}^{2}\left(  s\right)
=\left\langle \tilde{N}_{\ast}\left(  s\right)  \left\vert \tilde{N}_{\ast
}\left(  s\right)  \right.  \right\rangle $ can be expressed in a more
convenient form as%
\begin{equation}
\kappa_{\mathrm{AC}}^{2}\left(  s\right)  \overset{\text{def}}{=}\left\Vert
\mathrm{D}\left\vert T(s)\right\rangle \right\Vert ^{2}=\left\Vert
\mathrm{D}^{2}\left\vert \Psi\left(  s\right)  \right\rangle \right\Vert
^{2}\text{,} \label{peggio11}%
\end{equation}
where $\mathrm{D}\overset{\text{def}}{=}\mathrm{P}^{\left(  \Psi\right)
}d/ds=\left(  \mathrm{I}-\left\vert \Psi\right\rangle \left\langle
\Psi\right\vert \right)  d/ds$ such that $\mathrm{D}\left\vert
T(s)\right\rangle \overset{\text{def}}{=}\mathrm{P}^{\left(  \Psi\right)
}\left\vert T^{\prime}(s)\right\rangle =$ $\left\vert \tilde{N}_{\ast}\left(
s\right)  \right\rangle $ is the covariant derivative
\cite{carlocqg23,samuel88,paulPRA23}. As stated in Eq. (\ref{peggio11}), the
curvature coefficient $\kappa_{\mathrm{AC}}^{2}\left(  s\right)  $ is found to
be equivalent to the square of the magnitude of the second covariant
derivative of the state vector $\left\vert \Psi\left(  s\right)  \right\rangle
$, which delineates the quantum Schr\"{o}dinger trajectory in the context of
projective Hilbert space.

Ultimately, an expression that demonstrates computational efficiency for the
curvature coefficient $\kappa_{\mathrm{AC}}^{2}\left(  s\right)  $ in Eq.
(\ref{peggio11}) within any arbitrary nonstationary context is reduced to%
\begin{equation}
\kappa_{\mathrm{AC}}^{2}\left(  s\right)  =\left\langle (\Delta h)^{4}%
\right\rangle -\left\langle (\Delta h)^{2}\right\rangle ^{2}+\left[
\left\langle (\Delta h^{\prime})^{2}\right\rangle -\left\langle \Delta
h^{\prime}\right\rangle ^{2}\right]  +i\left\langle \left[  (\Delta
h)^{2}\text{, }\Delta h^{\prime}\right]  \right\rangle \text{.}
\label{curvatime11}%
\end{equation}
From Eq. (\ref{curvatime11}), it is noted that when the Hamiltonian \textrm{H}
is held constant, $\Delta h^{\prime}$ becomes the null operator, enabling us
to obtain the stationary limit $\left\langle (\Delta h)^{4}\right\rangle
-\left\langle (\Delta h)^{2}\right\rangle ^{2}$ for the curvature coefficient
$\kappa_{\mathrm{AC}}^{2}\left(  s\right)  $ \cite{alsing24A}. The expression
for $\kappa_{\mathrm{AC}}^{2}\left(  s\right)  $ in Eq. (\ref{curvatime11}) is
formulated through a method that relies on the calculation of expectation
values, which require a comprehension of the state vector $\left\vert
\psi\left(  t\right)  \right\rangle $ governed by the time-dependent
Schr\"{o}dinger's evolution equation. As elaborated in Ref. \cite{alsing24B},
this methodology of expectation values offers a significant statistical
meaning for $\kappa_{\mathrm{AC}}^{2}\left(  s\right)  $.

We are now prepared to introduce appropriate Hamiltonian models to which the
entanglement and geometric evolution quantifiers discussed in Sections II and
III, respectively, can be applied.

\section{Hamiltonian models}

In this section, we present the concepts of time optimal and time suboptimal
Hamiltonian evolutions applicable to any pair of quantum states that define
finite-dimensional quantum systems.

\subsection{Time optimal Hamiltonians}

In accordance with Refs. \cite{ali09,carlocqg23}, we consider a traceless and
stationary Hamiltonian \textrm{H} characterized by a spectral decomposition
expressed as \textrm{H}$\overset{\text{def}}{=}E_{1}\left\vert E_{1}%
\right\rangle \left\langle E_{1}\right\vert +E_{2}\left\vert E_{2}%
\right\rangle \left\langle E_{2}\right\vert $, where $\left\langle E_{1}%
|E_{2}\right\rangle =\delta_{12}$ and $E_{2}\geq E_{1}$. For the sake of
transparency, it is important to note that the sets $\left\{  \left\vert
E_{i}\right\rangle \right\}  _{i=1,2}$ and $\left\{  E_{i}\right\}  _{i=1,2}$
represent the eigenvectors and eigenvalues of the constant Hamiltonian
\textrm{H}, respectively. Additionally, $\delta_{ij}$ signifies the Kronecker
delta symbol, applicable for $1\leq i$\textbf{, }$j\leq2$. In scenarios that
are time-optimal, the focus is on the evolution of a state $\left\vert
A\right\rangle $, which may not be normalized, into a state $\left\vert
B\right\rangle $ in the least amount of time by maximizing the energy
uncertainty $\Delta E$,%
\begin{equation}
\Delta E\overset{\text{def}}{=}\left[  \frac{\left\langle A|\mathrm{H}%
^{2}\mathrm{|}A\right\rangle }{\left\langle A|A\right\rangle }-\left(
\frac{\left\langle A|\mathrm{H|}A\right\rangle }{\left\langle A|A\right\rangle
}\right)  ^{2}\right]  ^{1/2}\text{,} \label{deltaE11}%
\end{equation}
in such a manner that $\Delta E$ in Eq. (\ref{deltaE11}) is equal to $\Delta
E_{\max}$. The rationale behind maximizing the energy uncertainty $\Delta E$
is based on the observation that the rate of quantum evolution $ds/dt$ along
the trajectory connecting $\left\vert A\right\rangle $ to $\left\vert
B\right\rangle $ is directly proportional to the energy uncertainty $\Delta
E$, expressed as $ds/dt\propto\Delta E$. What is the maximum value of $\Delta
E_{\max}?$ To determine this value, we recognize that any unnormalized initial
state $\left\vert A\right\rangle $ can be expressed as $\left\vert
A\right\rangle =\alpha_{1}\left\vert E_{1}\right\rangle +\alpha_{2}\left\vert
E_{2}\right\rangle $, where $\alpha_{1}$ and $\alpha_{2}$ are generally
complex quantum amplitudes defined as $\alpha_{1}\overset{\text{def}}{=}$
$\left\langle E_{1}|A\right\rangle $ and $\alpha_{2}\overset{\text{def}}%
{=}\left\langle E_{2}|A\right\rangle $. By substituting this decomposition of
$\left\vert A\right\rangle $ into Eq. (\ref{deltaE11}), we obtain%
\begin{equation}
\Delta E=\frac{E_{2}-E_{1}}{2}\left[  1-\left(  \frac{\left\vert \alpha
_{1}\right\vert ^{2}-\left\vert \alpha_{2}\right\vert ^{2}}{\left\vert
\alpha_{1}\right\vert ^{2}+\left\vert \alpha_{2}\right\vert ^{2}}\right)
^{2}\right]  ^{1/2}\text{.} \label{chi41}%
\end{equation}
An examination of Eq. (\ref{chi41}) indicates that the peak value of $\Delta
E$ in Eq. (\ref{chi41}) occurs when $\left\vert \alpha_{1}\right\vert
=\left\vert \alpha_{2}\right\vert $. Specifically, it is equal to
\begin{equation}
\Delta E_{\max}\overset{\text{def}}{=}\left(  \frac{E_{2}-E_{1}}{2}\right)
\text{.}%
\end{equation}
A key principle in Mostafazadeh's methodology, as outlined in Ref.
\cite{ali09}, is the representation of \textrm{H}$\overset{\text{def}}{=}%
E_{1}\left\vert E_{1}\right\rangle \left\langle E_{1}\right\vert
+E_{2}\left\vert E_{2}\right\rangle \left\langle E_{2}\right\vert $ in
relation to the initial and final states $\left\vert A\right\rangle $ and
$\left\vert B\right\rangle $, respectively, while ensuring that $\Delta
E=\Delta E_{\max}$. In this framework, it is crucial to acknowledge that
$\left\vert A\right\rangle $ and $\left\vert B\right\rangle $ can be
represented as $\left\vert A\right\rangle =\alpha_{1}\left\vert E_{1}%
\right\rangle +\alpha_{2}\left\vert E_{2}\right\rangle $ and $\left\vert
B\right\rangle =\beta_{1}\left\vert E_{1}\right\rangle +\beta_{2}\left\vert
E_{2}\right\rangle $, respectively. Additionally, it is essential to enforce
$\left\vert \alpha_{1}\right\vert =\left\vert \alpha_{2}\right\vert $ and
$\left\vert \beta_{1}\right\vert =\left\vert \beta_{2}\right\vert $ or,
alternatively,%
\begin{equation}
\left\vert \alpha_{2}\right\vert ^{2}-\left\vert \alpha_{1}\right\vert
^{2}=0=\left\vert \beta_{2}\right\vert ^{2}-\left\vert \beta_{1}\right\vert
^{2}\text{,} \label{bruno111}%
\end{equation}
to satisfy the condition $\Delta E=\Delta E_{\max}$ and consequently guarantee
the minimum travel time $t_{AB}^{\min}\overset{\text{def}}{=}\hslash
\arccos\left(  \left\vert \left\langle A\left\vert B\right.  \right\rangle
\right\vert \right)  /\Delta E_{\max}$. After performing some straightforward
yet tedious algebra, one can demonstrate that the expression for the
time-optimal Hamiltonian \textrm{H}, which links $\left\vert A\right\rangle $
to $\left\vert B\right\rangle $ (assuming energy dispersion $\Delta E=E$, with
$E_{2}=-E_{1}=E$), is provided by \cite{carlocqg23}%
\begin{equation}
\mathrm{H}_{\mathrm{opt}}\overset{\text{def}}{=}i\Delta E\frac{\left\vert
\left\langle A\left\vert B\right.  \right\rangle \right\vert }{\sqrt
{1-\left\vert \left\langle A\left\vert B\right.  \right\rangle \right\vert
^{2}}}\left[  \frac{\left\vert B\right\rangle \left\langle A\right\vert
}{\left\langle A|B\right\rangle }-\frac{\left\vert A\right\rangle \left\langle
B\right\vert }{\left\langle B|A\right\rangle }\right]  \text{.} \label{amy11}%
\end{equation}
Recalling that quantum states differing only by a global phase are physically
indistinguishable, we emphasize for thoroughness that the unitary time
propagator $U_{\mathrm{opt}}\left(  t\right)  \overset{\text{def}}{=}%
e^{-\frac{i}{\hslash}\mathrm{H}_{\mathrm{opt}}t}$ satisfies $U_{\mathrm{opt}%
}(t_{\text{\textrm{opt}}})\left\vert A\right\rangle =\left\vert B\right\rangle
$ with $t_{\text{\textrm{opt}}}=t_{AB}^{\min}$ defined as%
\begin{equation}
t_{\text{\textrm{opt}}}\overset{\text{def}}{=}\frac{\hslash\arccos\left(
\left\vert \left\langle A\left\vert B\right.  \right\rangle \right\vert
\right)  }{\Delta E}\text{.} \label{amy222}%
\end{equation}
In conclusion, we note that for $\mathrm{H}_{\mathrm{opt}}$ as presented in
Eq. (\ref{amy11}), it is correctly derived that $\left\langle A|\mathrm{H}%
_{\mathrm{opt}}\mathrm{|}A\right\rangle /\left\langle A|A\right\rangle =0$ and
$\Delta E=\left[  \left\langle A|\mathrm{H}_{\mathrm{opt}}^{2}|A\right\rangle
/\left\langle A|A\right\rangle \right]  ^{1/2}=E=\Delta E_{\max}$.

Having covered the fundamentals of constructing optimal-time Hamiltonians, we
are now ready to investigate deviations from time-optimality in
two-dimensional subspaces.

\subsection{Time suboptimal Hamiltonians}

In the following discussion, we develop a one-parameter family of time
suboptimal, time-independent Hamiltonians that connect two arbitrary
nonorthogonal quantum states within any finite-dimensional quantum system. It
is important to note that regarding the eigenvectors of the Hamiltonian, the
initial and final states $\left\vert A\right\rangle $ and $\left\vert
B\right\rangle $ can be expressed as $\left\vert A\right\rangle =\alpha
_{1}\left\vert E_{1}\right\rangle +\alpha_{2}\left\vert E_{2}\right\rangle $,
and $\left\vert B\right\rangle =\beta_{1}\left\vert E_{1}\right\rangle
+\beta_{2}\left\vert E_{2}\right\rangle $, respectively. To guarantee a
non-minimum travel time $t_{\mathrm{AB}}\geq t_{\mathrm{AB}}^{\min}$ while
maintaining $\Delta E$ at a value consistently less than its maximum $\Delta
E_{\max}$, we establish $\delta\left\vert \alpha_{1}\right\vert =\left\vert
\alpha_{2}\right\vert $, and $\delta\left\vert \beta_{1}\right\vert
=\left\vert \beta_{2}\right\vert $. Consequently, we define $\alpha
_{2}\overset{\text{def}}{=}e^{i\varphi_{\alpha}}\delta\alpha_{1}$ and,
$\beta_{2}\overset{\text{def}}{=}e^{i\varphi_{\beta}}\delta\beta_{1}$, with
$\delta\in%
\mathbb{R}
_{+}\backslash\left\{  0\right\}  $. Then, $\left\vert A\right\rangle $ and
$\left\vert B\right\rangle $ can be expressed as%
\begin{equation}
\left\vert A\right\rangle =\alpha_{1}\left\vert E_{1}\right\rangle +\alpha
_{2}\left\vert E_{2}\right\rangle =\alpha_{1}\left\vert E_{1}\right\rangle
+e^{i\varphi_{\alpha}}\delta\alpha_{1}\left\vert E_{2}\right\rangle \text{,
and }\left\vert B\right\rangle =\beta_{1}\left\vert E_{1}\right\rangle
+\beta_{2}\left\vert E_{2}\right\rangle =\beta_{1}\left\vert E_{1}%
\right\rangle +e^{i\varphi_{\beta}}\delta\beta_{1}\left\vert E_{2}%
\right\rangle \text{,} \label{Jchi5b}%
\end{equation}
respectively. From Eq. (\ref{Jchi5b}), we conveniently introduce the
normalized states $\left\vert \mathcal{A}\right\rangle $ and $\left\vert
\mathcal{B}\right\rangle $ such that%
\begin{equation}
\left\vert E_{1}\right\rangle +e^{i\varphi_{\alpha}}\delta\left\vert
E_{2}\right\rangle =\alpha_{1}^{-1}\left\vert A\right\rangle \overset
{\text{def}}{=}\sqrt{1+\delta^{2}}\left\vert \mathcal{A}\right\rangle \text{,
and }\left\vert E_{1}\right\rangle +e^{i\varphi_{\beta}}\delta\left\vert
E_{2}\right\rangle =\beta_{1}^{-1}\left\vert B\right\rangle \overset
{\text{def}}{=}\sqrt{1+\delta^{2}}e^{-i\frac{\varphi_{\alpha}-\varphi_{\beta}%
}{2}}\left\vert \mathcal{B}\right\rangle \text{,} \label{Jchi6}%
\end{equation}
that is,
\begin{equation}
\left\vert \mathcal{A}\right\rangle \overset{\text{def}}{=}\frac{1}%
{\sqrt{1+\delta^{2}}}\left[  \left\vert E_{1}\right\rangle +e^{i\varphi
_{\alpha}}\delta\left\vert E_{2}\right\rangle \right]  \text{, and }\left\vert
\mathcal{B}\right\rangle \overset{\text{def}}{=}\frac{1}{\sqrt{1+\delta^{2}}%
}\left[  e^{i\frac{\varphi_{\alpha}-\varphi_{\beta}}{2}}\left\vert
E_{1}\right\rangle +e^{i\frac{\varphi_{\alpha}+\varphi_{\beta}}{2}}%
\delta\left\vert E_{2}\right\rangle \right]  \text{. } \label{Jchi8}%
\end{equation}
Evidently, the matrix representation illustrating the relationships between
the sets of states $\left\{  \left\vert \mathcal{A}\right\rangle \text{,
}\left\vert \mathcal{B}\right\rangle \right\}  $ and $\left\{  \left\vert
E_{1}\right\rangle \text{, }\left\vert E_{2}\right\rangle \right\}  $ in Eq.
(\ref{Jchi8}) produces%
\begin{equation}
\left(
\begin{array}
[c]{c}%
\left\vert \mathcal{A}\right\rangle \\
\left\vert \mathcal{B}\right\rangle
\end{array}
\right)  =\frac{1}{\sqrt{1+\delta^{2}}}\left(
\begin{array}
[c]{cc}%
1 & e^{i\varphi_{\alpha}}\delta\\
e^{i\frac{\varphi_{\alpha}-\varphi_{\beta}}{2}} & e^{i\frac{\varphi_{\alpha
}+\varphi_{\beta}}{2}}\delta
\end{array}
\right)  \left(
\begin{array}
[c]{c}%
\left\vert E_{1}\right\rangle \\
\left\vert E_{2}\right\rangle
\end{array}
\right)  \text{.} \label{Jchi10}%
\end{equation}
By inverting the matrix relation presented in Eq. (\ref{Jchi10}), we derive a
method to express $\left\{  \left\vert E_{1}\right\rangle \text{, }\left\vert
E_{2}\right\rangle \right\}  $ in terms of $\left\{  \left\vert \mathcal{A}%
\right\rangle \text{, }\left\vert \mathcal{B}\right\rangle \right\}  $.
Specifically, we achieve%
\begin{equation}
\left(
\begin{array}
[c]{c}%
\left\vert E_{1}\right\rangle \\
\left\vert E_{2}\right\rangle
\end{array}
\right)  =\frac{\sqrt{1+\delta^{2}}}{e^{i\frac{\varphi_{\alpha}+\varphi
_{\beta}}{2}}-e^{i\varphi_{\alpha}}e^{i\frac{\varphi_{\alpha}-\varphi_{\beta}%
}{2}}}\left(
\begin{array}
[c]{cc}%
e^{i\frac{\varphi_{\alpha}+\varphi_{\beta}}{2}} & -e^{i\varphi_{\alpha}}\\
-\frac{1}{\delta}e^{i\frac{\varphi_{\alpha}-\varphi_{\beta}}{2}} & \frac
{1}{\delta}%
\end{array}
\right)  \left(
\begin{array}
[c]{c}%
\left\vert \mathcal{A}\right\rangle \\
\left\vert \mathcal{B}\right\rangle
\end{array}
\right)  \text{.} \label{Jyo}%
\end{equation}
Consequently, by utilizing Eqs. (\ref{Jchi6}) and (\ref{Jyo}), we reach%
\begin{equation}
\left(
\begin{array}
[c]{c}%
\left\vert E_{1}\right\rangle \\
\left\vert E_{2}\right\rangle
\end{array}
\right)  =\frac{\sqrt{1+\delta^{2}}}{e^{i\frac{\varphi_{\alpha}+\varphi
_{\beta}}{2}}-e^{i\varphi_{\alpha}}e^{i\frac{\varphi_{\alpha}-\varphi_{\beta}%
}{2}}}\left(
\begin{array}
[c]{cc}%
e^{i\frac{\varphi_{\alpha}+\varphi_{\beta}}{2}} & -e^{i\varphi_{\alpha}}\\
-\frac{1}{\delta}e^{i\frac{\varphi_{\alpha}-\varphi_{\beta}}{2}} & \frac
{1}{\delta}%
\end{array}
\right)  \left(
\begin{array}
[c]{c}%
\frac{\alpha_{1}^{-1}}{\sqrt{1+\delta^{2}}}\left\vert A\right\rangle \\
\frac{\beta_{1}^{-1}}{\sqrt{1+\delta^{2}}}e^{i\frac{\varphi_{\alpha}%
-\varphi_{\beta}}{2}}\left\vert B\right\rangle
\end{array}
\right)  \text{.} \label{Jchichi}%
\end{equation}
Use of Eq. (\ref{Jyo}) leads to the expression for the quantum overlap
$\left\langle \mathcal{A}|\mathcal{B}\right\rangle $,%
\begin{equation}
\left\langle \mathcal{A}|\mathcal{B}\right\rangle =\cos\left(  \frac
{\varphi_{\alpha}-\varphi_{\beta}}{2}\right)  +i\frac{1-\delta^{2}}%
{1+\delta^{2}}\sin\left(  \frac{\varphi_{\alpha}-\varphi_{\beta}}{2}\right)
\text{,} \label{doit1}%
\end{equation}
with $\left\langle \mathcal{B}|\mathcal{A}\right\rangle =\left\langle
\mathcal{A}|\mathcal{B}\right\rangle ^{\ast}$. From Eq. (\ref{doit1}), the
probability amplitude $\left\vert \left\langle \mathcal{A}|\mathcal{B}%
\right\rangle \right\vert ^{2}$ reduces to%
\begin{equation}
\left\vert \left\langle \mathcal{A}|\mathcal{B}\right\rangle \right\vert
^{2}=\frac{\left\vert \left\langle A|B\right\rangle \right\vert ^{2}%
}{\left\langle A|A\right\rangle \left\langle B|B\right\rangle }=\frac
{1+2\delta^{2}\cos\left(  \varphi_{\alpha}-\varphi_{\beta}\right)  +\delta
^{4}}{\left(  1+\delta^{2}\right)  ^{2}}\text{.} \label{doit3}%
\end{equation}
As an additional remark, we emphasize that when $\delta=1$, Eq. (\ref{doit3})
simplifies to $\left\vert \left\langle \mathcal{A}|\mathcal{B}\right\rangle
\right\vert ^{2}=\cos^{2}\left[  (\varphi_{\alpha}-\varphi_{\beta})/2\right]
$. This equation represents what we anticipate in a time optimal context.
Moreover, to confirm the second relation in Eq. (\ref{doit3}), we observe that%
\begin{equation}
\left\langle A|A\right\rangle =\left\vert \alpha_{1}\right\vert ^{2}%
+\delta^{2}\left\vert \alpha_{1}\right\vert ^{2}\text{, }\left\langle
B|B\right\rangle =\left\vert \beta_{1}\right\vert ^{2}+\delta^{2}\left\vert
\beta_{1}\right\vert ^{2}\text{, and }\left\vert \left\langle A|B\right\rangle
\right\vert ^{2}=\left\vert \alpha_{1}^{\ast}\beta_{1}\left(  1+\delta
^{2}e^{-i\left(  \varphi_{\alpha}-\varphi_{\beta}\right)  }\right)
\right\vert ^{2}\text{.}%
\end{equation}
Therefore, after some algebra, we arrive at%
\begin{align}
\frac{\left\vert \left\langle A|B\right\rangle \right\vert ^{2}}{\left\langle
A|A\right\rangle \left\langle B|B\right\rangle }  &  =\frac{\left\vert
\alpha_{1}^{\ast}\beta_{1}\right\vert ^{2}\left\vert 1+\delta^{2}e^{-i\left(
\varphi_{\alpha}-\varphi_{\beta}\right)  }\right\vert ^{2}}{\left(  \left\vert
\alpha_{1}\right\vert ^{2}+\delta^{2}\left\vert \alpha_{1}\right\vert
^{2}\right)  \left(  \left\vert \beta_{1}\right\vert ^{2}+\delta^{2}\left\vert
\beta_{1}\right\vert ^{2}\right)  }\nonumber\\
&  =\frac{\left\vert 1+\delta^{2}e^{-i\left(  \varphi_{\alpha}-\varphi_{\beta
}\right)  }\right\vert ^{2}}{\left(  1+\delta^{2}\right)  ^{2}}\nonumber\\
&  =\frac{1+2\delta^{2}\cos\left(  \varphi_{\alpha}-\varphi_{\beta}\right)
+\delta^{4}}{\left(  1+\delta^{2}\right)  ^{2}}\nonumber\\
&  =\left\vert \left\langle \mathcal{A}|\mathcal{B}\right\rangle \right\vert
^{2}\text{.} \label{matter}%
\end{align}
It is worthwhile pointing out that Eq. (\ref{matter}) can be conveniently
recast as%
\begin{equation}
\frac{\left\vert \left\langle A|B\right\rangle \right\vert ^{2}}{\left\langle
A|A\right\rangle \left\langle B|B\right\rangle }=1-4\frac{\delta^{2}}{\left(
1+\delta^{2}\right)  ^{2}}\sin^{2}\left(  \frac{\varphi_{\alpha}%
-\varphi_{\beta}}{2}\right)  =\cos^{2}\left(  \frac{\theta_{AB}}{2}\right)
\text{,} \label{matter1}%
\end{equation}
with $\theta_{AB}$ being the geodesic distance between $\left\vert
A\right\rangle $ and $\left\vert B\right\rangle $,%
\begin{equation}
\theta_{AB}\overset{\text{def}}{=}2\arccos\left(  \frac{\left\vert
\left\langle A|B\right\rangle \right\vert }{\sqrt{\left\langle
A|A\right\rangle }\sqrt{\left\langle B|B\right\rangle }}\right)  \text{.}%
\end{equation}
We are now ready to construct the time suboptimal Hamiltonian in an explicit
manner. Observe that $\mathrm{H}=E\left[  -\left\vert E_{1}\right\rangle
\left\langle E_{1}\right\vert +\left\vert E_{2}\right\rangle \left\langle
E_{2}\right\vert \right]  $. Using Eq. (\ref{Jyo}), we can express $\left\vert
E_{1}\right\rangle \left\langle E_{1}\right\vert $ and $\left\vert
E_{2}\right\rangle \left\langle E_{2}\right\vert $ in terms of $\left\vert
\mathcal{A}\right\rangle \left\langle \mathcal{A}\right\vert $, $\left\vert
\mathcal{A}\right\rangle \left\langle \mathcal{B}\right\vert $, $\left\vert
\mathcal{B}\right\rangle \left\langle \mathcal{A}\right\vert $, and
$\left\vert \mathcal{B}\right\rangle \left\langle \mathcal{B}\right\vert
$.\ We have,
\begin{equation}
\left\vert E_{1}\right\rangle \left\langle E_{1}\right\vert =\frac
{1+\delta^{2}}{\left\vert e^{i\frac{\varphi_{\alpha}+\varphi_{\beta}}{2}%
}-e^{i\varphi_{\alpha}}e^{i\frac{\varphi_{\alpha}-\varphi_{\beta}}{2}%
}\right\vert ^{2}}\left[  \left\vert \mathcal{A}\right\rangle \left\langle
\mathcal{A}\right\vert -e^{-i\varphi_{\alpha}}e^{i\frac{\varphi_{\alpha
}+\varphi_{\beta}}{2}}\left\vert \mathcal{A}\right\rangle \left\langle
\mathcal{B}\right\vert -e^{i\varphi_{\alpha}}e^{-i\frac{\varphi_{\alpha
}+\varphi_{\beta}}{2}}\left\vert \mathcal{B}\right\rangle \left\langle
\mathcal{A}\right\vert +\left\vert \mathcal{B}\right\rangle \left\langle
\mathcal{B}\right\vert \right]  \text{,} \label{dawn1}%
\end{equation}
and,%
\begin{equation}
\left\vert E_{2}\right\rangle \left\langle E_{2}\right\vert =\frac
{1+\delta^{2}}{\delta^{2}\left\vert e^{i\frac{\varphi_{\alpha}+\varphi_{\beta
}}{2}}-e^{i\varphi_{\alpha}}e^{i\frac{\varphi_{\alpha}-\varphi_{\beta}}{2}%
}\right\vert ^{2}}\left[  \left\vert \mathcal{A}\right\rangle \left\langle
\mathcal{A}\right\vert -e^{i\frac{\varphi_{\alpha}-\varphi_{\beta}}{2}%
}\left\vert \mathcal{A}\right\rangle \left\langle \mathcal{B}\right\vert
-e^{-i\frac{\varphi_{\alpha}-\varphi_{\beta}}{2}}\left\vert \mathcal{B}%
\right\rangle \left\langle \mathcal{A}\right\vert +\left\vert \mathcal{B}%
\right\rangle \left\langle \mathcal{B}\right\vert \right]  \text{.}
\label{dawn2}%
\end{equation}
After some algebra, inserting Eqs. (\ref{dawn1}) and (\ref{dawn2}) into
$\mathrm{H}=E\left[  -\left\vert E_{1}\right\rangle \left\langle
E_{1}\right\vert +\left\vert E_{2}\right\rangle \left\langle E_{2}\right\vert
\right]  $, we get%
\begin{align}
\mathrm{H}  &  =E\frac{1-\delta^{2}}{\delta^{2}}\frac{1+\delta^{2}}{\left\vert
e^{i\frac{\varphi_{\alpha}+\varphi_{\beta}}{2}}-e^{i\varphi_{\alpha}}%
e^{i\frac{\varphi_{\alpha}-\varphi_{\beta}}{2}}\right\vert ^{2}}\left(
\left\vert \mathcal{A}\right\rangle \left\langle \mathcal{A}\right\vert
+\left\vert \mathcal{B}\right\rangle \left\langle \mathcal{B}\right\vert
\right)  +\nonumber\\
&  +E\frac{1+\delta^{2}}{\delta^{2}}\frac{1}{\left\vert e^{i\frac
{\varphi_{\alpha}+\varphi_{\beta}}{2}}-e^{i\varphi_{\alpha}}e^{i\frac
{\varphi_{\alpha}-\varphi_{\beta}}{2}}\right\vert ^{2}}\left[  \left(
\delta^{2}e^{-i\frac{\varphi_{\alpha}-\varphi_{\beta}}{2}}-e^{i\frac
{\varphi_{\alpha}-\varphi_{\beta}}{2}}\right)  \left\vert \mathcal{A}%
\right\rangle \left\langle \mathcal{B}\right\vert +\left(  \delta^{2}%
e^{i\frac{\varphi_{\alpha}-\varphi_{\beta}}{2}}-e^{-i\frac{\varphi_{\alpha
}-\varphi_{\beta}}{2})}\right)  \left\vert \mathcal{B}\right\rangle
\left\langle \mathcal{A}\right\vert \right]  \text{.} \label{GIA1}%
\end{align}
To further simplify the expression of $\mathrm{H}$ in Eq. (\ref{GIA1}) and
express the time suboptimal Hamiltonian in terms of $\left\vert A\right\rangle
\left\langle A\right\vert $, $\left\vert A\right\rangle \left\langle
B\right\vert $, $\left\vert B\right\rangle \left\langle A\right\vert $, and
$\left\vert B\right\rangle \left\langle B\right\vert $, we realize that%
\begin{equation}
\left\vert e^{i\frac{\varphi_{\alpha}+\varphi_{\beta}}{2}}-e^{i\varphi
_{\alpha}}e^{i\frac{\varphi_{\alpha}-\varphi_{\beta}}{2}}\right\vert
^{2}=4\sin^{2}\left(  \frac{\varphi_{\alpha}-\varphi_{\beta}}{2}\right)
\text{.} \label{yeye0}%
\end{equation}
From Eq. (\ref{Jchi6}), we note that $\left\vert \mathcal{A}\right\rangle
\left\langle \mathcal{A}\right\vert $ and $\left\vert \mathcal{B}\right\rangle
\left\langle \mathcal{B}\right\vert $ can be recast as%
\begin{equation}
\left\vert \mathcal{A}\right\rangle \left\langle \mathcal{A}\right\vert
=\frac{\left\vert A\right\rangle \left\langle A\right\vert }{\left\langle
A\left\vert A\right.  \right\rangle }\text{, and }\left\vert \mathcal{B}%
\right\rangle \left\langle \mathcal{B}\right\vert =\frac{\left\vert
B\right\rangle \left\langle B\right\vert }{\left\langle B\left\vert B\right.
\right\rangle }\text{,} \label{yeyemeno1}%
\end{equation}
respectively. In addition, $\left\vert \mathcal{A}\right\rangle \left\langle
\mathcal{B}\right\vert $ and $\left\vert \mathcal{B}\right\rangle \left\langle
\mathcal{A}\right\vert $ become
\begin{equation}
\left\vert \mathcal{A}\right\rangle \left\langle \mathcal{B}\right\vert
=\frac{1}{1+\delta^{2}}\frac{e^{-i\frac{\varphi_{\alpha}-\varphi_{\beta}}{2}}%
}{\alpha_{1}\beta_{1}^{\ast}}\left\vert A\right\rangle \left\langle
B\right\vert \text{, and }\left\vert \mathcal{B}\right\rangle \left\langle
\mathcal{A}\right\vert =\frac{1}{1+\delta^{2}}\frac{e^{i\frac{\varphi_{\alpha
}-\varphi_{\beta}}{2}}}{\alpha_{1}^{\ast}\beta_{1}}\left\vert B\right\rangle
\left\langle A\right\vert \text{,} \label{yeye}%
\end{equation}
respectively. To further simplify Eq. (\ref{yeye}), we notice from Eq.
(\ref{Jchi5b}) that%
\begin{equation}
\alpha_{1}^{\ast}\beta_{1}=\frac{e^{i\frac{\varphi_{\alpha}-\varphi_{\beta}%
}{2}}}{e^{i\frac{\varphi_{\alpha}-\varphi_{\beta}}{2}}+\delta^{2}%
e^{-i\frac{\varphi_{\alpha}-\varphi_{\beta}}{2}}}\left\langle A\left\vert
B\right.  \right\rangle \text{,} \label{yeye2}%
\end{equation}
with $\alpha_{1}\beta_{1}^{\ast}=\left(  \alpha_{1}^{\ast}\beta_{1}\right)
^{\ast}$. Then, making use of Eq. (\ref{yeye2}), the relations for $\left\vert
\mathcal{A}\right\rangle \left\langle \mathcal{B}\right\vert $ and $\left\vert
\mathcal{B}\right\rangle \left\langle \mathcal{A}\right\vert $ in Eq.
(\ref{yeye}) reduce to%
\begin{equation}
\left\vert \mathcal{A}\right\rangle \left\langle \mathcal{B}\right\vert
=\frac{e^{-i\frac{\varphi_{\alpha}-\varphi_{\beta}}{2}}+\delta^{2}%
e^{i\frac{\varphi_{\alpha}-\varphi_{\beta}}{2}}}{1+\delta^{2}}\frac{\left\vert
A\right\rangle \left\langle B\right\vert }{\left\langle B\left\vert A\right.
\right\rangle }\text{, and }\left\vert \mathcal{B}\right\rangle \left\langle
\mathcal{A}\right\vert =\frac{e^{i\frac{\varphi_{\alpha}-\varphi_{\beta}}{2}%
}+\delta^{2}e^{-i\frac{\varphi_{\alpha}-\varphi_{\beta}}{2}}}{1+\delta^{2}%
}\frac{\left\vert B\right\rangle \left\langle A\right\vert }{\left\langle
A\left\vert B\right.  \right\rangle }\text{,} \label{yeye4}%
\end{equation}
respectively. Finally, employing Eqs. (\ref{yeye0}), (\ref{yeyemeno1}), and
(\ref{yeye4}), the time suboptimal Hamiltonian in Eq. (\ref{GIA1}) can be
recast as%
\begin{align}
\mathrm{H}  &  =E\frac{1-\delta^{2}}{\delta^{2}}\frac{1+\delta^{2}}{4\sin
^{2}\left(  \frac{\varphi_{\alpha}-\varphi_{\beta}}{2}\right)  }\left(
\frac{\left\vert A\right\rangle \left\langle A\right\vert }{\left\langle
A\left\vert A\right.  \right\rangle }+\frac{\left\vert B\right\rangle
\left\langle B\right\vert }{\left\langle B\left\vert B\right.  \right\rangle
}\right)  +\nonumber\\
&  +E\frac{1+\delta^{2}}{\delta^{2}}\frac{1}{4\sin^{2}\left(  \frac
{\varphi_{\alpha}-\varphi_{\beta}}{2}\right)  }\left[
\begin{array}
[c]{c}%
\left(  \delta^{2}e^{-i\frac{\varphi_{\alpha}-\varphi_{\beta}}{2}}%
-e^{i\frac{\varphi_{\alpha}-\varphi_{\beta}}{2}}\right)  \frac{e^{-i\frac
{\varphi_{\alpha}-\varphi_{\beta}}{2}}+\delta^{2}e^{i\frac{\varphi_{\alpha
}-\varphi_{\beta}}{2}}}{1+\delta^{2}}\frac{\left\vert A\right\rangle
\left\langle B\right\vert }{\left\langle B\left\vert A\right.  \right\rangle
}\\
+\left(  \delta^{2}e^{i\frac{\varphi_{\alpha}-\varphi_{\beta}}{2}}%
-e^{-i\frac{\varphi_{\alpha}-\varphi_{\beta}}{2})}\right)  \frac
{e^{i\frac{\varphi_{\alpha}-\varphi_{\beta}}{2}}+\delta^{2}e^{-i\frac
{\varphi_{\alpha}-\varphi_{\beta}}{2}}}{1+\delta^{2}}\frac{\left\vert
B\right\rangle \left\langle A\right\vert }{\left\langle A\left\vert B\right.
\right\rangle }%
\end{array}
\right]  \text{.} \label{hsub}%
\end{align}
Following additional algebraic manipulation and the application of Eq.
(\ref{matter1}), the Hamiltonian $\mathrm{H}$ as presented in Eq. (\ref{hsub})
ultimately simplifies to
\begin{equation}
\mathrm{H}_{\mathrm{subopt}}=E\frac{\frac{1-\delta^{2}}{1+\delta^{2}}}%
{1-\cos^{2}\left(  \frac{\theta_{AB}}{2}\right)  }\left[  \frac{\left\vert
A\right\rangle \left\langle A\right\vert }{\left\langle A\left\vert A\right.
\right\rangle }+\frac{\left\vert B\right\rangle \left\langle B\right\vert
}{\left\langle B\left\vert B\right.  \right\rangle }-\left(  \frac{\left\vert
A\right\rangle \left\langle B\right\vert }{\left\langle B\left\vert A\right.
\right\rangle }+\frac{\left\vert B\right\rangle \left\langle A\right\vert
}{\left\langle A\left\vert B\right.  \right\rangle }\right)  \right]
+iE\cot\left(  \frac{\varphi_{\alpha}-\varphi_{\beta}}{2}\right)  \left(
\frac{\left\vert B\right\rangle \left\langle A\right\vert }{\left\langle
A\left\vert B\right.  \right\rangle }-\frac{\left\vert A\right\rangle
\left\langle B\right\vert }{\left\langle B\left\vert A\right.  \right\rangle
}\right)  \text{,} \label{R9}%
\end{equation}
where $\mathrm{H}=\mathrm{H}^{\dagger}$, \textrm{tr}$\left(  \mathrm{H}%
\right)  =0$, $\left\langle \mathrm{H}\right\rangle =\left[  \left(
\delta^{2}-1\right)  /\left(  1+\delta^{2}\right)  \right]  E$, $\Delta
E=\left[  2\delta/\left(  1+\delta^{2}\right)  \right]  E$ and, most
importantly, $\left\vert \left\langle A|B\right\rangle \right\vert
^{2}/\left[  \left\langle A|A\right\rangle \left\langle B|B\right\rangle
\right]  =1-\left[  4\delta^{2}/\left(  1+\delta^{2}\right)  ^{2}\right]
\sin^{2}\left[  \left(  \varphi_{\alpha}-\varphi_{\beta}\right)  /2\right]
=\cos^{2}\left(  \theta_{AB}/2\right)  $. In summary, for the sake of
thoroughness, we also highlight that when $\delta=1$, $\mathrm{H}%
_{\mathrm{subopt}}$ in Eq. (\ref{R9}) reduces to \textrm{H}$_{\mathrm{opt}}$
in Eq. (\ref{amy11}).

Considering the Hamiltonians presented in Eqs. (\ref{amy11}) and (\ref{R9}),
we can examine the quantum evolution across a wide range of initial and final
states, encompassing any level of entanglement. We will commence these
investigations in the following section.

\section{Applications}

We begin this section by examining two-qubit stationary evolutions that
transition between separable and maximally entangled states. Our study
encompasses four distinct types of Hamiltonian evolutions, utilizing both
geometric and entanglement quantifiers. The four scenarios under consideration
are: i) Time optimal evolutions between nonorthogonal quantum states; ii) Time
suboptimal quantum evolutions between nonorthogonal states; iii) Time optimal
evolutions between orthogonal quantum states; iv) Time suboptimal quantum
evolutions between orthogonal states. In Table III, we present a summary of
the characteristics pertaining to average entanglement, average entanglement
speed, and the nonlocal properties of the unitary time propagators that will
be examined in the subsequent subsections.

To gain preliminary physical insights, we restrict our attention to particular
Hamiltonian evolutions and to selected pairs of nonorthogonal and orthogonal
initial and final states. A comprehensive treatment would require a systematic
analysis of arbitrary Hamiltonians together with more general boundary
conditions. This extended investigation is beyond the scope of the present
work and is deferred to future research. \begin{table}[t]
\centering
\begin{tabular}
[c]{c|c|c|c|c}\hline\hline
\textbf{Evolution} & \textbf{Orthogonality} & \textbf{Average entanglement,
}$\mathrm{\bar{C}}$ & \textbf{Average entanglement speed, }$\bar
{v}_{\mathrm{C}}$ & \textbf{Nonlocal character, }$\varepsilon_{\mathrm{EP}%
}^{\mathrm{Yukalov}}$\\\hline
Optimal & No & Lower & Higher & Higher\\\hline
Suboptimal & No & Higher & Lower & High\\\hline
Optimal & Yes & Lower & Higher & High\\\hline
Suboptimal & Yes & Higher & Lower & Higher\\\hline
\end{tabular}
\caption{Summary of the characteristics of average entanglement $\mathrm{\bar
{C}}$, average entanglement speed $\bar{v}_{\mathrm{C}}$, and the nonlocal
properties of unitary time propagators quantified by $\varepsilon
_{\mathrm{EP}}^{\mathrm{Yukalov}}$. We consider both time-optimal and
time-suboptimal quantum evolutions between separable and maximally entangled
states. Furthermore, both scenarios in which this pair of states is either
nonorthogonal or orthogonal are examined.}%
\end{table}

\subsection{Time optimal evolution between nonorthogonal states}

In the initial application, we examine the time optimal evolution between a
separable quantum state and a maximally entangled quantum state that are not orthogonal.

Specifically, we wish to evolve from $\left\vert A\right\rangle \overset
{\text{def}}{=}\left\vert 00\right\rangle $ to $\left\vert B\right\rangle
\overset{\text{def}}{=}\left(  \left\vert 00\right\rangle +\left\vert
11\right\rangle \right)  /\sqrt{2}$ in a time-optimal manner, assuming $\Delta
E\overset{\text{def}}{=}E/\sqrt{2}$. The choice of taking $\Delta
E\overset{\text{def}}{=}E/\sqrt{2}$ is dictated by the fact that this choice
allows for an energetically fair comparison with the time suboptimal evolution
between the same nonorthogonal initial and final states $\left\vert
A\right\rangle $ and $\left\vert B\right\rangle $, respectively, that we
consider in the next example. Before finding the time optimal Hamiltonian, we
can observe that in this scenario, the geodesic distance $s_{0}$ between
$\left\vert A\right\rangle $ and $\left\vert B\right\rangle $ is
$s_{0}=2\arccos\left[  \left\vert \left\langle A\left\vert B\right.
\right\rangle \right\vert \right]  =\pi/2$. Therefore, assuming $\Delta
E\overset{\text{def}}{=}E/\sqrt{2}$, the optimal time to arrive at $\left\vert
B\right\rangle $ from $\left\vert A\right\rangle $ is given by
$t_{\mathrm{opt}}=\left[  \hslash\left(  \pi/4\right)  \right]  /(E/\sqrt
{2})=\left(  \hslash\pi\right)  /(2\sqrt{2}E)$. Having stated that, the matrix
representation of the time optimal Hamiltonian in the canonical basis
$\mathcal{B}_{\mathcal{H}_{2}^{2}}\overset{\text{def}}{=}\left\{  \left\vert
00\right\rangle \text{, }\left\vert 01\right\rangle \text{, }\left\vert
10\right\rangle \text{, }\left\vert 11\right\rangle \right\}  $ of
$\mathcal{H}_{2}^{2}$ (i.e., the Hilbert space of two-qubit quantum states) is
given by
\begin{equation}
\mathrm{H}_{\mathrm{opt}}=\frac{E}{\sqrt{2}}\left(
\begin{array}
[c]{cccc}%
0 & 0 & 0 & -i\\
0 & 0 & 0 & 0\\
0 & 0 & 0 & 0\\
i & 0 & 0 & 0
\end{array}
\right)  \text{.} \label{amore1}%
\end{equation}
For completeness, we note from Eq. (\ref{amore1}) that $\mathrm{H}%
_{\mathrm{opt}}=\mathrm{H}_{\mathrm{opt}}^{\dagger}$, \textrm{tr}%
$(\mathrm{H}_{\mathrm{opt}})=0$, $\Delta E^{2}=\left\langle \mathrm{H}%
_{\mathrm{opt}}^{2}\right\rangle -\left\langle \mathrm{H}_{\mathrm{opt}%
}\right\rangle ^{2}=E^{2}/2$, $\eta_{\mathrm{geo}}=1$ (since $s=s_{0}$),
$\eta_{\mathrm{Uzdin}}=1$, and $\kappa_{\mathrm{AC}}^{2}=0$. After
diagonalizing the matrix in Eq. (\ref{amore1}), we find that the corresponding
unitary time propagator $U_{\mathrm{opt}}(t)=e^{-\frac{i}{\hslash}%
\mathrm{H}_{\mathrm{opt}}t}$ is given by%
\begin{equation}
U_{\mathrm{opt}}(t)=\left(
\begin{array}
[c]{cccc}%
\cos(\frac{1}{\sqrt{2}}\frac{E}{\hslash}t) & 0 & 0 & -\sin(\frac{1}{\sqrt{2}%
}\frac{E}{\hslash}t)\\
0 & 1 & 0 & 0\\
0 & 0 & 1 & 0\\
\sin(\frac{1}{\sqrt{2}}\frac{E}{\hslash}t) & 0 & 0 & \cos(\frac{1}{\sqrt{2}%
}\frac{E}{\hslash}t)
\end{array}
\right)  \text{,} \label{prop1}%
\end{equation}
with $U_{\mathrm{opt}}(t)U_{\mathrm{opt}}^{\dagger}(t)=U_{\mathrm{opt}%
}^{\dagger}(t)U_{\mathrm{opt}}(t)=\mathbf{1}_{4\times4}$, and $U_{\mathrm{opt}%
}(t_{\mathrm{opt}})\left\vert A\right\rangle =\left\vert B\right\rangle $.
From an entanglement standpoint, we note that the entanglement $\mathrm{C}%
\left(  \gamma\left(  t\right)  \right)  $ of the path $\gamma\left(
t\right)  :t\mapsto\left\vert \psi\left(  t\right)  \right\rangle
\overset{\text{def}}{=}U_{\mathrm{opt}}(t)\left\vert A\right\rangle $ is given
by $\mathrm{C}\left(  t\right)  =\mathrm{C}\left[  \left\vert \psi\left(
t\right)  \right\rangle \right]  =\left\vert \sin(\sqrt{2}\frac{E}{\hslash
}t)\right\vert =\sin(\sqrt{2}\frac{E}{\hslash}t)$ for $0\leq t\leq\left(
\hslash\pi\right)  /(2\sqrt{2}E)$. Moreover, the average path entanglement
during the evolution defined as%
\begin{equation}
\mathrm{\bar{C}}=\frac{1}{t_{\mathrm{opt}}}\int_{0}^{t_{\mathrm{opt}}%
}\mathrm{C}\left(  t\right)  dt\text{,}%
\end{equation}
is equal to $2/\pi\simeq0.64$. Finally, to quantify the nonlocal and
entangling character of the time propagator in Eq. (\ref{prop1}), we calculate
the entanglement production $\varepsilon_{\mathrm{EP}}^{\mathrm{Yukalov}%
}\left(  t\right)  $ in Eq. (\ref{russia}). After some algebra, we find%
\begin{equation}
\varepsilon_{\mathrm{EP}}^{\mathrm{Yukalov}}\left(  t\right)  =\frac{1}{2}%
\log\left\{  \frac{4}{\left[  1+\cos(\frac{1}{\sqrt{2}}\frac{E}{\hslash
}t)\right]  ^{2}}\right\}  \text{.} \label{chiappa1}%
\end{equation}
In the short-time limit, $\varepsilon_{\mathrm{EP}}^{\mathrm{Yukalov}}\left(
t\right)  $ exhibits a polynomially quadratic growth specified by the
relation
\begin{equation}
\varepsilon_{\mathrm{EP}}^{\mathrm{Yukalov}}\left(  t\right)  =\frac{1}%
{8}\left(  \frac{E}{\hslash}\right)  ^{2}t^{2}+\frac{1}{384}\left(  \frac
{E}{\hslash}\right)  ^{4}t^{4}+\mathcal{O}\left(  t^{6}\right)  \text{.}
\label{chiappa2}%
\end{equation}
We observe, as expected, that $\varepsilon_{\mathrm{EP}}^{\mathrm{Yukalov}%
}\left(  0\right)  =0$ in Eq. (\ref{chiappa1}).

\subsection{Time suboptimal evolution between nonorthogonal states}

In the second application, we analyze the time suboptimal evolution that takes
place between a separable quantum state and a maximally entangled quantum
state that are not orthogonal.

We want to go from $\left\vert A\right\rangle \overset{\text{def}}%
{=}\left\vert 00\right\rangle $ to $\left\vert B\right\rangle \overset
{\text{def}}{=}\left(  \left\vert 00\right\rangle +\left\vert 11\right\rangle
\right)  /\sqrt{2}$ in a suboptimal way with an Hamiltonian with energy
variance $\Delta E^{2}\overset{\text{def}}{=}(1/2)E^{2}$. Recall that the
formal expression of a suboptimal Hamiltonian that connects $\left\vert
A\right\rangle $ to $\left\vert B\right\rangle $ is given in Eq. (\ref{R9}),
with $\left\vert \left\langle A\left\vert B\right.  \right\rangle \right\vert
^{2}$ satisfying the condition in Eq. (\ref{matter1}). Note that $\theta_{AB}$
in Eq. (\ref{matter1}) is the geodesic distance $s_{0}$ which, in our case,
equals $\pi/2$. Furthermore, since the variance $\Delta E^{2}$ of
$\mathrm{H}_{\mathrm{subopt}}$ in Eq. (\ref{R9}) is $\left[  4\delta
^{2}/(1+\delta^{2})^{2}\right]  E^{2}$, we set in Eq. (\ref{matter1})
$\delta\overset{\text{def}}{=}1+\sqrt{2}$ and $\varphi_{\alpha}-\varphi
_{\beta}\overset{\text{def}}{=}\pi$. This way, with $\theta_{AB}=\pi/2$,
$\delta\overset{\text{def}}{=}1+\sqrt{2}$, and $\varphi_{\alpha}%
-\varphi_{\beta}\overset{\text{def}}{=}\pi$, Eq. (\ref{matter1}) is properly
satisfied. Having said that, the matrix representation of the time suboptimal
Hamiltonian in the canonical basis $\mathcal{B}_{\mathcal{H}_{2}^{2}}%
\overset{\text{def}}{=}\left\{  \left\vert 00\right\rangle \text{, }\left\vert
01\right\rangle \text{, }\left\vert 10\right\rangle \text{, }\left\vert
11\right\rangle \right\}  $ of $\mathcal{H}_{2}^{2}$ is given by%
\begin{equation}
\mathrm{H}_{\mathrm{subopt}}=\frac{E}{\sqrt{2}}\left(
\begin{array}
[c]{cccc}%
1 & 0 & 0 & 1\\
0 & 0 & 0 & 0\\
0 & 0 & 0 & 0\\
1 & 0 & 0 & -1
\end{array}
\right)  \text{.} \label{amore2}%
\end{equation}
For thoroughness, we note from Eq. (\ref{amore2}) that $\mathrm{H}%
_{\mathrm{subopt}}=\mathrm{H}_{\mathrm{subopt}}^{\dagger}$, \textrm{tr}%
$(\mathrm{H}_{\mathrm{subopt}})=0$, $\Delta E^{2}=\left\langle \mathrm{H}%
_{\mathrm{subopt}}^{2}\right\rangle -\left\langle \mathrm{H}_{\mathrm{subopt}%
}\right\rangle ^{2}=E^{2}/2$, $\eta_{\mathrm{geo}}=1/\sqrt{2}<1$ (since
$s=\sqrt{2}s_{0}$), $\eta_{\mathrm{Uzdin}}=1/\sqrt{2}<1$, and the curvature
coefficient in Eq. (\ref{curvatime11}) does not depend on the energy $E$ since
it equals $\kappa_{\mathrm{AC}}^{2}=4$. After diagonalizing the matrix in Eq.
(\ref{amore2}), we find that the corresponding unitary time propagator
$U_{\mathrm{subopt}}(t)=e^{-\frac{i}{\hslash}\mathrm{H}_{\mathrm{subopt}}t}$
is given by%
\begin{equation}
U_{\mathrm{subopt}}(t)=\left(
\begin{array}
[c]{cccc}%
\cos(\frac{E}{\hslash}t)-\frac{i}{\sqrt{2}}\sin(\frac{E}{\hslash}t) & 0 & 0 &
-\frac{i}{\sqrt{2}}\sin(\frac{E}{\hslash}t)\\
0 & 1 & 0 & 0\\
0 & 0 & 1 & 0\\
-\frac{i}{\sqrt{2}}\sin(\frac{E}{\hslash}t) & 0 & 0 & \cos(\frac{E}{\hslash
}t)+\frac{i}{\sqrt{2}}\sin(\frac{E}{\hslash}t)
\end{array}
\right)  \text{,} \label{prop2}%
\end{equation}
with $U_{\mathrm{subopt}}(t)U_{\mathrm{subopt}}^{\dagger}%
(t)=U_{\mathrm{subopt}}^{\dagger}(t)U_{\mathrm{subopt}}(t)=\mathbf{1}%
_{4\times4}$, and $U_{\mathrm{subopt}}(t_{\mathrm{\ast}})\left\vert
A\right\rangle =\left\vert B\right\rangle $ with $t_{\mathrm{\ast}}=\left(
\hslash\pi\right)  /(2E)=\sqrt{2}t_{\mathrm{opt}}>t_{\mathrm{opt}}$. From an
entanglement standpoint, we note that the entanglement $\mathrm{C}\left(
\gamma\left(  t\right)  \right)  $ of the path $\gamma\left(  t\right)
:t\mapsto\left\vert \psi\left(  t\right)  \right\rangle \overset{\text{def}%
}{=}U_{\mathrm{subopt}}(t)\left\vert A\right\rangle $ is given by
$\mathrm{C}\left(  t\right)  =\mathrm{C}\left[  \left\vert \psi\left(
t\right)  \right\rangle \right]  =2\sqrt{\frac{3}{16}-\frac{1}{16}\cos
^{2}\left(  2\frac{E}{\hslash}t\right)  -\frac{1}{8}\cos\left(  2\frac
{E}{\hslash}t\right)  }$ for $0\leq t\leq\left(  \hslash\pi\right)  /(2E)$.
Moreover, the average path entanglement during the evolution, defined as
$\mathrm{\bar{C}}\overset{\text{def}}{=}(1/t_{\mathrm{\ast}})\int
_{0}^{t_{\mathrm{\ast}}}\mathrm{C}\left(  t\right)  dt$, is equal to $\left[
2\sqrt{2}+\cosh^{-1}(3)\right]  /(2\pi)$ $\simeq0.73$ where \textquotedblleft%
$\cosh^{-1}$\textquotedblright\ is the inverse hyperbolic cosine function.
Finally, to quantify the nonlocal entangling character of the time propagator
in Eq. (\ref{prop2}), we calculate the entanglement production $\varepsilon
_{\mathrm{EP}}^{\mathrm{Yukalov}}\left(  t\right)  $ in Eq. (\ref{russia}).
After some algebra, we arrive at%
\begin{equation}
\varepsilon_{\mathrm{EP}}^{\mathrm{Yukalov}}\left(  t\right)  =\frac{1}{2}%
\log\left\{  \frac{4\left[  2\cos(\frac{E}{\hslash}t)+2\right]  ^{2}}{\left[
\cos^{2}(\frac{E}{\hslash}t)+4\cos(\frac{E}{\hslash}t)+3\right]  ^{2}%
}\right\}  \text{.} \label{chiappa3}%
\end{equation}
In the short-time limit, $\varepsilon_{\mathrm{EP}}^{\mathrm{Yukalov}}\left(
t\right)  $ displays a polynomially quadratic growth defined by the
relationship
\begin{equation}
\varepsilon_{\mathrm{EP}}^{\mathrm{Yukalov}}\left(  t\right)  =\frac{1}%
{8}\left(  \frac{E}{\hslash}\right)  ^{2}t^{2}-\frac{1}{384}\left(  \frac
{E}{\hslash}\right)  ^{4}t^{4}+\mathcal{O}\left(  t^{6}\right)  \text{.}%
\end{equation}
It is noted, as anticipated, that $\varepsilon_{\mathrm{EP}}^{\mathrm{Yukalov}%
}\left(  0\right)  =0$ in Eq. (\ref{chiappa3}).

\subsection{Time optimal evolution between orthogonal states}

In the third application, we investigate the time optimal evolution between a
separable state and a maximally entangled state that are orthogonal.

Specifically, we wish to evolve from $\left\vert A\right\rangle \overset
{\text{def}}{=}\left\vert 00\right\rangle $ to $\left\vert B\right\rangle
\overset{\text{def}}{=}\left(  \left\vert 01\right\rangle +\left\vert
10\right\rangle \right)  /\sqrt{2}$ in a time-optimal manner, assuming $\Delta
E\overset{\text{def}}{=}\sqrt{5/2}E$. The choice of taking $\Delta
E\overset{\text{def}}{=}\sqrt{5/2}E$ is dictated by the fact that this choice
allows for an energetically fair comparison with the time suboptimal evolution
between the same orthogonal initial and final states $\left\vert
A\right\rangle $ and $\left\vert B\right\rangle $, respectively, that we
consider in the next example. Before finding the time optimal Hamiltonian, we
can observe that in this scenario, the geodesic distance $s_{0}$ between
$\left\vert A\right\rangle $ and $\left\vert B\right\rangle $ is
$s_{0}=2\arccos\left[  \left\vert \left\langle A\left\vert B\right.
\right\rangle \right\vert \right]  =\pi$. Therefore, assuming $\Delta
E\overset{\text{def}}{=}\sqrt{5/2}E$, the optimal time to arrive at
$\left\vert B\right\rangle $ from $\left\vert A\right\rangle $ is given by
$t_{\mathrm{opt}}=\left[  \hslash\left(  \pi/2\right)  \right]  /(\sqrt
{5/2}E)=\left(  \hslash\pi\right)  /(\sqrt{10}E)$. \ Having mentioned that,
the matrix representation of the time optimal Hamiltonian $\mathrm{H}%
_{\mathrm{opt}}\overset{\text{def}}{=}i\Delta E\left[  \left\vert
B\right\rangle \left\langle A\right\vert -\left\vert A\right\rangle
\left\langle B\right\vert \right]  $ in the canonical basis $\mathcal{B}%
_{\mathcal{H}_{2}^{2}}\overset{\text{def}}{=}\left\{  \left\vert
00\right\rangle \text{, }\left\vert 01\right\rangle \text{, }\left\vert
10\right\rangle \text{, }\left\vert 11\right\rangle \right\}  $ of
$\mathcal{H}_{2}^{2}$ is,%
\begin{equation}
\mathrm{H}_{\mathrm{opt}}=E\frac{\sqrt{5}}{2}\left(
\begin{array}
[c]{cccc}%
0 & -i & -i & 0\\
i & 0 & 0 & 0\\
i & 0 & 0 & 0\\
0 & 0 & 0 & 0
\end{array}
\right)  \text{.} \label{amore3}%
\end{equation}
For exhaustiveness, we note from Eq. (\ref{amore3}) that $\mathrm{H}%
_{\mathrm{opt}}=\mathrm{H}_{\mathrm{opt}}^{\dagger}$, \textrm{tr}%
$(\mathrm{H}_{\mathrm{opt}})=0$, $\Delta E^{2}=\left\langle \mathrm{H}%
_{\mathrm{opt}}^{2}\right\rangle -\left\langle \mathrm{H}_{\mathrm{opt}%
}\right\rangle ^{2}=\left(  5/2\right)  E^{2}$, $\eta_{\mathrm{geo}}=1$ (since
$s=s_{0}=\pi$), $\eta_{\mathrm{Uzdin}}=1$, and $\kappa_{\mathrm{AC}}^{2}=0$.
After diagonalizing the matrix in Eq. (\ref{amore3}), we find that the
corresponding unitary time propagator $U_{\mathrm{opt}}(t)=e^{-\frac
{i}{\hslash}\mathrm{H}_{\mathrm{opt}}t}$ is given by%
\begin{equation}
U_{\mathrm{opt}}(t)=\left(
\begin{array}
[c]{cccc}%
\cos(\frac{\sqrt{10}}{2}\frac{E}{\hslash}t) & -\frac{1}{\sqrt{2}}\sin
(\frac{\sqrt{10}}{2}\frac{E}{\hslash}t) & -\frac{1}{\sqrt{2}}\sin(\frac
{\sqrt{10}}{2}\frac{E}{\hslash}t) & 0\\
\frac{1}{\sqrt{2}}\sin(\frac{\sqrt{10}}{2}\frac{E}{\hslash}t) & \cos^{2}%
(\frac{\sqrt{10}}{2}\frac{E}{\hslash}t) & -\sin^{2}(\frac{\sqrt{10}}{2}%
\frac{E}{\hslash}t) & 0\\
\frac{1}{\sqrt{2}}\sin(\frac{\sqrt{10}}{2}\frac{E}{\hslash}t) & -\sin
^{2}(\frac{\sqrt{10}}{2}\frac{E}{\hslash}t) & \cos^{2}(\frac{\sqrt{10}}%
{2}\frac{E}{\hslash}t) & 0\\
0 & 0 & 0 & 1
\end{array}
\right)  \text{,} \label{prop3}%
\end{equation}
with $U_{\mathrm{opt}}(t)U_{\mathrm{opt}}^{\dagger}(t)=U_{\mathrm{opt}%
}^{\dagger}(t)U_{\mathrm{opt}}(t)=\mathbf{1}_{4\times4}$, and $U_{\mathrm{opt}%
}(t_{\mathrm{opt}})\left\vert A\right\rangle =\left\vert B\right\rangle $
where $t_{\mathrm{opt}}\overset{\text{def}}{=}\left(  \hslash\pi\right)
/(\sqrt{10}E)$. From an entanglement standpoint, we note that the entanglement
$\mathrm{C}\left(  \gamma\left(  t\right)  \right)  $ of the path
$\gamma\left(  t\right)  :t\mapsto\left\vert \psi\left(  t\right)
\right\rangle \overset{\text{def}}{=}U_{\mathrm{opt}}(t)\left\vert
A\right\rangle $ is given by $\mathrm{C}\left(  t\right)  =\mathrm{C}\left[
\left\vert \psi\left(  t\right)  \right\rangle \right]  =\sin^{2}(\frac
{\sqrt{10}}{2}\frac{E}{\hslash}t)$ for $0\leq t\leq$ $t_{\mathrm{opt}}$.
Moreover, the average path entanglement during the evolution is $\mathrm{\bar
{C}}=1/2$. Finally, to quantify the nonlocal entangling character of the time
propagator in Eq. (\ref{prop3}), we calculate the entanglement production
$\varepsilon_{\mathrm{EP}}^{\mathrm{Yukalov}}\left(  t\right)  $ in Eq.
(\ref{russia}). After some algebraic manipulations, we obtain%
\begin{equation}
\varepsilon_{\mathrm{EP}}^{\mathrm{Yukalov}}\left(  t\right)  =\frac{1}{2}%
\log\left\{  \frac{4\left[  \allowbreak2\cos(\frac{\sqrt{10}}{2}\frac
{E}{\hslash}t)+2\right]  ^{2}}{\left[  \frac{3}{2}\cos^{2}(\frac{\sqrt{10}}%
{2}\frac{E}{\hslash}t)+3\cos(\frac{\sqrt{10}}{2}\frac{E}{\hslash}t)+\frac
{7}{2}\right]  ^{2}}\right\}  \text{.} \label{chiappa3b}%
\end{equation}
In the short-time limit, $\varepsilon_{\mathrm{EP}}^{\mathrm{Yukalov}}\left(
t\right)  $ shows a polynomially quadratic growth characterized by the
relation
\begin{equation}
\varepsilon_{\mathrm{EP}}^{\mathrm{Yukalov}}\left(  t\right)  =\frac{5}%
{16}\left(  \frac{E}{\hslash}\right)  ^{2}t^{2}+\mathcal{O}\left(
t^{4}\right)  \text{.}%
\end{equation}
It is observed, as foreseen, that $\varepsilon_{\mathrm{EP}}^{\mathrm{Yukalov}%
}\left(  0\right)  =0$ in Eq. (\ref{chiappa3b}).

\subsection{Time suboptimal evolution between orthogonal states}

In the fourth application, we study the time suboptimal evolution that takes
place between a separable quantum state and a maximally entangled quantum
state, which are orthogonal.

To construct a suboptimal stationary Hamiltonian that connects two orthogonal
states, it is necessary to exit the two-dimensional space spanned by initial
and final states $\left\vert A\right\rangle $ and $\left\vert B\right\rangle
$, respectively \cite{brody03}. Therefore, we cannot use our proposed
suboptimal Hamiltonian construction in Section IV. Instead, we need to build
an example by hand. We proceed as follows. Let us consider an Hamiltonian
whose spectral decomposition is given by,%
\begin{equation}
\mathrm{H}\overset{\text{def}}{=}\sum_{i=1}^{4}E_{i}\left\vert E_{i}%
\right\rangle \left\langle E_{i}\right\vert \text{,} \label{try1}%
\end{equation}
with $E_{1}\overset{\text{def}}{=}-2E$, $E_{2}\overset{\text{def}}{=}-E$,
$E_{3}\overset{\text{def}}{=}E$, $E_{4}\overset{\text{def}}{=}2E$, and
$\left\langle E_{i}\left\vert E_{j}\right.  \right\rangle =\delta_{ij}$ for
any $1\leq i$, $j\leq4$. We note that $\mathrm{H}$ in Eq. (\ref{try1}) is
capable of evolving $\left\vert A\right\rangle \overset{\text{def}}%
{=}(1/2)\left[  \left\vert E_{1}\right\rangle +\left\vert E_{2}\right\rangle
+\left\vert E_{3}\right\rangle +\left\vert E_{4}\right\rangle \right]  $ into
$\left\vert B\right\rangle \overset{\text{def}}{=}(1/2)\left[  \left\vert
E_{1}\right\rangle -\left\vert E_{2}\right\rangle -\left\vert E_{3}%
\right\rangle +\left\vert E_{4}\right\rangle \right]  $, with $\left\langle
A\left\vert B\right.  \right\rangle =0$ and $\Delta E=\sqrt{5/2}E$, in a time
$t_{\ast}=(\pi\hslash)/E>t_{\mathrm{opt}}=(\pi\hslash)/(\sqrt{10}E)$. Since we
wish to have $\left\vert A\right\rangle \overset{\text{def}}{=}\left\vert
00\right\rangle $ and $\left\vert B\right\rangle \overset{\text{def}}%
{=}\left(  \left\vert 01\right\rangle +\left\vert 10\right\rangle \right)
/\sqrt{2}$, we apply the Gram-Schmidt orthogonalization procedure to
ultimately find a suitable matrix transformation that allows us to transition
from the eigenvector basis $\mathcal{B}_{\mathrm{diag}}\overset{\text{def}}%
{=}\left\{  \left\vert E_{1}\right\rangle \text{, }\left\vert E_{2}%
\right\rangle \text{, }\left\vert E_{3}\right\rangle \text{, }\left\vert
E_{4}\right\rangle \right\}  $ to the canonical computational basis
$\mathcal{B}_{\mathrm{can}}\overset{\text{def}}{=}\left\{  \left\vert
00\right\rangle \text{, }\left\vert 01\right\rangle \text{, }\left\vert
10\right\rangle \text{, }\left\vert 11\right\rangle \right\}  $. Specifically,
we arrive at the following two matricial relations%
\begin{equation}
\left(
\begin{array}
[c]{c}%
\left\vert 00\right\rangle \\
\left\vert 01\right\rangle \\
\left\vert 10\right\rangle \\
\left\vert 11\right\rangle
\end{array}
\right)  =\left(
\begin{array}
[c]{cccc}%
\frac{1}{2} & \frac{1}{2} & \frac{1}{2} & \frac{1}{2}\\
\frac{1}{\sqrt{2}} & -\frac{1}{\sqrt{2}} & 0 & 0\\
0 & 0 & -\frac{1}{\sqrt{2}} & \frac{1}{\sqrt{2}}\\
\frac{1}{2} & \frac{1}{2} & -\frac{1}{2} & -\frac{1}{2}%
\end{array}
\right)  \left(
\begin{array}
[c]{c}%
\left\vert E_{1}\right\rangle \\
\left\vert E_{2}\right\rangle \\
\left\vert E_{3}\right\rangle \\
\left\vert E_{4}\right\rangle
\end{array}
\right)  \text{, and }\left(
\begin{array}
[c]{c}%
\left\vert E_{1}\right\rangle \\
\left\vert E_{2}\right\rangle \\
\left\vert E_{3}\right\rangle \\
\left\vert E_{4}\right\rangle
\end{array}
\right)  =\left(
\begin{array}
[c]{cccc}%
\frac{1}{2} & \frac{1}{\sqrt{2}} & 0 & \frac{1}{2}\\
\frac{1}{2} & -\frac{1}{\sqrt{2}} & 0 & \frac{1}{2}\\
\frac{1}{2} & 0 & -\frac{1}{\sqrt{2}} & -\frac{1}{2}\\
\frac{1}{2} & 0 & \frac{1}{\sqrt{2}} & -\frac{1}{2}%
\end{array}
\right)  \left(
\begin{array}
[c]{c}%
\left\vert 00\right\rangle \\
\left\vert 01\right\rangle \\
\left\vert 10\right\rangle \\
\left\vert 11\right\rangle
\end{array}
\right)  \text{.} \label{care}%
\end{equation}
From the second relation in Eq. (\ref{care}), the eigenvector matrix
$M_{\mathrm{H}}$ and its inverse $M_{\mathrm{H}}^{\dagger}$ are given by%
\begin{equation}
M_{\mathrm{H}}=\left(
\begin{array}
[c]{cccc}%
\frac{1}{2} & \frac{1}{2} & \frac{1}{2} & \frac{1}{2}\\
\frac{1}{\sqrt{2}} & -\frac{1}{\sqrt{2}} & 0 & 0\\
0 & 0 & -\frac{1}{\sqrt{2}} & \frac{1}{\sqrt{2}}\\
\frac{1}{2} & \frac{1}{2} & -\frac{1}{2} & -\frac{1}{2}%
\end{array}
\right)  \text{,} \label{baby}%
\end{equation}
and $M_{\mathrm{H}}^{\dagger}$ being the transpose of $M_{\mathrm{H}}$ in Eq.
(\ref{baby}), respectively. Finally, we arrive at the matrix representation of
our time suboptimal Hamiltonian in Eq. (\ref{try1}) with respect to the
computational basis $\mathcal{B}_{\mathrm{can}}$. We obtain,%
\begin{equation}
\left[  \mathrm{H}_{\mathrm{subopt}}\right]  _{\mathcal{B}_{\mathrm{can}}%
}=M_{\mathrm{H}}\left[  \mathrm{H}_{\mathrm{subopt}}\right]  _{\mathcal{B}%
_{\mathrm{diag}}}M_{\mathrm{H}}^{\dagger}=\mathrm{H}=E\left(
\begin{array}
[c]{cccc}%
0 & -\frac{1}{2\sqrt{2}} & \frac{1}{2\sqrt{2}} & -\frac{3}{2}\\
-\frac{1}{2\sqrt{2}} & -\frac{3}{2} & 0 & -\frac{1}{2\sqrt{2}}\\
\frac{1}{2\sqrt{2}} & 0 & \frac{3}{2} & -\frac{1}{2\sqrt{2}}\\
-\frac{3}{2} & -\frac{1}{2\sqrt{2}} & -\frac{1}{2\sqrt{2}} & 0
\end{array}
\right)  \text{.} \label{memory}%
\end{equation}
In what follows, to avoid a cumbersome notation, we simply denote $\left[
\mathrm{H}_{\mathrm{subopt}}\right]  _{\mathcal{B}_{\mathrm{can}}}$ in Eq.
(\ref{memory}) as $\mathrm{H}_{\mathrm{subopt}}$. For the sake of
completeness, we notice from Eq. (\ref{memory}) that $\mathrm{H}%
_{\mathrm{subopt}}=\mathrm{H}_{\mathrm{subopt}}^{\dagger}$, \textrm{tr}%
$(\mathrm{H}_{\mathrm{subopt}})=0$, $\Delta E^{2}=\left\langle \mathrm{H}%
_{\mathrm{opt}}^{2}\right\rangle -\left\langle \mathrm{H}_{\mathrm{opt}%
}\right\rangle ^{2}=\left(  5/2\right)  E^{2}$, $\eta_{\mathrm{geo}}<1$ (since
$s=\sqrt{10}\pi>s_{0}=\pi$), $\eta_{\mathrm{Uzdin}}=(1/2)(\sqrt{5/2})<1$, and
$\kappa_{\mathrm{AC}}^{2}=9/25$. After diagonalizing the matrix in Eq.
(\ref{memory}), we find that the corresponding unitary time propagator
$U_{\mathrm{subopt}}(t)=e^{-\frac{i}{\hslash}\mathrm{H}_{\mathrm{subopt}}t}$
reduces to%
\begin{equation}
U_{\mathrm{subopt}}\left(  t\right)  \allowbreak=\left(
\begin{array}
[c]{cccc}%
\frac{\cos(\frac{E}{\hslash}t)+\cos(2\frac{E}{\hslash}t)}{2} & \frac{i}%
{\sqrt{2}}e^{i\frac{3}{2}\frac{E}{\hslash}t}\sin(\frac{1}{2}\frac{E}{\hslash
}t) & -\frac{i}{\sqrt{2}}e^{-i\frac{3}{2}\frac{E}{\hslash}t}\sin(\frac{1}%
{2}\frac{E}{\hslash}t) & i\frac{\sin(\frac{E}{\hslash}t)+\sin(2\frac
{E}{\hslash}t)}{2}\\
\frac{i}{\sqrt{2}}e^{i\frac{3}{2}\frac{E}{\hslash}t}\sin(\frac{1}{2}\frac
{E}{\hslash}t) & e^{i\frac{3}{2}\frac{E}{\hslash}t}\cos(\frac{1}{2}\frac
{E}{\hslash}t) & 0 & \frac{i}{\sqrt{2}}e^{i\frac{3}{2}\frac{E}{\hslash}t}%
\sin(\frac{1}{2}\frac{E}{\hslash}t)\\
-\frac{i}{\sqrt{2}}e^{-i\frac{3}{2}\frac{E}{\hslash}t}\sin(\frac{1}{2}\frac
{E}{\hslash}t) & 0 & e^{-i\frac{3}{2}\frac{E}{\hslash}t}\cos(\frac{1}{2}%
\frac{E}{\hslash}t) & \frac{i}{\sqrt{2}}e^{-i\frac{3}{2}\frac{E}{\hslash}%
t}\sin(\frac{1}{2}\frac{E}{\hslash}t)\\
i\frac{\sin(\frac{E}{\hslash}t)+\sin(2\frac{E}{\hslash}t)}{2} & \frac{i}%
{\sqrt{2}}e^{i\frac{3}{2}\frac{E}{\hslash}t}\sin(\frac{1}{2}\frac{E}{\hslash
}t) & \frac{i}{\sqrt{2}}e^{-i\frac{3}{2}\frac{E}{\hslash}t}\sin(\frac{1}%
{2}\frac{E}{\hslash}t) & \frac{\cos(\frac{E}{\hslash}t)+\cos(2\frac{E}%
{\hslash}t)}{2}%
\end{array}
\right)  \text{,} \label{fire}%
\end{equation}
with $U_{\mathrm{subopt}}(t)U_{\mathrm{subopt}}^{\dagger}%
(t)=U_{\mathrm{subopt}}^{\dagger}(t)U_{\mathrm{subopt}}(t)=\mathbf{1}%
_{4\times4}$, and $U_{\mathrm{subopt}}(t_{\mathrm{\ast}})\left\vert
A\right\rangle =\left\vert B\right\rangle $ with $t_{\ast}=(\pi\hslash
)/E>t_{\mathrm{opt}}=(\pi\hslash)/(\sqrt{10}E)$. From an entanglement
standpoint, we note that the entanglement $\mathrm{C}\left(  \gamma\left(
t\right)  \right)  $ of the path $\gamma\left(  t\right)  :t\mapsto\left\vert
\psi\left(  t\right)  \right\rangle \overset{\text{def}}{=}U_{\mathrm{subopt}%
}(t)\left\vert A\right\rangle $ is given by $\mathrm{C}\left(  t\right)
=\mathrm{C}\left[  \left\vert \psi\left(  t\right)  \right\rangle \right]
=2\sqrt{\left[  \frac{1}{8}\sin(2\frac{E}{\hslash}t)+\frac{1}{4}\sin(3\frac
{E}{\hslash}t)+\frac{1}{8}\sin(4\frac{E}{\hslash}t)\right]  ^{2}+\left[
\frac{1}{4}\cos(\frac{E}{\hslash}t)-\frac{1}{4}\right]  ^{2}}$ for $0\leq
t\leq(\pi\hslash)/E$. Moreover, the average path entanglement during the
evolution, defined as $\mathrm{\bar{C}}\overset{\text{def}}{=}%
(1/t_{\mathrm{\ast}})\int_{0}^{t_{\mathrm{\ast}}}\mathrm{C}\left(  t\right)
dt$, is approximately equal to $0.71$. Finally, to quantify the nonlocal
entangling character of the time propagator in Eq. (\ref{fire}), we calculate
the entanglement production $\varepsilon_{\mathrm{EP}}^{\mathrm{Yukalov}%
}\left(  t\right)  $ in Eq. (\ref{russia}). Following a series of algebraic
manipulations, we find%
\begin{equation}
\varepsilon_{\mathrm{EP}}^{\mathrm{Yukalov}}\left(  t\right)  =\frac{1}{2}%
\log\left\{  \frac{\mathcal{N}\left(  t\right)  }{\mathcal{D}_{1}%
(t)\mathcal{D}_{2}(t)}\right\}  \text{,} \label{chiappachist}%
\end{equation}
where,%
\begin{equation}
\left\{
\begin{array}
[c]{c}%
\mathcal{N}\left(  t\right)  \overset{\text{def}}{=}4\left[  2\cos(\frac
{E}{\hslash}t)+2\cos(2\frac{E}{\hslash}t)\right]  ^{2}\text{,}\\
\mathcal{D}_{1}(t)\overset{\text{def}}{=}\frac{3}{2}\cos(\frac{E}{\hslash
}t)+\frac{1}{4}\cos(\frac{2E}{\hslash}t)+\frac{5}{2}\cos(\frac{3E}{\hslash
}t)+\frac{1}{4}\cos(\frac{4E}{\hslash}t)+\frac{7}{2}\text{,}\\
\mathcal{D}_{2}(t)\overset{\text{def}}{=}\frac{3}{2}\cos(\frac{E}{\hslash
}t)+\frac{5}{4}\cos(\frac{2E}{\hslash}t)+\frac{1}{2}\cos(\frac{3E}{\hslash
}t)+\frac{5}{4}\cos(\frac{4E}{\hslash}t)+\frac{7}{2}\text{.}%
\end{array}
\right.
\end{equation}
In the short-time limit, $\varepsilon_{\mathrm{EP}}^{\mathrm{Yukalov}}\left(
t\right)  $ exhibits a polynomially quadratic growth specified by the
relation
\begin{equation}
\varepsilon_{\mathrm{EP}}^{\mathrm{Yukalov}}\left(  t\right)  =\frac{5}%
{8}\left(  \frac{E}{\hslash}\right)  ^{2}t^{2}+\mathcal{O}\left(
t^{4}\right)  \text{.}%
\end{equation}
We observe, as expected, that $\varepsilon_{\mathrm{EP}}^{\mathrm{Yukalov}%
}\left(  0\right)  =0$ in Eq. (\ref{chiappachist}).\begin{figure}[t]
\centering
\includegraphics[width=0.5\textwidth] {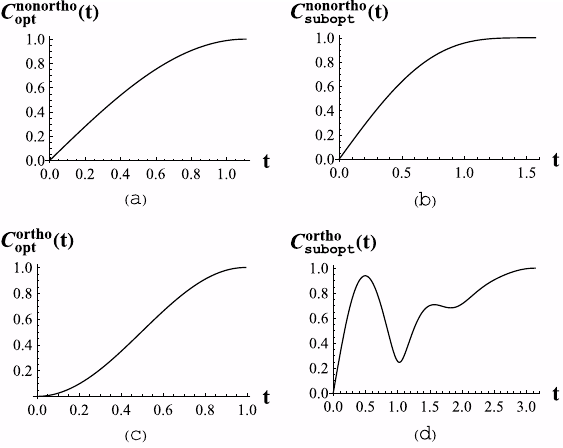}\caption{In (a), there is a plot
of the temporal behavior of the entanglement of the path that connects
nonorthogonal initial (separable) and final (maximally entangled) states
$\left\vert A\right\rangle \overset{\text{def}}{=}\left\vert 00\right\rangle $
and $\left\vert B\right\rangle \overset{\text{def}}{=}\left[  \left\vert
00\right\rangle +\left\vert 11\right\rangle \right]  /\sqrt{2}$, respectively,
when the evolution is time optimal. Entanglement is specified by the
concurrence $\mathrm{C}_{\mathrm{opt}}^{\mathrm{nonortho}}\left(  t\right)  $.
Moreover, the unitary evolution occurs with energy dispersion $\Delta
E=E/\sqrt{2}$ and $0\leq t\leq\left(  \pi\hslash\right)  /(2\sqrt{2}E)$. In
(b), there is a plot of the temporal behavior of the entanglement of the path
that connects $\left\vert A\right\rangle \overset{\text{def}}{=}\left\vert
00\right\rangle $ and $\left\vert B\right\rangle \overset{\text{def}}%
{=}\left[  \left\vert 00\right\rangle +\left\vert 11\right\rangle \right]
/\sqrt{2}$ when the evolution is time suboptimal. Entanglement is specified by
the concurrence $\mathrm{C}_{\mathrm{subopt}}^{\mathrm{nonortho}}\left(
t\right)  $. Moreover, the unitary evolution occurs with energy dispersion
$\Delta E=E/\sqrt{2}$ and $0\leq t\leq\left(  \pi\hslash\right)  /(2E)$. In
(c), there is a plot of the temporal behavior of the entanglement of the path
that connects orthogonal initial (separable) and final (maximally entangled)
states $\left\vert A\right\rangle \overset{\text{def}}{=}\left\vert
00\right\rangle $ and $\left\vert B\right\rangle \overset{\text{def}}%
{=}\left[  \left\vert 01\right\rangle +\left\vert 10\right\rangle \right]
/\sqrt{2}$, respectively, when the evolution is time optimal. Entanglement is
specified by the concurrence $\mathrm{C}_{\mathrm{opt}}^{\mathrm{ortho}%
}\left(  t\right)  $. Moreover, the unitary evolution occurs with energy
dispersion $\Delta E=(\sqrt{5/2})E$ and $0\leq t\leq\left(  \pi\hslash\right)
/(\sqrt{10}E)$. In (d), there is a plot of the temporal behavior of the
entanglement of the path that connects $\left\vert A\right\rangle
\overset{\text{def}}{=}\left\vert 00\right\rangle $ and $\left\vert
B\right\rangle \overset{\text{def}}{=}\left[  \left\vert 01\right\rangle
+\left\vert 10\right\rangle \right]  /\sqrt{2}$ when the evolution is time
suboptimal. Entanglement is specified by the concurrence $\mathrm{C}%
_{\mathrm{subopt}}^{\mathrm{ortho}}\left(  t\right)  $. Moreover, the unitary
evolution occurs with energy dispersion $\Delta E=(\sqrt{5/2})E$ and $0\leq
t\leq\left(  \pi\hslash\right)  /E$. In all plots, we set $E=1$ and, finally,
physical units are chosen using $\hslash=1$.}%
\end{figure}

The temporal behavior of the entanglement associated with the evolution paths
that define our four illustrative examples are presented in Fig. $1$. When
comparing time optimal (a) and time suboptimal (b) evolutions between
nonorthogonal states $\left\vert 00\right\rangle $ (separable) and $\left(
\left\vert 00\right\rangle +\left\vert 11\right\rangle \right)  /\sqrt{2}$
(maximally entangled), it is observed that the time optimal evolutions
conclude in a shorter duration, without any energy waste, and without any
bending of the trajectory. Additionally, from the perspective of entanglement
analysis, optimal evolutions appear to occur with lower average path
entanglement values, higher average entanglement speed $\bar{v}_{\mathrm{C}%
}\overset{\text{def}}{=}1/\Delta t_{\left\vert A\right\rangle \rightarrow
\left\vert B\right\rangle }$, and, ultimately, a greater degree of nonlocal
entangling character in their respective unitary time propagators. Similar
findings are evident when contrasting time optimal (c) and time suboptimal (d)
evolutions between orthogonal states $\left\vert 00\right\rangle $ (separable)
and $\left(  \left\vert 01\right\rangle +\left\vert 10\right\rangle \right)
/\sqrt{2}$ (maximally entangled). Nevertheless, several critical differences
must be highlighted. Firstly, in the comparison of (c) and (d), only the time
optimal evolution (c) occurs within a two-dimensional subspace of the complete
four-dimensional Hilbert space. In contrast, the time suboptimal evolution (d)
transpires within the entire four-dimensional space. Secondly, unlike the
situation observed when comparing (a) and (b), in the analysis between (c) and
(d), the nonlocal entangling character of the unitary time propagator
associated with the time suboptimal Hamiltonian is more pronounced than that
which characterizes the time optimal evolution. Lastly, when comparing the
time suboptimal evolutions in (b) and (d), it is noted that while the time
suboptimal evolution between orthogonal states (d) occurs along longer paths
with smaller curvature and greater energy resource waste, the time suboptimal
evolution between nonorthogonal states (b) takes place along shorter paths
with higher curvature and reduced energy resource waste. We direct your
attention to Table III for a summary of the insights we have gathered.

We have not employed Zanardi's entangling power $\varepsilon_{\mathrm{EP}%
}^{\mathrm{Zanardi}}\left(  U\right)  $ in Eq. (\ref{epower}) in our four
illustrative examples discussed so far. The main reason for this absence is
because of computational challenges and is due to the fact that none of the
unitary time propagators in Eqs. (\ref{prop1}), (\ref{prop2}), (\ref{prop3}),
and (\ref{fire}) appears in the canonical form specified by Eq. (\ref{Q3}). To
clarify, we emphasize that it is not strictly necessary to have the unitary
evolution operator in its canonical form in order to numerically estimate
Zanardi's entangling power for a bipartite quantum system. Nevertheless,
expressing it in its canonical form typically simplifies the calculation and
enhances transparency. This is particularly apparent for general two-qubit
unitary gates (see Eqs. (9) in Refs. \cite{reza04} and \cite{bala10}, for
instance). However, obtaining the canonical form of a two-qubit unitary
operator necessitates knowledge of the Weyl chamber coordinates $\left(
c_{1}\text{, }c_{2}\text{, }c_{3}\right)  $. Deriving these coordinates in a
closed analytical form is often a tedious task from a computational
perspective. For a comprehensive method on how to constructively determine
Weyl chamber coordinates, along with the four single qubit unitary operators
in \textrm{SU}$\left(  2\right)  $ that define the canonical form of any
unitary operator in \textrm{SU}$\left(  4\right)  $, we recommend the analysis
conducted by Kraus and Cirac in Appendix A of Ref. \cite{KC}.

In the following section, we circumvent these computational difficulties
(which we will address in future numerical investigations) by presenting
innovative and equally enlightening applications where we uphold analytical
computations and focus on the connections between time optimality, the
nonlocal characteristics of unitary time propagators, and the entanglement
capability of time optimal evolutions with different degrees of energy wastefulness.

\section{Time optimality, nonlocality, and entangling power}

In this section, with the understanding that our primary goal is to acquire
insight rather than to achieve generality, we examine the entanglement
characteristics of three distinct unitary propagators that do not necessarily
correspond to stationary Hamiltonians constructed in Section IV. Each of these
evolutions is time-optimal and transitions from a separable state to a
maximally entangled state. Additionally, the initial and final states are
considered to be non-orthogonal. Notably, we demonstrate through explicit
examples that: ii) $\varepsilon_{\mathrm{EP}}^{\mathrm{Zanardi}}\left(
U_{1}\right)  =\varepsilon_{\mathrm{EP}}^{\mathrm{Zanardi}}\left(
U_{2}\right)  $ does not necessarily indicate that $U_{1}$ and $U_{2}$ are
part of the same equivalence class; ii) $\varepsilon_{\mathrm{EP}%
}^{\mathrm{Yukalov}}\left(  U_{1}\right)  =\varepsilon_{\mathrm{EP}%
}^{\mathrm{Yukalov}}\left(  U_{2}\right)  $ does not imply that $U_{1}$ and
$U_{2}$ belong to the same equivalence class; iii) Time-optimal evolutions
that are also energetically efficient (i.e., exhibiting high speed efficiency
$\eta_{\mathrm{Uzdin}}$) do not have to achieve Zanardi's entangling power
values as elevated as those associated with inefficient time-optimal
evolutions to attain the maximally entangled target state.

\subsection{Example 1}

We begin by examining the time optimal evolution between a separable quantum
state and a maximally entangled quantum state that are not orthogonal.
Specifically, we evolve from $\left\vert A\right\rangle \overset{\text{def}%
}{=}\left\vert 01\right\rangle $ to $\left\vert B\right\rangle \overset
{\text{def}}{=}\frac{1+i}{2}\left\vert 01\right\rangle +\frac{1-i}%
{2}\left\vert 10\right\rangle $ in a time-optimal manner, with $\Delta E=E$
and $t_{\mathrm{opt}}\overset{\text{def}}{=}\left(  \hslash\pi\right)  /(4E)$.
This is accomplished by the non-local two-qubit Hamiltonian given by,%
\begin{equation}
\mathrm{H}\overset{\text{def}}{=}E(\sigma_{x}^{\left(  1\right)  }%
\otimes\sigma_{x}^{\left(  2\right)  })+E(\sigma_{z}^{\left(  1\right)
}\otimes\sigma_{z}^{\left(  2\right)  })\text{.} \label{mylove2}%
\end{equation}
The matrix representation of the time optimal Hamiltonian in Eq.
(\ref{mylove2}) with respect to the canonical basis $\mathcal{B}%
_{\mathcal{H}_{2}^{2}}\overset{\text{def}}{=}\left\{  \left\vert
00\right\rangle \text{, }\left\vert 01\right\rangle \text{, }\left\vert
10\right\rangle \text{, }\left\vert 11\right\rangle \right\}  $ of
$\mathcal{H}_{2}^{2}$ is,%
\begin{equation}
\mathrm{H}=\left(
\begin{array}
[c]{cccc}%
E & 0 & 0 & E\\
0 & -E & E & 0\\
0 & E & -E & 0\\
E & 0 & 0 & E
\end{array}
\right)  \text{.} \label{tazza}%
\end{equation}
For completeness, we note from Eq. (\ref{tazza}) that $\mathrm{H}%
=\mathrm{H}^{\dagger}$, \textrm{tr}$(\mathrm{H})=0$, $\Delta E^{2}%
=\left\langle \mathrm{H}^{2}\right\rangle -\left\langle \mathrm{H}%
\right\rangle ^{2}=E^{2}$, $\eta_{\mathrm{geo}}=1$ (since $s=s_{0}=\pi/2$),
$\eta_{\mathrm{Uzdin}}=1/2<1$, and $\kappa_{\mathrm{AC}}^{2}=0$. After
diagonalizing the matrix in Eq. (\ref{tazza}), we find that the corresponding
unitary time propagator $U(t)=e^{-\frac{i}{\hslash}\mathrm{H}t}$ is given by%
\begin{equation}
U(t)=\left(
\begin{array}
[c]{cccc}%
e^{-\frac{i}{\hslash}Et}\cos(\frac{E}{\hslash}t) & 0 & 0 & -ie^{-\frac
{i}{\hslash}Et}\sin(\frac{E}{\hslash}t)\\
0 & e^{\frac{i}{\hslash}Et}\cos(\frac{E}{\hslash}t) & -ie^{\frac{i}{\hslash
}Et}\sin(\frac{E}{\hslash}t) & 0\\
0 & -ie^{\frac{i}{\hslash}Et}\sin(\frac{E}{\hslash}t) & e^{\frac{i}{\hslash
}Et}\cos(\frac{E}{\hslash}t) & 0\\
-ie^{-\frac{i}{\hslash}Et}\sin(\frac{E}{\hslash}t) & 0 & 0 & e^{-\frac
{i}{\hslash}Et}\cos(\frac{E}{\hslash}t)
\end{array}
\right)  \text{,} \label{u1c}%
\end{equation}
with $U(t)U^{\dagger}(t)=U^{\dagger}(t)U(t)=\mathbf{1}_{4\times4}$, and
$U(t_{\mathrm{opt}})\left\vert A\right\rangle =\left\vert B\right\rangle $.
From an entanglement standpoint, we note that the entanglement $\mathrm{C}%
\left(  \gamma\left(  t\right)  \right)  $ of the path $\gamma\left(
t\right)  :t\mapsto\left\vert \psi\left(  t\right)  \right\rangle
\overset{\text{def}}{=}U(t)\left\vert A\right\rangle $ is given by
$\mathrm{C}\left(  t\right)  =\mathrm{C}\left[  \left\vert \psi\left(
t\right)  \right\rangle \right]  =\left\vert \sin(\frac{2E}{\hslash
}t)\right\vert =\sin(\frac{2E}{\hslash}t)$ for $0\leq t\leq\left(  \hslash
\pi\right)  /(4E)$. Moreover, the average path entanglement equals
$\mathrm{\bar{C}}=2/\pi\simeq0.64$. To quantify the nonlocal entangling
character of the time propagator in Eq. (\ref{u1c}), we evaluate the
entanglement production $\varepsilon_{\mathrm{EP}}^{\mathrm{Yukalov}}\left(
t\right)  $ in Eq. (\ref{russia}). After some algebra, we get%
\begin{equation}
\varepsilon_{\mathrm{EP}}^{\mathrm{Yukalov}}\left(  t\right)  =\frac{1}{2}%
\log\left\{  \frac{4\left[  2+2\cos\left(  \frac{2E}{\hslash}t\right)
\right]  ^{2}}{\left[  8\cos^{4}\left(  \frac{E}{\hslash}t\right)  \right]
^{2}}\right\}  \text{.} \label{KK1}%
\end{equation}
In the short-time limit, $\varepsilon_{\mathrm{EP}}^{\mathrm{Yukalov}}\left(
t\right)  $ possesses a polynomially quadratic growth specified by the
relation
\begin{equation}
\varepsilon_{\mathrm{EP}}^{\mathrm{Yukalov}}\left(  t\right)  =\left(
\frac{E}{\hslash}\right)  ^{2}t^{2}+\frac{1}{6}\left(  \frac{E}{\hslash
}\right)  ^{4}t^{4}+\mathcal{O}\left(  t^{5}\right)  \text{.}%
\end{equation}
We note, as expected, that $\varepsilon_{\mathrm{EP}}^{\mathrm{Yukalov}%
}\left(  0\right)  =0$ in Eq. (\ref{KK1}). Finally, inspecting the unitary
evolution operator $U(t)$ in Eq. (\ref{u1c}), we observe that its geometrical
point is characterized by the vector $\mathbf{c}$ given by
\begin{equation}
\mathbf{c}\overset{\text{def}}{=}\left(  c_{1}\text{, }c_{2}\text{, }%
c_{3}\right)  =\left(  \frac{E}{\hslash}t\text{, }0\text{, }\frac{E}{\hslash
}t\right)  \text{.} \label{donald1}%
\end{equation}
From Eq. (\ref{donald1}), we also note that the entangling power
$\varepsilon_{\mathrm{EP}}^{\mathrm{Zanardi}}\left(  U\right)  $ in Eq.
(\ref{cami}) of this unitary operator $U$ in Eq. (\ref{u1c}) is equal to%
\begin{equation}
\varepsilon_{\mathrm{EP}}^{\mathrm{Zanardi}}\left(  t\right)  =\frac{1}%
{18}\left[  3-\cos^{2}(\frac{4E}{\hslash}t)-2\cos(\frac{4E}{\hslash}t)\right]
\text{.} \label{ass1}%
\end{equation}
Interestingly, we observe from Eq. (\ref{ass1}) that $\varepsilon
_{\mathrm{EP}}^{\mathrm{Zanardi}}\left(  0\right)  =0$ and $\varepsilon
_{\mathrm{EP}}^{\mathrm{Zanardi}}\left(  t_{\mathrm{opt}}\right)  =2/9$, where
$t_{\mathrm{opt}}\overset{\text{def}}{=}\left(  \hslash\pi\right)  /(4E)$.

\subsection{Example 2}

Following the previous example, we continue by studying the optimal evolution
between a separable quantum state and a maximally entangled quantum state that
are not orthogonal. Specifically, we evolve from $\left\vert A\right\rangle
\overset{\text{def}}{=}\left\vert 01\right\rangle $ to $\left\vert
B\right\rangle \overset{\text{def}}{=}\frac{1+i}{2}\left\vert 01\right\rangle
+\frac{1-i}{2}\left\vert 10\right\rangle $ in a time-optimal manner, with
$\Delta E=2E$ and $t_{\mathrm{opt}}\overset{\text{def}}{=}\left(  \hslash
\pi\right)  /(8E)$. This is accomplished by the non-local two-qubit
Hamiltonian defined as,%
\begin{equation}
\mathrm{H}\overset{\text{def}}{=}E(\sigma_{x}^{\left(  1\right)  }%
\otimes\sigma_{x}^{\left(  2\right)  })+E(\sigma_{y}^{\left(  1\right)
}\otimes\sigma_{y}^{\left(  2\right)  })+E(\sigma_{z}^{\left(  1\right)
}\otimes\sigma_{z}^{\left(  2\right)  })\text{.} \label{servire}%
\end{equation}
The matrix representation of $\mathrm{H}$ in the computational basis of the
Hilbert space of two-qubit states is%
\begin{equation}
\mathrm{H}=\left(
\begin{array}
[c]{cccc}%
E & 0 & 0 & 0\\
0 & -E & 2E & 0\\
0 & 2E & -E & 0\\
0 & 0 & 0 & E
\end{array}
\right)  \text{.} \label{tazza2}%
\end{equation}
For thoroughness, we note from Eq. (\ref{tazza2}) that $\mathrm{H}%
=\mathrm{H}^{\dagger}$, \textrm{tr}$(\mathrm{H})=0$, $\Delta E^{2}%
=\left\langle \mathrm{H}^{2}\right\rangle -\left\langle \mathrm{H}%
\right\rangle ^{2}=4E^{2}$, $\eta_{\mathrm{geo}}=1$ (since $s=s_{0}=\pi/2$),
$\eta_{\mathrm{Uzdin}}=2/3<1$, and $\kappa_{\mathrm{AC}}^{2}=0$. After
diagonalizing the matrix in Eq. (\ref{tazza2}), we find that the corresponding
unitary time propagator $U(t)=e^{-\frac{i}{\hslash}\mathrm{H}t}$ is given by%
\begin{equation}
U(t)=\left(
\begin{array}
[c]{cccc}%
e^{-\frac{i}{\hslash}Et} & 0 & 0 & 0\\
0 & e^{\frac{i}{\hslash}Et}\cos(\frac{2E}{\hslash}t) & -ie^{\frac{i}{\hslash
}Et}\sin(\frac{2E}{\hslash}t) & 0\\
0 & -ie^{\frac{i}{\hslash}Et}\sin(\frac{2E}{\hslash}t) & e^{\frac{i}{\hslash
}Et}\cos(\frac{2E}{\hslash}t) & 0\\
0 & 0 & 0 & e^{-\frac{i}{\hslash}Et}%
\end{array}
\right)  \text{,} \label{u2d}%
\end{equation}
with $U(t)U^{\dagger}(t)=U^{\dagger}(t)U(t)=\mathbf{1}_{4\times4}$, and
$U(t_{\mathrm{opt}})\left\vert A\right\rangle =\left\vert B\right\rangle $.
From an entanglement standpoint, we note that the entanglement $\mathrm{C}%
\left(  \gamma\left(  t\right)  \right)  $ of the path $\gamma\left(
t\right)  :t\mapsto\left\vert \psi\left(  t\right)  \right\rangle
\overset{\text{def}}{=}U(t)\left\vert A\right\rangle $ is given by
$\mathrm{C}\left(  t\right)  =\mathrm{C}\left[  \left\vert \psi\left(
t\right)  \right\rangle \right]  =\left\vert \sin(\frac{4E}{\hslash
}t)\right\vert =\sin(\frac{4E}{\hslash}t)$ for $0\leq t\leq\left(  \hslash
\pi\right)  /(8E)$. Moreover, the average path entanglement equals
$\mathrm{\bar{C}}=2/\pi\simeq0.64$. To quantify the nonlocal entangling
character of the time propagator in Eq. (\ref{u2d}), we calculate the
entanglement production $\varepsilon_{\mathrm{EP}}^{\mathrm{Yukalov}}\left(
t\right)  $ in Eq. (\ref{russia}). After some algebraic manipulations, we
arrive at%
\begin{equation}
\varepsilon_{\mathrm{EP}}^{\mathrm{Yukalov}}\left(  t\right)  =\frac{1}{2}%
\log\left\{  \frac{4\left[  10+6\cos(\frac{4E}{\hslash}t)\right]  }{\left[
3\cos(\frac{4E}{\hslash}t)+5\right]  ^{2}}\right\}  \label{KK2}%
\end{equation}
In the short-time limit, $\varepsilon_{\mathrm{EP}}^{\mathrm{Yukalov}}\left(
t\right)  $ exhibits a polynomially quadratic growth specified by the
relation
\begin{equation}
\varepsilon_{\mathrm{EP}}^{\mathrm{Yukalov}}\left(  t\right)  =\frac{3}%
{2}\left(  \frac{E}{\hslash}\right)  ^{2}t^{2}+\frac{1}{4}\left(  \frac
{E}{\hslash}\right)  ^{4}t^{4}+\mathcal{O}\left(  t^{6}\right)  \text{.}%
\end{equation}
We observe, as expected, that $\varepsilon_{\mathrm{EP}}^{\mathrm{Yukalov}%
}\left(  0\right)  =0$ in Eq. (\ref{KK2}). Finally, from $U(t)$ in Eq.
(\ref{u2d}), we observe that the geometrical point is specified by the vector
$\mathbf{c}$ given by
\begin{equation}
\mathbf{c}\overset{\text{def}}{=}\left(  c_{1}\text{, }c_{2}\text{, }%
c_{3}\right)  =\left(  \frac{E}{\hslash}t\text{, }\frac{E}{\hslash}t\text{,
}\frac{E}{\hslash}t\right)  \text{.} \label{donaldb}%
\end{equation}
From Eq. (\ref{donaldb}), we also note that the entangling power
$\varepsilon_{\mathrm{EP}}^{\mathrm{Zanardi}}\left(  U\right)  $ in Eq.
(\ref{cami}) of this unitary operator $U$ in Eq. ((\ref{u2d}) is equal to%
\begin{equation}
\varepsilon_{\mathrm{EP}}^{\mathrm{Zanardi}}\left(  t\right)  =\frac{1}{6}%
\sin^{2}(\frac{4E}{\hslash}t)\text{.} \label{ass2}%
\end{equation}
We note from Eq. (\ref{ass2}) that $\varepsilon_{\mathrm{EP}}%
^{\mathrm{Zanardi}}\left(  0\right)  =0$ and $\varepsilon_{\mathrm{EP}%
}^{\mathrm{Zanardi}}\left(  t_{\mathrm{opt}}\right)  =1/6$, where
$t_{\mathrm{opt}}\overset{\text{def}}{=}\left(  \hslash\pi\right)  /(8E)$.

\subsection{Example 3}

In this last example, we wish to evolve from $\left\vert A\right\rangle
\overset{\text{def}}{=}\left\vert 01\right\rangle $ to $\left\vert
B\right\rangle \overset{\text{def}}{=}\frac{1+i}{2}\left\vert 01\right\rangle
+\frac{1-i}{2}\left\vert 10\right\rangle $ in a time-optimal manner, with
$\Delta E=2E$ and $t_{\mathrm{opt}}\overset{\text{def}}{=}\left(  \hslash
\pi\right)  /(8E)$. This was accomplished in the second example. However, we
want to construct an alternative time-independent Hamiltonian here. We proceed
as follows. Consider the Hamiltonian $\mathrm{H}$ given by Eq. (\ref{amy11})
with $\Delta E\overset{\text{def}}{=}2E$, $\left\vert A\right\rangle
\overset{\text{def}}{=}\left\vert 01\right\rangle $, and $\left\vert
B\right\rangle \overset{\text{def}}{=}\frac{1+i}{2}\left\vert 01\right\rangle
+\frac{1-i}{2}\left\vert 10\right\rangle $. After some algebra, the matrix
representation of this $\mathrm{H}$ in the computational basis of the Hilbert
space of two-qubit states becomes%
\begin{equation}
\mathrm{H}=2E\left(
\begin{array}
[c]{cccc}%
0 & 0 & 0 & 0\\
0 & 0 & 1 & 0\\
0 & 1 & 0 & 0\\
0 & 0 & 0 & 0
\end{array}
\right)  \text{.} \label{tazza3}%
\end{equation}
For the sake of completeness, we point out from Eq. (\ref{tazza3}) that
$\mathrm{H}=\mathrm{H}^{\dagger}$, \textrm{tr}$(\mathrm{H})=0$, $\Delta
E^{2}=\left\langle \mathrm{H}^{2}\right\rangle -\left\langle \mathrm{H}%
\right\rangle ^{2}=4E^{2}$, $\eta_{\mathrm{geo}}=1$ (since $s=s_{0}=\pi/2$),
$\eta_{\mathrm{Uzdin}}=1$, and $\kappa_{\mathrm{AC}}^{2}=0$. After
diagonalizing the matrix in Eq. (\ref{tazza2}), we find that the corresponding
unitary time propagator $U(t)=e^{-\frac{i}{\hslash}\mathrm{H}t}$ becomes%
\begin{equation}
U(t)=\left(
\begin{array}
[c]{cccc}%
1 & 0 & 0 & 0\\
0 & \cos\left(  \frac{2E}{\hslash}t\right)  & -i\sin(\frac{2E}{\hslash}t) &
0\\
0 & -i\sin(\frac{2E}{\hslash}t) & \cos\left(  \frac{2E}{\hslash}t\right)  &
0\\
0 & 0 & 0 & 1
\end{array}
\right)  \text{,} \label{w1}%
\end{equation}
with $U(t)U^{\dagger}(t)=U^{\dagger}(t)U(t)=\mathbf{1}_{4\times4}$, and
$U(t_{\mathrm{opt}})\left\vert A\right\rangle =\left\vert B\right\rangle $.
From an entanglement standpoint, we note that the entanglement $\mathrm{C}%
\left(  \gamma\left(  t\right)  \right)  $ of the path $\gamma\left(
t\right)  :t\mapsto\left\vert \psi\left(  t\right)  \right\rangle
\overset{\text{def}}{=}U(t)\left\vert A\right\rangle $ is given by
$\mathrm{C}\left(  t\right)  =\mathrm{C}\left[  \left\vert \psi\left(
t\right)  \right\rangle \right]  =\left\vert \sin(\frac{4E}{\hslash
}t)\right\vert =\sin(\frac{4E}{\hslash}t)$ for $0\leq t\leq\left(  \hslash
\pi\right)  /(8E)$. Moreover, the average path entanglement equals
$\mathrm{\bar{C}}=2/\pi\simeq0.64$. \ To characterize the nonlocal entangling
character of the time propagator in Eq. (\ref{w1}), we calculate the
entanglement production $\varepsilon_{\mathrm{EP}}^{\mathrm{Yukalov}}\left(
t\right)  $ in Eq. (\ref{russia}). After some algebra, we find%
\begin{equation}
\varepsilon_{\mathrm{EP}}^{\mathrm{Yukalov}}\left(  t\right)  =\frac{1}{2}%
\log\left\{  \frac{4\left[  2+2\cos\left(  \frac{2E}{\hslash}t\right)
\right]  ^{2}}{\left[  8\cos^{4}\left(  \frac{E}{\hslash}t\right)  \right]
^{2}}\right\}  \label{KK3}%
\end{equation}
In the short-time limit, $\varepsilon_{\mathrm{EP}}^{\mathrm{Yukalov}}\left(
t\right)  $ shows a polynomially quadratic growth specified by the relation
\begin{equation}
\varepsilon_{\mathrm{EP}}^{\mathrm{Yukalov}}\left(  t\right)  =\left(
\frac{E}{\hslash}\right)  ^{2}\allowbreak t^{2}+\frac{1}{6}\left(  \frac
{E}{\hslash}\right)  ^{4}t^{4}+\mathcal{O}\left(  t^{5}\right)  \text{.}%
\end{equation}
We observe, as expected, that $\varepsilon_{\mathrm{EP}}^{\mathrm{Yukalov}%
}\left(  0\right)  =0$ in Eq. (\ref{KK1}). Finally, from $U(t)$ in Eq.
(\ref{u2d}), we observe that the geometrical point is specified by the vector
$\mathbf{c}$ given by%
\begin{equation}
\mathbf{c}\overset{\text{def}}{=}\left(  c_{1}\text{, }c_{2}\text{, }%
c_{3}\right)  =\left(  \frac{E}{\hslash}t\text{, }\frac{E}{\hslash}t\text{,
}0\right)  \text{.} \label{donald2bb}%
\end{equation}
From Eq. (\ref{donald2bb}), we also note that the entangling power of this
unitary operator is equal to%
\begin{equation}
\varepsilon_{\mathrm{EP}}^{\mathrm{Zanardi}}\left(  t\right)  =\frac{1}%
{18}\left[  3-\cos^{2}(\frac{4E}{\hslash}t)-2\cos(\frac{4E}{\hslash}t)\right]
\text{.} \label{asso}%
\end{equation}
Interestingly, we observe from Eq. (\ref{asso}) that $\varepsilon
_{\mathrm{EP}}^{\mathrm{Zanardi}}\left(  0\right)  =0$, $\varepsilon
_{\mathrm{EP}}^{\mathrm{Zanardi}}\left(  t_{\mathrm{opt}}\right)  =1/6$, and
$\varepsilon_{\mathrm{EP}}^{\mathrm{Zanardi}}\left(  2t_{\mathrm{opt}}\right)
=2/9$, where $t_{\mathrm{opt}}\overset{\text{def}}{=}\left(  \hslash
\pi\right)  /(8E)$.

We observe that all three Hamiltonians defined in Eqs. (\ref{mylove2}),
(\ref{servire}), and (\ref{tazza3}) generate geodesic trajectories connecting
the same initial and final states. However, the corresponding travel times
coincide only for the Hamiltonians in Eqs. (\ref{servire}) and (\ref{tazza3}),
as both share the same energy uncertainty, $\Delta E=2E$. By contrast, the
geodesic evolution driven by the Hamiltonian in Eq. (\ref{mylove2}) requires a
longer time, equal to\textbf{ }$\hslash\pi/(4E)$\textbf{, }which is twice the
duration of the other two cases (with travel time equal to $\hslash\pi/(8E)$).
This difference arises because its energy uncertainty is smaller\textbf{,
}$\Delta E=E$\textbf{,} i.e., half that of the other Hamiltonians.
Importantly, even though all three scenarios exhibit unit geodesic efficiency
(and vanishing curvature coefficients), they differ in their speed efficiency.
Only the Hamiltonian in Eq. (\ref{tazza3}) achieves unit speed efficiency,
$\eta_{\mathrm{Uzdin}}=1$,\textbf{ }meaning that the full available energy-
quantified by the spectral norm- is converted into the system's evolution
speed. In contrast, the Hamiltonians in Eqs. (\ref{mylove2}) and
(\ref{servire}) produce trajectories that are less energy-efficient, as
reflected by their lower speed efficiencies of\textbf{ }$\eta_{\mathrm{Uzdin}%
}=1/2$ and $\eta_{\mathrm{Uzdin}}=2/3$, respectively.

It is also evident that all unitary operators within the same equivalence
class exhibit an equal level of entangling power. However, the reverse is not
necessarily the case. For instance, when we compare Example $1$ with Example
$3$, we observe that the unitary time propagators represented in Eqs.
(\ref{u1c}) and (\ref{w1}) do not belong to the same equivalence class, as the
parameter $\mathbf{c}$ in Eq. (\ref{donald1}) differs from that in Eq.
(\ref{donald2bb}). Nevertheless, these operators correspond to quantum
evolutions that yield identical entanglement productions $\varepsilon
_{\mathrm{EP}}^{\mathrm{Yukalov}}$ (with Eqs. (\ref{KK1}) and (\ref{KK3} being
identical) and entangling powers $\varepsilon_{\mathrm{EP}}^{\mathrm{Zanardi}%
}$ (with Eqs. (\ref{ass1}) and (\ref{asso}) being the same). Additionally,
when we compare Example $2$ to Example $3$, we discover that time-optimal
evolutions that are also energetically efficient (as seen in Eq. (\ref{w1}),
Example $3$) do not necessarily need to achieve entangling power values of
$\varepsilon_{\mathrm{EP}}^{\mathrm{Zanardi}}$ that are as high as those
linked to energetically inefficient time-optimal evolutions (as indicated in
Eq. (\ref{u2d}), Example $2$) in order to reach maximally entangled target
states. A comparative analysis of the temporal behaviors of path entanglement,
Yukalov's entanglement production, and Zanardi's entangling power for the
quantum evolutions studied in Examples $2$ and $3$ is illustrated in Fig. $2$.

We are now prepared to present our summary of results and final observations.

\begin{figure}[t]
\centering
\includegraphics[width=1\textwidth] {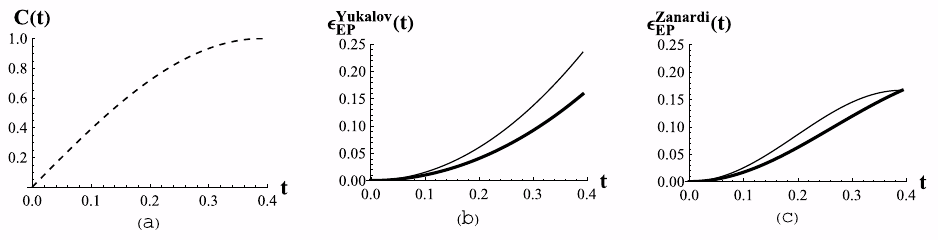}\caption{Entanglement-based
comparative analysis of two distinct optimal time evolutions (Example 2 and
Example 3) from an initial (separable) state to a final (maximally entangled)
state $\left\vert A\right\rangle \overset{\text{def}}{=}\left\vert
01\right\rangle $ and $\left\vert B\right\rangle \overset{\text{def}}{=}%
\frac{1+i}{2}\left\vert 01\right\rangle $ $+\frac{1-i}{2}\left\vert
10\right\rangle $, respectively, with $\left\vert \left\langle A\left\vert
B\right.  \right\rangle \right\vert \neq0$. In (a), there is a plot of the
temporal behavior of the entanglement of the paths that connect $\left\vert
A\right\rangle $ and $\left\vert B\right\rangle $. State entanglement is
specified by the concurrence \textrm{C}$\left(  t\right)  $ which exhibits an
identical time behavior in both cases. Both unitary time evolutions occur with
energy dispersion $\Delta E=2E$ and $0\leq t\leq\left(  \pi\hslash\right)
/(8E)$. In (b), the nonlocal character of the unitary time propagators is
captured by Yukalov's entanglement production $\varepsilon_{\mathrm{EP}%
}^{\mathrm{Yukalov}}\left(  t\right)  $ versus time $t$ for Example 2 (thin
solid line) and Example 3 (thick solid line). In (c), the entanglement
capability of the unitary time propagators is characterized by Zanardi's
entanglement power $\varepsilon_{\mathrm{EP}}^{\mathrm{Zanardi}}\left(
t\right)  $ versus time $t$ for Example 2 (thin solid line) and Example 3
(thick solid line). Observe that $\left[  \varepsilon_{\mathrm{EP}%
}^{\mathrm{Zanardi}}\left(  \left(  \pi\hslash\right)  /(8E)\right)  \right]
_{\text{\textrm{Example}-}2}=\left[  \varepsilon_{\mathrm{EP}}%
^{\mathrm{Zanardi}}\left(  \left(  \pi\hslash\right)  /(8E)\right)  \right]
_{\text{\textrm{Example}-}3}=1/6$. In all plots, we set $E=1$ and, finally,
physical units are chosen using $\hslash=1$.}%
\end{figure}

\section{Final Remarks}

In this paper, we studied relevant geometric properties of entanglement that
emerge from time-independent nonlocal Hamiltonian evolutions transitioning
from separable to maximally entangled two-qubit quantum states. From a
geometric viewpoint, each evolution is defined by geodesic efficiency (Eq.
(\ref{efficiency11})), speed efficiency (Eq. (\ref{se111})), and curvature
coefficient (Eq. (\ref{curvatime11})). On the other hand, from the perspective
of entanglement, these evolutions are assessed using various metrics,
including concurrence (Eq. (\ref{conc1})), entanglement power (Eq.
(\ref{russia})), and entangling capability (Eq. (\ref{epower})).

\subsection{Summary of results}

Our principal results and the accompanying physical insights can be summarized
as follows.

\smallskip

\begin{enumerate}
\item[{[i]}] First, we demonstrated that one can construct parametric families
of stationary Hamiltonians, both optimal (Eq. (\ref{amy11})) and suboptimal
(Eq. (\ref{R9})), that drive the evolution from any separable quantum state to
a maximally entangled state, provided the dynamics are restricted to the
two-dimensional subspace spanned by the initial and target states.

\item[{[ii]}] Second, we were able to construct and study \textit{ad hoc}
examples of suboptimal evolutions with constant Hamiltonians between
orthogonal states in higher-dimensional subspaces of the complete Hilbert
space for two-qubit quantum states (Eq. (\ref{fire})), given that such
evolutions are not permitted in two dimensions \cite{james26}.

\item[{[iii]}] Third, by examining evolutions between identical pairs of
\emph{nonorthogonal} separable and maximally entangled states (subsections A
and B in Section V), we showed that time-optimal evolutions are not
necessarily generated by unitary propagators with a greater degree of
nonlocality (expressed in terms of Yukalov's entanglement production) than
those associated with suboptimal evolutions. Moreover, contrary to intuitive
expectations, such time-optimal trajectories do not necessarily exhibit higher
average path entanglement- quantified via concurrence- than their
time-suboptimal counterparts\textbf{.}

\item[{[iv]}] Fourth, by analyzing evolutions between identical pairs of
\emph{orthogonal} separable and maximally entangled states (subsections C and
D in Section V), we found- contrary to intuitive expectations- that
time-suboptimal dynamics may be driven by unitary propagators exhibiting a
higher degree of nonlocality, as quantified by Yukalov's measure of
entanglement production, than those governing optimal evolutions. Furthermore,
these suboptimal trajectories can display a greater average path entanglement-
measured via concurrence- than their time-optimal counterparts.

\item[{[v]}] Fifth, when comparing the time suboptimal evolutions between
nonorthogonal states (subsection B in\ Section V) with time suboptimal
evolutions between orthogonal states (subsection D in Section V), it is noted
that while the time suboptimal evolution between orthogonal states can occur
along longer paths with smaller curvature and greater energy resource waste,
the time suboptimal evolution between nonorthogonal states can take place
along shorter paths with higher curvature and reduced energy resource waste.
In contrast to the results in [iii] and [iv], the finding in [v] appears to be
consistent with the fact that suboptimal evolutions governed by constant
Hamiltonians between orthogonal states are not allowed in two-dimensional settings.

\item[{[vi]}] Finally, contrary to intuitive expectations, we showed (Section
VI) that time-optimal evolutions from a separable state to a maximally
entangled state that are also energetically efficient (i.e., characterized by
high speed efficiency) need not attain values of Zanardi's entangling power as
high as those associated with energetically inefficient time-optimal
evolutions in order to reach the same maximally entangled target state.
\end{enumerate}

\medskip

When analyzing time optimal (Eq. (\ref{prop1})) and time suboptimal (Eq.
(\ref{prop2})) evolutions between separable and maximally entangled
\emph{nonorthogonal} states (first scenario), we noted that the time optimal
evolutions conclude in a shorter time frame, without any energy loss, and
without any deviation from the geodesic trajectory (i.e., zero curvature).
Furthermore, from the standpoint of entanglement analysis, optimal evolutions
seem to occur with lower average path entanglement values, higher average
entanglement speed, and ultimately, a greater degree of nonlocal entangling
character in their respective unitary time propagators. Similar observations
are apparent when comparing time optimal (Eq. (\ref{prop3})) and time
suboptimal (Eq. (\ref{fire})) evolutions between separable and maximally
entangled \emph{orthogonal} states (second scenario). However, several
significant differences must be emphasized. Firstly, in this second scenario,
only the time optimal evolution takes place within a two-dimensional subspace
of the complete four-dimensional Hilbert space. In contrast, the time
suboptimal evolution occurs throughout the entire four-dimensional space.
Secondly, unlike the first scenario, in the second scenario, the nonlocal
entangling character of the unitary time propagator associated with the time
suboptimal Hamiltonian is more pronounced than that which characterizes the
time optimal evolution. Finally, when analyzing the time suboptimal evolutions
in both the first and second scenarios, it is observed that the time
suboptimal evolution between orthogonal states occurs along longer
trajectories characterized by lower curvature and increased energy resource
expenditure, whereas the time suboptimal evolution between nonorthogonal
states transpires along shorter trajectories with higher curvature and
diminished energy resource expenditure. These findings concerning the first
part of our study are presented in Section V and are partially summarized in
Table III and Fig. $1$.

\medskip

It is clear that all unitary operators within the same equivalence class
demonstrate an equivalent degree of entangling power. However, the converse is
not necessarily true. In fact, we provided specific examples where the unitary
time propagators (Eqs. (\ref{u1c}) and (\ref{w1})) do not belong to the same
equivalence class, yet these operators are associated with quantum evolutions
that produce identical levels of entanglement production and entangling power.
Furthermore, we found that time-optimal evolutions that are also energetically
efficient (Eq. (\ref{w1})) do not have to achieve entangling power values as
high as those associated with energetically inefficient (Eq. (\ref{u2d}))
time-optimal evolutions to attain maximally entangled target states within the
same timeframe. The results of this second part of our study are detailed in
Section VI and are partially encapsulated in Fig. $2$. In the interest of
thoroughness, we wish to emphasize that the formulation of the time suboptimal
Hamiltonian presented in Eq. (\ref{R9}) constitutes another important finding
disclosed in this paper, which builds upon our earlier results documented in
Ref. \cite{carlocqg23}.

\subsection{Limitations and future investigations}

Although we thoroughly examined points [i] (Eq. (\ref{R9})) and [ii] (Eq.
(\ref{fire})) as discussed in the Introduction, our findings yielded insights
rather than a comprehensive understanding regarding points [iii]-[vi]
(Sections V and VI).

Our work is subject to three main limitations: (a) it is restricted to
stationary Hamiltonian evolutions, (b) it focuses on particular choices of
initial and final states, and (c) it evaluates entangling power only for
unitary time propagators that are already expressed in their canonical form.
This limitation to canonical forms is mainly for simplicity of analytical
calculations. Indeed, in addition to the remarks presented at the end of
Section II, we note that the Weyl chamber coordinates of an arbitrary
\textrm{SU}$\left(  4\right)  $ two-qubit gate (or, more generally, of any
unitary time propagator originating from an arbitrary nonstationary two-qubit
Hamiltonian) may also be determined via the so-called KAK decomposition
\cite{glaser01}. In brief, the KAK decomposition proceeds through the
following steps: i) Normalization of the unitary operator $U$ so to arrive at
$\tilde{U}=U/\left[  \det\left(  U\right)  \right]  ^{1/4}$;\textbf{ }ii)
Finding the matrix representation of $\tilde{U}$ in the magic basis,
$\tilde{U}\rightarrow\tilde{U}_{B}\overset{\text{def}}{=}Q^{\dagger}\tilde
{U}Q$, where $Q$ is the magic basis matrix; iii) Finding the eigenvalues
$\left\{  \lambda_{i}\right\}  $ of the unitary and symmetric\textbf{ }matrix
$M\overset{\text{def}}{=}\tilde{U}_{B}^{\intercal}\tilde{U}_{B}$, with
\textquotedblleft$\intercal$\textquotedblright\ denoting the transpose
operation; iv) Extracting the Weyl chamber coordinates $\left\{
c_{i}\right\}  $ from the phases $\left\{  \mu_{i}\right\}  $, with $\mu
_{i}\propto\arg\left(  \lambda_{i}\right)  $ that emerge from the eigenvalues
of\textbf{ }$M$. A detailed treatment of this approach is deferred to future
work. That said, although the present analysis is explicitly carried out for
two-qubit systems governed by stationary Hamiltonians, it can, in principle,
be extended to higher-dimensional settings, including qudits and multipartite
quantum systems with dynamics driven by nonstationary Hamiltonians. The key
notions introduced here- such as geodesic and speed efficiencies, the
curvature coefficient, and measures of entanglement production and entangling
power- admit natural generalizations to these broader contexts. Indeed, the
concepts of geodesic efficiency, speed efficiency, and the curvature
coefficient of a quantum evolution can be applied to any pure quantum state
space endowed with a Fubini-Study metric. While it is formally possible to
extend these notions to mixed quantum states, such generalizations are not
unique due to the freedom in choosing positive-definite Riemannian metric
structures on the space of mixed states \cite{karol06}. Furthermore, the
concept of concurrence in the Wootters original formulation is applicable to
both pure and mixed states, but only in two-qubit systems. For qudit or
multiqubit systems, one must consider generalized forms of concurrence
\cite{rungta01,li22} or alternative measures such as logarithmic negativity or
entanglement entropy \cite{horodecki09}, although the latter is restricted to
bipartitions and does not apply to mixed states. Finally, while Yukalov's
entanglement production can be applied to pure qudit and multiqubit systems as
well as to open quantum systems in mixed states, Zanardi's entangling power is
defined only for bipartite $(d_{1}\times d_{2})$-dimensional quantum systems
undergoing unitary evolution. Consequently, in its original form, it cannot be
applied to open quantum systems and is restricted to bipartitions of closed
systems. For notable attempts to extend the notion of entangling power to
qudit and multiqubit systems, as well as to nonunitary quantum evolutions, we
refer the reader to Refs. \cite{wang03,serrano25,kong15}, respectively.

\smallskip

In general, we anticipate that as the geometry of Hilbert space becomes
increasingly intricate with a growing number of qubits, the interplay among
efficiency, curvature, and entanglement in quantum evolution will become even
more nuanced than what is observed in the present study. Exploring these
deeper connections, while systematically addressing the current limitations,
forms an important direction for our future work. Nevertheless, despite these
constraints, we believe that the results presented here are sufficiently
significant to facilitate further geometric explorations of quantum evolutions
in higher-dimensional systems, where efficiency, curvature, and entanglement
are crucial for the quantum-mechanical manipulation of these physical systems
\cite{hete23,cc23,cc24,dd21}.

\begin{acknowledgments}
C.C. is grateful to the Griffiss Institute (Rome-NY) and to the United States
Air Force Research Laboratory (AFRL) Visiting Faculty Research Program (VFRP)
for providing support for this work. J.S. acknowledges support from the AFRL.
The authors wish to convey their appreciation for the insightful discussions
conducted with P. M. Alsing. Any opinions, findings and conclusions or
recommendations expressed in this material are those of the authors and do not
necessarily reflect the views of the AFRL. The authors thank two anonymous
referees for very useful comments leading to an improved version of this manuscript.
\end{acknowledgments}


\begin{thebibliography}{99}                                                                                               %


\bibitem {peres91}A. Peres and W. K. Wootters, \emph{Optical detection of
quantum information}, Phys. Rev. Lett. \textbf{66}, 1119 (1991).

\bibitem {bennett99}C. H. Bennett, D. P. DiVincenzo, Ch. A. Fuchs, T. Mor, E.
Rains, P. W. Shor, J. A. Smolin, and W. K. Wootters, \emph{Quantum nonlocality
without entanglement}, Phys. Rev. \textbf{A59}, 1070 (1999).

\bibitem {halder19}S. Halder, M. Banik, S. Agrawal, and S. Bandyopadhyay,
\emph{Strong quantum nonlocality without entanglement}, Phys. Rev. Lett.
\textbf{122}, 040403 (2019).

\bibitem {banik20}S. Bhattacharya, S. Saha, T. Guha, and M. Banik,
\emph{Nonlocality without entanglement: Quantum theory and beyond}, Phys. Rev.
Research \textbf{2}, 012068(R) (2020).

\bibitem {bell64}J. S. Bell, \emph{On the Einstein Podolsky Rosen paradox},
Physics \textbf{1}, 195 (1964).

\bibitem {clauser69}J. F. Clauser, M. A. Horne, A.\ Shimony, and R. A. Holt,
\emph{Proposed experiment to test local hidden variable theories}, Phys. Rev.
Lett. \textbf{23}, 880 (1969).

\bibitem {aspect82}A. Aspect, P. Grangier, and G. Roger, \emph{Experimental
realization of Einstein-Podoslky-Rosen-Bohm Gedankenexperiment: A new
violation of Bell's inequalities}, Phys. Rev. Lett. \textbf{49}, 91 (1982).

\bibitem {cafaroijtp}C. Cafaro, Ch. Corda, P. Cairns, and A. Bingolbali,
\emph{Violation of Bell's inequality in the Clauser-Horne-Shimony Holt form
with entangled quantum states revisited}, Int. J. Theor. Phys. \textbf{63},
112 (2024).

\bibitem {collins01}D. Collins, N. Linden, and S. Popescu, \emph{Nonlocal
content of quantum operations}, Phys. Rev. \textbf{A64}, 032302 (2001).

\bibitem {dur01}W. Dur, G. Vidal, J. I.\ Cirac, N. Linden, and S. Popescu,
\emph{Entanglement capabilities of nonlocal Hamiltonians}, Phys. Rev. Lett.
\textbf{87,} 137901 (2001).

\bibitem {maccone03}V. Giovannetti, S. lloyd, and L. Maccone, \emph{The role
of entanglement in dynamical evolution}, Europhys. Lett. \textbf{62}, 615 (2003).

\bibitem {maccone03b}V. Giovannetti, S. lloyd, and L. Maccone, \emph{Quantum
limits to dynamical evolution}, Phys. Rev. \textbf{A67}, 052109 (2003).

\bibitem {maccone04}V. Giovannetti, S. lloyd, and L. Maccone, \emph{The speed
limit of quantum unitary evolution}, J. Opt. B: Quantum Semiclass. Opt.
\textbf{6}, S807 (2004).

\bibitem {batle05}J. Batle, M. Casas, A. Plastino, and A. R. Plastino,
\emph{Connection between entanglement and the speed of quantum evolution},
Phys. Rev. \textbf{A72}, 032337 (2005).

\bibitem {batle06}J. Batle, M. Casas, A. Plastino, and A. R. Plastino,
\emph{Erratum}: \emph{Connection between entanglement and the speed of quantum
evolution}, Phys. Rev. \textbf{A73}, 049904(E) (2006).

\bibitem {borras06}A. Borras, M. Casas, A. R. Plastino, and A. Plastino,
\emph{Entanglement and the lower bounds on the speed of quantum evolution},
Phys. Rev. \textbf{A74}, 022326 (2006).

\bibitem {curilef06}S. Curilef, C. Zander, and A. R. Plastino, \emph{Two
particles in a double well: Illustrating the connection between entanglement
and the speed of quantum evolution}, Eur. J.\ Phys. \textbf{27}, 1193 (2006).

\bibitem {zander07}C. Zander, A. R. Plastino, A. Plastino, and M. Casas,
\emph{Entanglement and the speed of evolution of multi-partite quantum
systems}, J. Phys. A: Math. Theor. \textbf{40}, 2861 (2007).

\bibitem {curilef08}S. Curilef, C. Zander, and A. R. Plastino, \emph{Speed of
quantum evolution of entangled two qubit states: Local vs. global evolution},
J. Phys.: Conf. Ser. \textbf{134}, 012003 (2008).

\bibitem {chau10}H. F. Chau, \emph{Comment on} \textquotedblleft%
\emph{Connection between entanglement and the speed of quantum evolution}%
\textquotedblright, Phys. Rev. \textbf{A82}, 056301 (2010).

\bibitem {reznik08}J. Kupferman and B. Reznik, \emph{Entanglement and the
speed of evolution in mixed states}, Phys. Rev. \textbf{A78}, 042305 (2008).

\bibitem {frowis12}F. Frowis, \emph{Kind of entanglement that speeds up
quantum evolution}, Phys. Rev. \textbf{A85}, 052127 (2012).

\bibitem {rudniki21}L. Rudnicki, \emph{Quantum speed limit and geometric
measure of entanglement}, Phys. Rev. \textbf{A104}, 032417 (2021).

\bibitem {pati22}D. Shrimali, S. Bhowmick, V. Pandey, and A. K. Pati,
\emph{Capacity of entanglement for a nonlocal Hamiltonian}, Phys. Rev.
\textbf{A106}, 042419 (2022).

\bibitem {pati23}V. Pandey, D. Shrimali, B. Mohan, S. Das, and A. K. Pati,
\emph{Speed limits on correlations in bipartite quantum systems}, Phys. Rev.
\textbf{A107}, 052419 (2023).

\bibitem {pandey24}V. Pandey, S. Bhowmick, B. Mohan, Sohail, and U. Sen,
\emph{Fundamental speed limits on entanglement dynamics of bipartite quantum
systems}, Phys. Rev. \textbf{A110}, 052420 (2024).

\bibitem {deb16}P. Deb, \emph{Geometry of quantum state space and quantum
correlations}, Quantum Inf. Process. \textbf{15}, 1629 (2016).

\bibitem {deb19}P. Bej and P. Deb, \emph{Geometry of quantum state space and
entanglement}, Quantum Inf. Process. \textbf{18}, 72 (2019).

\bibitem {luo03}S. Luo, \emph{Wigner-Yanase skew information and uncertainty
relations}, Phys. Rev. Lett. \textbf{91}, 180403 (2003).

\bibitem {kuzmak19}A. M. Frydryszak, M. Gieysztor, and A. Kuzmak,
\emph{Probing the geometry of two-qubit state space by evolution}, Quantum
Inf. Process. \textbf{18}, 84 (2019).

\bibitem {saleem25}Z. H. Saleem et \textit{al}., \emph{Quantum Fisher
information and the curvature of entanglement}, arXiv:quant-ph/2504.13729 (2025).

\bibitem {anandan90}J. Anandan and Y. Aharonov, \emph{Geometry of quantum
evolution}, Phys. Rev. Lett. \textbf{65}, 1697 (1990).

\bibitem {carlini06}A. Carlini, A. Hosoya, T. Koike, and Y. Okudaira,
\emph{Time-optimal quantum evolution}, Phys. Rev. Lett. \textbf{96}, 060503 (2006).

\bibitem {carlini07}A. Carlini, A. Hosoya, T. Koike, and Y. Okudaira,
\emph{Time-optimal unitary operations}, Phys. Rev. \textbf{A75}, 042308 (2007).

\bibitem {carlocqg23}C. Cafaro and P. M. Alsing, \emph{Qubit geodesics on the
Bloch sphere from optimal-speed Hamiltonian evolutions}, Class. Quantum Grav.
\textbf{40}, 115005 (2023).

\bibitem {rossetti24}L. Rossetti, C. Cafaro, and N. Bahreyni,
\emph{Constructions of optimal-speed quantum evolutions: A comparative study},
Physica Scripta \textbf{99}, 095121 (2024).

\bibitem {uzdin12}R. Uzdin, U. G\"{u}nther, S. Rahav, and N. Moiseyev,
\emph{Time-dependent Hamiltonians with 100\% evolution speed efficiency}, J.
Phys. A: Math. Theor. \textbf{45}, 415304 (2012).

\bibitem {rossetti25}L. Rossetti, C. Cafaro, and P. M. Alsing,
\emph{Deviations from geodesic evolutions and energy waste on the Bloch
sphere}, Phys. Rev. \textbf{A111}, 022441 (2025).

\bibitem {alsing24A}P. M.\ Alsing and C. Cafaro, \emph{From the classical
Frenet--Serret apparatus to the curvature and torsion of quantum-mechanical
evolutions. Part I. Stationary Hamiltonians}, Int. J. Geom. Methods Mod. Phys.
\textbf{21}, 2450152 (2024).

\bibitem {alsing24B}P. M.\ Alsing and C. Cafaro, \emph{From the classical
Frenet--Serret apparatus to the curvature and torsion of quantum-mechanical
evolutions. Part II. Nonstationary Hamiltonians}, Int. J. Geom. Methods Mod.
Phys. \textbf{21}, 2450151 (2024).

\bibitem {cafaropra25}C. Cafaro, L. Rossetti, and P. M. Alsing,
\emph{Curvature of quantum evolutions for qubits in time-dependent magnetic
fields}, Phys. Rev. \textbf{A111}, 012408 (2025).

\bibitem {wootters98}W. K. Wootters, \emph{Entanglement of formation of an
arbitrary state of two qubits}, Phys. Rev. Lett. \textbf{80}, 2245 (1998).

\bibitem {shimony95}A. Shimony, \emph{Degree of entanglement}, Annals of the
New York Academy of Sciences \textbf{755}, 675 (1995).

\bibitem {yukalov15}V. I. Yukalov and E. P. Yukalova, \emph{Evolutional
entanglement production}, Phys. Rev. \textbf{A92}, 052121 (2015).

\bibitem {zanardi00}P. Zanardi, C. Zalka, and L. Faoro, \emph{Entangling power
of quantum evolutions}, Phys. Rev. \textbf{A62}, 030301(R) (2000).

\bibitem {wootters97}S. A. Hill and W. K. Wootters, \emph{Entanglement of a
pair of quantum bits}, Phys. Rev. Lett. \textbf{78}, 5022 (1997).

\bibitem {wootters01}W. K. Wootters, \emph{Entanglement of formation and
concurrence}, Quantum Information and Computation \textbf{1}, 27 (2001).

\bibitem {collins11}C. P. Williams, \emph{Explorations in Quantum Computing},
Springer-Verlag London (2011).

\bibitem {caves94}S. L. Braunstein and C. M. Caves, \emph{Statistical distance
and the geometry of quantum states}, Phys. Rev. Lett. \textbf{72}, 3439 (1994).

\bibitem {gold03}T.-C. Wei and P. M. Goldbart, \emph{Geometric measure of
entanglement and applications to bipartite and multipartite quantum states},
Phys. Rev. \textbf{A68}, 042307 (2003).

\bibitem {yukalov03}V. I. Yukalov, \emph{Entanglement measure for composite
systems}, Phys. Rev. Lett. \textbf{90}, 167905 (2003).

\bibitem {yukalov03b}V. I. Yukalov, \emph{Quantifying entanglement production
of quantum operators}, Phys. Rev. \textbf{A68}, 022109 (2003).

\bibitem {yukalov17}V. I. Yukalov and E. P. Yukalova, \emph{Entanglement
production by evolution operator}, Journal of Physics: Conf. Series
\textbf{826}, 012021 (2017).

\bibitem {coleman00}A. J. Coleman and V. I. Yukalov,\emph{\ Reduced Density
Matrices}, Springer-Verlag (2000).

\bibitem {yukalov02}V. I. Yukalov, \emph{Matrix order indices in statistical
mechanics}, Physica \textbf{A310}, 413 (2002).

\bibitem {witte99}C. Witte and M. Trucks, \emph{A new entanglement measure
induced by the Hilbert-Schmidt norm}, Phys. Lett. \textbf{A257}, 14 (1999).

\bibitem {ovrum06}J. Magne Leinaas, J. Myrheim, and E. Ovrum,
\emph{Geometrical aspects of entanglement}, Phys. Rev. \textbf{A74}, 012313 (2006).

\bibitem {yuka20}V. I. Yukalov, \emph{Order indices and entanglement
production in quantum systems}, Entropy \textbf{22}, 565 (2020).

\bibitem {pandya20}P. Pandya, O. Sakarya, and M. Wiesniak,
\emph{Hilbert-Schmidt distance and entanglement witnessing}, Phys. Rev.
\textbf{A102}, 012409 (2020).

\bibitem {siewert22}J. Siewert, \emph{On orthogonal bases in the
Hilbert-Schmidt space of matrices}, J. Phys. Commun. \textbf{6}, 055014 (2022).

\bibitem {zanardi01}P. Zanardi, \emph{Entanglement of quantum evolutions},
Phys. Rev. \textbf{A63}, 040304(R) (2001).

\bibitem {wang02}X. Wang and P.\ Zanardi, \emph{Quantum entanglement of
unitary operators on bi-partite systems}, Phys. Rev. \textbf{A66}, 044303 (2002).

\bibitem {reza04}A. T. Rezakhani, \emph{Characterization of two-qubit perfect
entanglers}, Phys. Rev. \textbf{A70}, 052313 (2004).

\bibitem {bala10}S. Balakrishnan and R. Sankaranarayanan, \emph{Entangling
power and local invariants of two-qubit gates}, Phys. Rev. \textbf{A82},
034301 (2010).

\bibitem {mike98}M.\ A. Nielsen, \emph{Quantum Information Theory}, Ph.D.
thesis, University of New Mexico,\ Albuquerque (1998).

\bibitem {stefano}C. Cafaro and S. Mancini, \emph{A geometric algebra
perspective on quantum computational gates and universality in quantum
computing}, Advances in Applied Clifford Algebras \textbf{21}, 493 (2011).

\bibitem {cafaro20}C. Cafaro, S. Ray, and P. M. Alsing, \emph{Geometric
aspects of analog quantum search evolutions}, Phys. Rev. \textbf{A102, }052607 (2020).

\bibitem {MT}L. Mandelstam and Ig. Tamm, \emph{The uncertainty relation
between energy and time in non-relativistic quantum mechanics}, J. Phys. USSR
\textbf{9}, 249-254 (1945).

\bibitem {alsing24C}P. M. Alsing and C. Cafaro, \emph{Upper limit on the
acceleration of a quantum evolution in projective Hilbert space}, Int. J.
Geom. Methods Mod. Phys. \textbf{21}, 2440009 (2024).

\bibitem {samuel88}J. Samuel and R. Bhandari, \emph{General setting for
Berry's phase}, Phys. Rev. Lett. \textbf{60}, 2339 (1988).

\bibitem {paulPRA23}P. M. Alsing, C. Cafaro, O. Luongo, C. Lupo, S. Mancini,
and H. Quevedo, \emph{Comparing metrics for mixed quantum states: Sj\"{o}qvist
and Bures}, Phys. Rev. \textbf{A107}, 052411 (2023).

\bibitem {ali09}A. Mostafazadeh, \emph{Hamiltonians generating optimal-speed
evolutions}, Phys. Rev. \textbf{A79}, 014101 (2009).

\bibitem {brody03}D. C. Brody, \emph{Elementary derivation for passage times},
J. Phys.: Math. Gen. \textbf{36}, 5587 (2003).

\bibitem {KC}B. Kraus and J. I. Cirac, \emph{Optimal creation of entanglement
using a two-qubit gate}, Phys. Rev. \textbf{A63}, 062309 (2001).

\bibitem {james26}C. Cafaro and J. Schneeloch, \emph{Symmetry-based
perspectives on Hamiltonian quantum search algorithms and Schr\"{o}dinger's
dynamics between orthogonal states}, Symmetry \textbf{18}, 422 (2026).

\bibitem {glaser01}N. Khaneja and S. Glaser, \emph{Cartan decomposition of}
\textrm{SU}$\left(  2^{n}\right)  $ \emph{and control of spin systems},
Chemical Physics \textbf{267}, 11 (2001).

\bibitem {karol06}I. Bengtsson and K. Zyczkowski, \emph{Geometry of Quantum
States}, Cambridge University Press (2006).

\bibitem {rungta01}P. Rungta, V. Buzek, C.\ M. Caves, M. Hillery, and G. J.
Milburn, \emph{Universal state inversion and concurrence in arbitrary
dimensions}, Phys. Rev. \textbf{A64}, 042315 (2001).

\bibitem {li22}Y. Li and J. Shang, \emph{Geometric mean of bipartite
concurrences as a genuine multipartite entanglement measure}, Phys. Rev.
Research \textbf{4}, 023059 (2022).

\bibitem {horodecki09}R. Horodecki, P. Horodecki, M. Horodecki, and K.
Horodecki, \emph{Quantum entanglement}, Rev. Mod. Phys. \textbf{81}, 865 (2009).

\bibitem {wang03}X. Wang, B. C. Sanders, and D. W. Berry, \emph{Entangling
power and operator entanglement in qudit systems}, Phys. Rev. \textbf{A67},
042323 (2003).

\bibitem {serrano25}E. Serrano-Ensastiga, D. M. Galindo, J.\ A. Maytorena, and
C. Chryssomalakos, \emph{Entangling power for symmetric multiqubit systems: A
geometrical approach}, Ann. Phys. \textbf{481}, 170143 (2025).

\bibitem {kong15}F.-Z. Kong, J.-L. Zhao, M. Yang, and Z.-L. Cao,
\emph{Entangling power and operator entanglement of nonunitary quantum
evolutions}, Phys. Rev. \textbf{A92}, 012127 (2015).

\bibitem {hete23}B. Hetenyi and P. Levay, \emph{Fluctuations, uncertainty
relations, and the geometry of quantum state manifolds}, Phys. Rev.
\textbf{A108}, 032218 (2023).

\bibitem {cc23}C. Chryssomalakos et \textit{al}., \emph{Curves in quantum
state space, geometric phases, and the brachistophase}, J. Phys. A: Math.
Theor. \textbf{56}, 285301 (2023).

\bibitem {cc24}C. Chryssomalakos et \textit{al}., \emph{Speed axcess and total
acceleration: A kinematical approach to entanglement}, Phys. Scr. \textbf{99},
125116 (2024).

\bibitem {dd21}D. D'Alessandro,\emph{ Introduction to Quantum Control and
Dynamics}, Chapman and Hall/CRC (2021).
\end{thebibliography}
\end{document}